\documentclass[11pt]{article}

\usepackage[dvips]{graphicx}
\usepackage{epsfig,amsmath,amssymb,verbatim,mathrsfs,array,layout,textcomp,latexsym, slashed,graphicx,bbm,physics,tensor,cite}

\usepackage[normalem]{ulem}
\usepackage{appendix}
\usepackage{enumitem}
\usepackage{rotating}
\usepackage[hidelinks]{hyperref}
\usepackage[usenames,dvipsnames]{color} 

\usepackage{soul}
\usepackage{dsfont}

\usepackage{tikz}
\usepackage{tikz-feynman}
\tikzset{
	cross/.style={path picture={\draw[black]
			(path picture bounding box.south east) -- (path picture bounding box.north west)
			(path picture bounding box.south west) -- (path picture bounding box.north east);}}
}

\allowdisplaybreaks

\makeatletter

\newenvironment{bottompar}{\par\vspace*{\fill}}{\clearpage}
\newcommand*{\xdash}[1][3em]{\rule[0.5ex]{#1}{0.55pt}}
\def\mysection#1{{\bf #1.} }

\def\mysection#1{{\bf #1.}}

\newcommand\mydots{\hbox to 1.1em{$\,\cdot\hss\cdot\hss\cdot\,$}}

\begin{document}

\begin{titlepage}
	\setcounter{page}{1} \baselineskip=15.5pt 
	\thispagestyle{empty}
	$\quad$
	{\raggedleft IFT UAM-CSIC 25-168\par}
	\vskip 60 pt
	
	\begin{center}
		{\fontsize{18}{18} \bf Finite parts of inflationary loops II:\\ \vspace{0.4cm} A streamlined UV in-in algorithm \\  \vspace{0.2cm} and distinguishable signatures}
	\end{center}

	\vskip 20pt
	\begin{center}
		\noindent
		{\fontsize{13}{30}\selectfont  Guillermo Ballesteros$^{1,2}$, Jes\'us Gamb\'in Egea$^{2}$, and Flavio Riccardi$^{3}$}
	\end{center}

	\begin{center}
		\vskip 4pt
		\textit{ $^1${Departamento de F\'{\i}sica Te\'{o}rica, Universidad Aut\'{o}noma de Madrid (UAM), \\Campus de Cantoblanco, 28049 Madrid, Spain}
		}
		\vskip 5pt
		\textit{ $^2${Instituto de F\'{\i}sica Te\'{o}rica UAM-CSIC,  Campus de Cantoblanco, 28049 Madrid, Spain}
		}
		\vskip 5pt
		\textit{ $^{3}${Dipartimento di Fisica, Sapienza Università di Roma, Piazzale Aldo Moro 5, 00185, Roma, Italy}
		}
	\end{center}
	
	\vspace{0.4cm}
	\centerline{\bf Abstract}
	\vspace{0.3cm}
	\noindent 
	
	We introduce a streamlined method for evaluating in-in loop integrals using dimensional regularization for diagrams with an arbitrary number of external legs and vertices, which complements earlier work and facilitates the extraction of the ultraviolet contributions. The method leads us to identify an apparent difficulty to renormalize with Hamiltonian counterterms within the in-in formalism. We also discuss the importance of the finite parts of loop corrections that can be distinguished from their associated counterterm contributions. As an application, we examine the one-loop primordial bispectrum in the context of the effective field theory of inflation, considering a specific set of interactions, and identifying a contribution distinguishable from its tree-level counterpart.
		
	\begin{bottompar}
		\noindent\xdash[15em]\\
		\small{
			guillermo.ballesteros@uam.es\\
			j.gambin@csic.es\\
			flavio.riccardi@uniroma1.it}
	\end{bottompar}
\end{titlepage}

\setcounter{page}{2}
\newpage

\tableofcontents

\section{Introduction} \label{sec: Intro}

Computing ultraviolet (UV) loop corrections to cosmological correlators using the in-in formalism has long been considered notably difficult. In \cite{Ballesteros:2024cef} we put forward a method for solving in-in loop integrals in dimensional regularization. The method, which is a direct application of dimensional regularization, splits momentum integrals into UV and infrared (IR) pieces and attacks the computation of the UV part by expanding the integrand in the large-momentum limit, which allows to identify the single poles {at} $\delta=0$ in $3+\delta$ spatial dimensions. In \cite{Ballesteros:2024cef}, we illustrated the application of the method for single-field inflation by computing, in the de Sitter limit, the one-loop power spectrum of tensor modes generated by scalar fluctuations running in the loop; obtaining for the first time the finite contribution coming from the loop at any time. It has also been recently used in \cite{Braglia:2025cee,Braglia:2025qrb,Kristiano:2025ajj} to explore scalar correlators. 

\subsection{Simplifying the calculation of UV contributions to in-in correlators}

Despite its convenience, the method (as introduced in \cite{Ballesteros:2024cef}) is limited by the complication, inherent to the in-in formalism, of computing time integrals along the Schwinger-Keldysh contour \cite{Schwinger:1960qe,Bakshi:1962dv,Bakshi:1963bn,Keldysh:1964ud,Weinberg:2005vy}. This can seem particularly involved if the modes running in the loop do not have a simple time dependence \cite{Weinberg:2010wq}.
In this paper, we push further the program of simplifying the full computation of loop level in-in correlators by noticing that the structure of the inflationary modes in the large-momentum limit can be exploited to decouple momentum and time integrals and, also, to trade (at least partially) time integrals {for} time derivatives (which are much simpler to compute). This simplifies the method proposed in \cite{Ballesteros:2024cef}, making it possible to tackle harder problems. 

In this paper we apply this (improved) method for cosmological correlators to explore the one-loop correction to the primordial scalar bispectrum of curvature fluctuations. The motivation for choosing this quantity is two-fold. Firstly, we are interested in {\it distinguishable loop effects}, meaning corrections that, given their functional form, can be distinguished from counterterm contributions.
As we will now explain, in general the power spectrum is not necessarily sufficient for finding distinguishable loop corrections. Secondly, the bispectrum allows us to provide a concrete example of a difficulty that can arise trying to renormalize loop contributions with two or more insertions of the interaction Hamiltonian. 

\subsection{Distinguishable loop effects}

A central question in the context of effective field theories is whether loop corrections to a given observable encode genuinely new, intrinsic effects, or whether they can be fully captured by suitable counterterms, in which case explicit loop computations are, strictly speaking, not required.\footnote{Unfortunately, in practice it may not be obvious if a specific observable features distinguishable loop corrections without computing them.} By counterterms we mean local operators that belong to the action of an effective field theory, and whose coefficients are adjusted to cancel the UV divergences of loop diagrams. The diagrams obtained using these operators define the corresponding counterterm contributions to a given observable.

Throughout this work we focus on One-Particle Irreducible (1PI) one-loop diagrams\footnote{The discussion of non-1PI diagrams follows cutting them.} and on the associated tree-level, one-insertion counterterm contributions that renormalize them. We call a loop contribution {\it distinguishable} if its momentum dependence cannot be reproduced by any choice of the finite parts of the counterterm coefficients, and indistinguishable otherwise. Since loop and counterterm contributions enter at the same perturbative order, we are interested in knowing whether an observable computed at that order receives a distinguishable or an indistinguishable loop correction.\footnote{It may occur that non-divergent loop corrections to an observable arise at a certain order in perturbation theory as, generically, increasing the number of vertices increases the degree of convergence. In that case, our definition of distinguishable loop contributions may appear to lose its meaning, as no counterterms are needed for renormalization. However, we can always compare those loop corrections with the tree-level ones that appear at the same order, and the definition holds its value.} In the indistinguishable case, the renormalized loop can be fully mimicked by tree-level counterterm operators: {its effect is entirely {\it scheme-dependent} and it} carries no additional physical information.\footnote{The arbitrariness in the finite parts of the counterterm coefficients is removed by imposing as many experimental ({\it renormalization}) conditions, at selected values of the external momenta, as there are free parameters in the observable under consideration. Once fixed, the counterterm coefficients are uniquely determined for all observables to which the corresponding operators contribute. Therefore, no scheme-dependent contribution remains after renormalization. This does not affect the notion of a distinguishable loop effect, which by definition cannot be reproduced by tree-level counterterm insertions.}

To make this more concrete, let us consider an observable $\mathcal{O}$ --typically a primordial correlator-- computed at one loop: 
\begin{equation}
	\mathcal{O}(t,\{k_i\}) = g\,\mathcal{O}^{\rm tree}(t,\{k_i\}) + g^2\,\mathcal{O}^{\rm 1\text{-}loop}(t,\{k_i\}) + g^2\,\mathcal{O}^{\rm cts}(t,\{k_i\})+\mathcal{O}(g^3) \,,
\end{equation}
where $\mathcal{O}^{\rm tree}$ is the tree-level contribution, $\mathcal{O}^{\rm 1\text{-}loop}$ the one-loop contribution, and $\mathcal{O}^{\rm cts}$ the contribution from (tree-level) counterterm diagrams. The coupling $g\ll1$ (which we have pulled out of the different contributions for convenience) organizes the perturbative expansion. If the loop is indistinguishable, there exists a choice of the finite parts of the counterterm coefficients such that
$\mathcal{O}^{\rm 1\text{-}loop}(t,\{k_i\}) + \mathcal{O}^{\rm cts}(t,\{k_i\}) = 0$ for all $\{k_i\}$ (the momenta external to the loop), so that the loop carries no new
physical information with respect to a higher-order tree-level computation. This is because in an effective field theory --and using dimensional regularization, which respects the symmetries of the action-- the counterterms are operators that (already) belong in the effective theory.

Let us consider for illustrative purposes the case of a primordial (either scalar or tensor) power spectrum in the de Sitter limit. In that case, the quantity 
\begin{equation}
	\mathcal{P} - \mathcal{P}^{\rm tree} = \mathcal{P}^{\rm loop} + \mathcal{P}^{\rm cts} 
\end{equation}
is, in general, not zero. If the effect of the loops, $\mathcal{P}^{\rm loop}$, is indistinguishable from that of the counterterms $\mathcal{P}^{\rm cts}$ (i.e.\ they share the same scale dependence), the freedom in the coefficients of the latter prevents {that physical information is extracted} from the loop. This role of the counterterms has sometimes been overlooked in the literature, with attempts to read off properties of the theory from finite (but scheme-dependent) loop results, leading to erroneous conclusions. See e.g.\ \cite{Braglia:2025cee}, and also \cite{Kristiano:2022maq}, which we will discuss below. 

In \cite{Ballesteros:2024cef} we computed the late-time one-loop (plus counterterms) correction to the primordial spectrum of tensor modes in the de Sitter limit with scalar fluctuations of a canonical scalar field running in the loop. We found this one-loop correction to be of the form $\mathcal{P}_h(k)\propto (H/M_P)^4\log(H/\mu)$ (where $\mu$ is the renormalization scale) plus a (scheme-dependent) constant piece. In this particular case, this contribution is featureless, in the sense of being scale-independent, i.e.\ independent of the comoving momentum $k$. The one-loop contribution is therefore indistinguishable from the tree-level counterterm contribution, $\mathcal{P}_h^{\rm cts} \propto (H/ M_P)^4$.\footnote{We point out that, in the de Sitter limit, the loop and counterterm contributions cannot be separated either from the tree-level one at leading order, which is $(H/ (\pi M_P))^2$.} This result occurs thanks to the absence of late-time divergences. If there were late-time divergences we could have a scale-dependent power spectrum of the form $\log (k/\mu a(\tau))$, being $\tau$ conformal time and $a(\tau)$ the scale factor of the Universe (which goes to infinity in the late-time limit). Besides, there are other quantities with dimensions of energy (such as $H$ and the reduced Planck mass $M_P$) which could seemingly lead to terms like e.g.\ $\log(k/H)$, inducing a scale dependence \cite{Weinberg:2005vy}. The fact that this does not happen is due to a symmetry of the problem \cite{Senatore:2009cf}, as we explain now. 

The background FLRW metric, 
\begin{equation} \label{FRWM}
	\dd s^2 = a^2(\tau)(-\dd \tau^2 +  \dd \vb{x}^2) \,,
\end{equation}
is invariant under the scale redefinition
\begin{align} \label{scalings}
		a \to \lambda \, a \,, \quad \vb{x} \to \lambda^{-1}\, \vb{x} \quad {\rm and}\quad \tau \to \lambda^{-1}\, \tau \,.
\end{align}
This symmetry strongly constrains the type of loop contributions that can arise in inflation. We distinguish between physical energy scales (which are invariant under this transformation), and comoving ones. The latter transform according to: $E \to \lambda \, E$. An observable $\mathcal{O}(\tau, k)$ (such as the dimensionless power spectrum) must be invariant under these transformations. Both $\tau^{-1}$ and $k$ behave as comoving scales, while $H$, $M_P$ and $\mu$ are physical energy scales. In the absence of additional comoving scales, the invariance of $\mathcal{O}(\tau, k)$ under Eq.\ \eqref{scalings} implies that it can only be a function of the product $k \tau$. In the limit $\tau \to 0$, there are only two options: either $\mathcal{O}(\tau, k)$ is constant (in time) or it exhibits a late-time divergence \cite{Burgess:2009bs,Chen:2016nrs,Gorbenko:2019rza}. Therefore, in the absence of late-time divergences, it must be constant \cite{Ballesteros:2024cef,Senatore:2009cf,Braglia:2025cee}. These conclusions apply order by order in perturbation theory, so they extend separately to tree-level, and to loop and counterterm contributions. They are also robust under the introduction of other physical scales, such as field masses \cite{Chen:2016nrs,Firouzjahi:2024tyt,Cespedes:2025dnq}, for instance. In these situations with no late-time divergences and no additional comoving scales, the effects of the loops are expected to be indistinguishable from those of the counterterms. In turn, this implies that the loop corrections cannot have relevant observational implications (as all the physics is contained in tree-level contributions, including counterterms).

As we have mentioned above, in \cite{Ballesteros:2024cef} we computed the one-loop contribution to the dimensionless power spectrum of tensor modes (with scalar modes running in the loop) in the de Sitter limit, finding it to be scale invariant in the late-time limit. Moreover, in \cite{Ballesteros:2024zdp} the same property for the late-time, large-scale, power spectrum of scalar modes (again, with scalars inside the loop) was also found in a more complicated ultra slow-roll scenario. This analysis was motivated by the work presented in \cite{Kristiano:2022maq}, which originally contained a claim of a significant loop effect at large scales from (UV modes running in) the loop in a somewhat less sophisticated ultra slow-roll toy model. Such a claim appeared to contradict the basic notion of decoupling between low and high energies and lead to a burst of activity around the topic. In reality, the one-loop and tree-level (including counterterms) contributions to the scalar power spectrum must be indistinguishable at large scales in models of that kind. The renormalized one-loop contribution at large scales cannot be measurably larger than the tree-level one, as it was incorrectly initially claimed in \cite{Kristiano:2022maq}.
We also think that it is not exactly zero in a scheme-independent way, as it appears to have been argued later in other works, see for instance \cite{Fumagalli:2023zzl, Tada:2023rgp, Inomata:2024lud,Fumagalli:2024jzz,Inomata:2025pqa,Inomata:2025bqw}.\footnote{In fact, some of these studies draw conclusions from a formally divergent, unrenormalized loop, omitting counterterms.} This discussion highlights the importance of clarifying the circumstances under which non-trivial loop effects may arise. 

One could think that, in the absence of late-time divergences, an extra comoving scale, $k_e$, would be needed to obtain a (scale-dependent) distinguishable one-loop power spectrum, that would be a function of the ratio $k/k_e$, such as e.g.\ $\sim\log(k/k_e)$. While this idea sounds superficially reasonable in spirit, in practice it is not guaranteed that a single comoving scale can lead to a distinguishable one-loop contribution to the primordial power spectrum (either for scalars or tensors). 
The reason is that before shouting scale-dependence one needs to verify that the effect that is sought after cannot be generated by tree-level counterterms. To illustrate this point, let us first consider the effect of the background metric evolution including slow roll corrections, which will generically produce scale-dependent effects on the tree-level power spectrum that mimic the ones that might be expected at loop-level. Let us consider a small deviation from de Sitter parametrized by a single number, the slow-roll parameter $\epsilon = - H'/(a \,H^2)$, where the prime denotes differentiation with respect to conformal time. The Hubble function can be expanded in this case around a constant value as $H(\tau)=H_*(1+\epsilon \log (-k_*\tau)+\mathcal{O}(\epsilon^2))$. At tree-level this will lead to a power spectrum $\mathcal{P}\propto (k/k_*)^\alpha = 1+\alpha\log(k/k_*)+\mathcal{O}(\epsilon^2)$, where $\alpha$ is some constant linearly dependent on $\epsilon$. {At one loop} level, the same kind of correction could be expected because, working perturbatively in $\epsilon$, the only source of $k_*$ in the loop is through the combination $\epsilon \log (k / k_*)$, see \cite{delRio:2018vrj} for a related discussion.
Therefore, even if it is true that introducing an extra comoving scale may lead to a scale-dependent loop, this alone is not sufficient to ensure that the loop cannot be mimicked by counterterms.

This argument is sufficient to rule out any naive contribution of the type $\log(k/k_e)$ from loops as a distinguishable part, since its form coincides with that of slow-roll corrections at tree-level. Furthermore, it emphasizes the importance of doing a proper treatment of the effect of counterterms in order to extract intrinsic information from the loop.
In Appendix \ref{app: Power spectrum} we show with a concrete example how this situation can occur. There, we consider the effective theory of inflation \cite{Cheung:2007st} in the decoupling limit, where the background remains purely de Sitter. In this case, an auxiliary comoving scale can be introduced via the time dependence of the couplings of suppressed operators. This preserves the simplicity of the de Sitter mode functions while, one could naively think, possibly allowing for a distinguishable one-loop contribution to the two-point function. We will focus on the operator $M_3^4(\tau)(\delta g^{00})^3$ in the unitary gauge, where $\delta g^{00}$ is the fluctuation of the time-time component of the inverse metric and we will expand $M_3^4(\tau)$ around a value of $\tau$ whose reciprocal plays the role of the extra comoving scale. This mechanism alone is insufficient to generate a truly non-trivial running. 

A possible origin for distinguishable loop effects arises instead naturally for correlators that depend on more than one external comoving scale, the prime example of which is the primordial bispectrum. Indeed, an observable $\mathcal{O}(\tau, \{\vb{k}_i\})$ can feature a distinguishable one-loop effect, even in the absence of late-time divergences. Such an observable can only depend on $\vb{k}_i \tau$ and $k_i / k_j$ and in the limit $\tau \to 0$ it is either independent of $\tau$ (and therefore a function of $k_i / k_j$ and constant in time), or it exhibits a late-time divergence. To illustrate the appearance of a distinguishable loop effect in the late-time de Sitter limit, we will consider in Section \ref{sec: One-Loop Bispectrum} the primordial scalar bispectrum in the context of the effective theory of inflation with $M_3^4(\tau)(\delta g^{00})^3$, this time for constant $M_3^4$, which is sufficient for our purposes. 

We also stress that there can be situations in which the loop correction is indistinguishable from the counterterms, while the leading order tree-level contribution can be discriminated from them. An example is given by a constant tree-level power spectrum and scale-dependent one-loop and counterterm contributions that are indistinguishable from each other. The key point here is the different scale dependence between the lowest order (at tree-level) and the higher order contributions (one-loop and counterterms). A situation such as this could, in principle, be of phenomenological interest (provided that the observational sensitivity were good enough to access the higher order corrections).\footnote{If both the leading tree-level contribution and the combined one-loop and counterterm correction share the same momentum dependence, one might naively wonder whether the theory may lose predictivity and the Hubble scale $H$ or the slow-roll parameters could become unmeasurable. Nevertheless, assuming that perturbation theory is valid, the observable is dominated by the tree-level contribution, so predictivity at leading order is preserved, while the magnitude of the loop correction can still be consistently estimated.} See Appendix~\ref{app: Power spectrum} for a concrete case in the de Sitter limit.

\subsection{Renormalization of in-in correlators with several vertices}

Let us also mention another motivation for considering the one-loop scalar bispectrum. As we will discuss later in Section \ref{sec: Renorm Problem}, the method we propose to compute the UV part of cosmological correlators allows us to identify an apparent difficulty to renormalize correlators with more than one insertion of the interaction Hamiltonian. In essence, these correlators contain terms whose field structure is substantially different from the one of the possible Hamiltonian tree-level counterterms. The bispectrum calculation that we do in detail in Section \ref{sec: One-Loop Bispectrum} will help us to illustrate the issue with a concrete example. We hope to overcome this difficulty in a future work. 

\subsection{Summary and structure of the paper}

The paper focuses on three points. The first one is the presentation of a useful simplification of the method we introduced in \cite{Ballesteros:2024cef} for applying dimensional regularization to compute loop-level in-in correlators. The second point is a discussion of the conditions under which such loop corrections can be relevant, in the sense of being distinguishable from counterterm effects. The third and final point draws attention to a seemingly general difficulty to renormalize multi-point loop correlators. 

In Section \ref{sec: Dim Reg} we give a brief review of the method of \cite{Ballesteros:2024cef} and then discuss how to improve it using the UV mode structure. Then, in Section \ref{sec: Renorm Problem} we use the results of Section \ref{sec: Dim Reg} to highlight the difficulty we find to renormalize multi-vertex loops. In Section \ref{sec: model} we present the model we use to illustrate the main points of the paper. In Section \ref{sec: One-Loop Bispectrum} we proceed to explore the one-loop bispectrum of that model to show the appearance of distinguishable loop effects and the renormalization issue. We present our conclusions in Section \ref{sec: Discussion}. The paper contains three appendices. The first one, Appendix \ref{app: HI} gives some general details about the interaction Hamiltonian in the interaction picture. Appendix \ref{app: Bispectrum formulas} provides a set of various lengthy formulae that are needed to write our result for the one-loop contribution to the bispectrum. Finally, Appendix \ref{app: Power spectrum} considers the one-loop scalar power spectrum for the model of Section \ref{sec: model}, showing that, even adding an additional comoving scale, no distinguishable loop effects are possible for this observable. 

\section{Dimensional regularization: computing the UV contribution} \label{sec: Dim Reg}

We start this section summarizing the essence of the method we will employ to evaluate loop integrals regularized via dimensional regularization, which preserves the symmetries of the theory.\footnote{Except for symmetries that are related to the dimensionality of the problem \cite{Siegel:1979wq,Siegel:1980qs,Capper:1979ns}.} See \cite{Bollini:1972bi,tHooft:1972tcz,Cicuta:1972jf,Ashmore:1972uj} for seminal references about dimensional regularization and e.g.\ \cite{Weinberg:2005vy,Senatore:2009cf,Chen:2016nrs} for earlier applications in the context of inflation. For further details about the method we discuss immediately below, we refer to our previous work \cite{Ballesteros:2024cef}.

The vacuum expectation value of a generic (time-dependent) Hermitian operator $\mathcal{O}(\tau)$ in the Heisenberg picture (composed of fields and their conjugate momenta and where $\tau$ denotes conformal time) can be obtained using the in-in formalism \cite{Weinberg:2005vy,Weinberg:2006ac,Chen:2010xka,Wang:2013zva}:
\begin{equation} \label{inin}
	\expval{\mathcal{O}(\tau)} = \bra{0} F^{-1}(\tau,-\infty_+) \mathcal{O}_I(\tau) F(\tau,-\infty_-) \ket{0} \eval_{\rm no \, bubbles} \,.
\end{equation}
In this expression, 
\begin{equation}
	F(\tau,-\infty_-) \equiv T\exp \Bigg\lbrace -i \int_{-\infty_-}^{\tau} \dd \tau' H_I(\tau') \Bigg\rbrace \,,
\end{equation}
where $T$ denotes time ordering and both $\mathcal{O}_I$ and the interaction Hamiltonian $H_I$ belong to the interaction picture (see Appendix \ref{app: HI}); that is, the fields that compose them evolve under the free Hamiltonian.
We employ the $i\epsilon$ prescription ($\tau_\pm \equiv \tau (1\pm i \epsilon)$), which guarantees the projection of the vacuum in the Heisenberg picture $\ket{\Omega}$ onto the interaction vacuum $\ket{0}$ in the limit $\tau \to -\infty$ \cite{Adshead:2009cb}.
As indicated in Eq.\ \eqref{inin}, bubble diagrams do not contribute to $\expval{\mathcal{O}(\tau)}$. This occurs by cancellation with the overlap $\braket{\Omega}{0}$, as demonstrated in \cite{Ballesteros:2024cef} (see also \cite{Senatore:2016aui} for a discussion about this). 

We can expand $\expval{\mathcal{O}(\tau)}$ perturbatively in powers of the interaction picture Hamiltonian, $H_I$, obtaining the following first three orders:
\begin{align} \label{eq: In-In rule 1H}
	\expval{\mathcal{O}(\tau)}^{(0)} =& \expval{\mathcal{O}_I(\tau)} \,, \quad
	\expval{\mathcal{O}(\tau)}^{(1)} = 2\Im{ \int_{-\infty_-}^\tau \dd \tau' \expval{\mathcal{O}_I(\tau) H_I(\tau')}} \,,\\ \nonumber
	\expval{\mathcal{O}(\tau)}^{(2)} =& \int_{-\infty_+}^\tau \dd \tau' \int_{-\infty_-}^\tau \dd \tau'' \expval{H_I(\tau') \mathcal{O}_I(\tau) H_I(\tau'')} \\ \label{eq: In-In rule 2H}
	&- 2\Re{ \int_{-\infty_-}^\tau \dd \tau' \int_{-\infty_-}^{\tau'} \dd \tau'' \expval{ \mathcal{O}_I(\tau) H_I(\tau') H_I(\tau'')}}\,. 
\end{align}
These rules allow to calculate correlation functions at one loop, which involves computing momentum integrals that are often divergent in the ultraviolet (UV). In dimensional regularization, the momentum integrals that arise in this context are typically of the following form:
\begin{equation}
	I(\delta) = \int _0^\infty \dd p \, p^\delta f(\delta,p)\,,
\end{equation}
where the function $f(\delta,p)$ encodes all model-dependent information, including interaction couplings, the dynamics of both external and internal modes to the loop, and the structure of time integrals associated with loop diagrams. The factor $p^\delta$ originates from the Fourier volume element in $3+\delta$ spatial dimensions and serves as the regulator that ensures UV convergence. 

The method we presented in \cite{Ballesteros:2024cef} to compute this kind of integral is a direct application of dimensional regularization based on the following decomposition into an infrared (IR) and a UV part:
\begin{equation} \label{decomp}
	I(\delta) =  \int _0^L \dd p \, f(0,p) + \int _L^\infty \dd p \, p^\delta f(\delta,p)+\mathcal{O}(\delta)\,,
\end{equation}
where  $L$ is an arbitrarily large comoving scale. More specifically, this comoving scale has to be chosen to be much larger than any other comoving scale of the problem.  By construction, the infrared contribution over the domain $p\in [0,L]$ is finite in the absence of IR divergences (which we will assume).\footnote{The generalization to the case with IR divergences is straightforward, once again cutting the loop integral on an arbitrarily small comoving scale $L_{\rm IR}$ and using the freedom in $\delta$ to impose the convergence in the limit $p\to 0$. See \cite{Braglia:2025cee} for an example.} Therefore, we have set $\delta = 0$ in the first addend of Eq.~\eqref{decomp} neglecting $\order{ \delta}$ corrections, as no regularization is required for that part of the integral. In contrast, the UV contribution over $p\in [L,\infty)$ needs to be regularized and thus has to be computed with $\delta \neq 0$. Therefore, obtaining the UV contribution to $I(\delta)$ requires knowing the dynamics of the fields (as well as the dependence on the couplings) in $3+\delta$ spatial dimensions. 

The idea used in \cite{Ballesteros:2024cef} to compute the UV contribution to $I(\delta)$ consists in expanding $f(\delta,p)$ around $p = \infty$. 
For the clarity of the argument, let us consider the case where $f(\delta,p)$ can be expanded in a Laurent series around $p = \infty$. Assuming that infinity is not an essential singularity, the series expansion stops at a given power $p^N$, corresponding to the maximal divergence of the integrand:\footnote{This UV behavior of the integrand $f(\delta,p)$ is to be expected in inflationary scenarios due to the UV behavior of the modes running into the loop (see Eq.\ \ref{wkbexp}). However, the procedure presented for calculating in dimensional regularization is general and applies to more exotic UV behaviors (see \cite{Ballesteros:2024cef} for details).}
\begin{equation}
	f(\delta,p) = \sum_{n = -\infty}^N p^n \, c_n(\delta) \,,
\end{equation}
where $c_n(\delta)$ are dimensionful coefficients which may depend on the external momenta, and which are analytic in $\delta = 0$.
After that, and integrating in $p$ assuming that $\delta$ is such that UV convergence is ensured (effectively regularizing the integral) the result is analytically continued in the complex plane including $\delta = 0$. The limit $\delta \rightarrow 0$ is finally taken to obtain the (regularized) value of the integral in three spatial dimensions:
\begin{equation} \label{eq: Dim Reg First Paper}
		I(\delta) = \lim_{L\to \infty} \left( \int_0^L \dd p \, f(0,p) - c_{-1}(0) \log L - \sum_{n = 0}^N \dfrac{L^{n+1}}{n+1} c_n(0) \right) - \dfrac{c_{-1}(0)}{\delta} - \dfrac{\dd c_{-1}(\delta)}{\dd \delta} \eval_{\delta = 0} + \order{\delta} \,.
\end{equation}
Importantly, this procedure preserves the independence of $I(\delta)$ on $L$, through the mutual cancellation of $L$-dependent IR and UV terms. In practice, the procedure can be further simplified by linearly expanding $f(\delta,p)$ around $\delta=0$, since $I(\delta)$ only has single poles at $\delta =0$ and the terms of $\mathcal{O}(\delta^2)$ (and higher) vanish in the limit $\delta\rightarrow 0$, see \cite{Ballesteros:2024cef}. 

The procedure we have just summarized focuses on the computation of the momentum integrals appearing in the in-in formalism but makes no mention of the time integrals, such as the ones in Eqs.~\eqref{eq: In-In rule 1H} and \eqref{eq: In-In rule 2H}. These may seem difficult to compute analytically, which could impede obtaining $\langle\mathcal{O}(\tau)\rangle$ at any order in the interaction Hamiltonian above zero. However, there is a way to deal with these time integrals, which allows to perform them in a straightforward way, significantly easing the calculation of renormalized cosmological correlators at loop level. The idea, which we introduce in the present work, exploits the fact that the auxiliary scale $L$ can be taken arbitrarily large. This allows to consider the Fourier modes that run within the loop (in the UV part of Eq.~\eqref{decomp}) in the high momentum limit, where they are simpler, complementing the method presented in \cite{Ballesteros:2024cef} and making it into a more efficient computational tool. 

\subsection{The high-momentum limit of the loops} \label{highm}

In order to describe the simplification of the time integrals that we intend to do, let us start by analyzing the large momentum limit for a real scalar field, $\phi(x)$, in an arbitrary number ($3+\delta$) of spatial dimensions. In the interaction picture, 
\begin{equation}
	\phi(x) = \int \dfrac{\dd^{3+\delta} \vb{k}}{(2\pi)^{(3+\delta)/2}} e^{i \vb{k}\cdot \vb{x}} \phi_{\vb{k}}(\tau) \quad {\rm where} \quad \phi_{\vb{k}}(\tau) = \phi_{k}(\tau) a_{\vb{k}}+ \phi_{k}^*(\tau) a^\dagger_{-\vb{k}} \quad {\rm with} \quad \comm{a_{\vb{k}}}{a^\dagger_{\vb{p}}} = \delta(\vb{k} - \vb{p})\,.
\end{equation}
From $\phi(x)$, we can define a canonically normalized field $u(x) \equiv c(\tau) \phi(x)$,\footnote{The Lagrangian for $u(x)$ will contain the term $ \dot u^2(x)/2$, so that this field satisfies Bunch-Davies initial conditions.} where $c(\tau)$ is a model dependent function of time. The Fourier counterpart of $u(x)$ satisfies the equation of motion determined by the free Lagrangian and with Bunch-Davies initial conditions: 
\begin{equation}
	u''_k(\tau) + \omega^2_k(\tau) u_k(\tau) = 0  \quad {\rm and} \quad u_k(\tau) \xrightarrow{\tau \to - \infty} \dfrac{e^{-i k \tau}}{\sqrt{2k}} \,,
\end{equation}	
where primes indicate derivatives with respect to $\tau$ (conformal time) and the frequency of the modes depends on an effective mass (squared), $m_{\rm eff}^2(\tau)$:
\begin{equation}
\omega^2_k(\tau) = k^2 + m_{\rm eff}^2(\tau) \,.
\end{equation}
We note that although from an EFT point of view $m_{\rm eff}$ could depend on $k$, in the in-in formalism such an effect can be considered as part of the interaction terms, leaving the free action unchanged. 

Bunch-Davies initial conditions guarantee that the usual commutation rules between field and conjugate momentum are satisfied.
Following the spirit of the WKB method \cite{1926ZPhy...38..518W,1926ZPhy...39..828K,Brillouin:1926blg}, in order to describe the dynamics of $u_k(\tau)$ in the limit $k \to \infty$ we can assume the {\it ansatz}:
\begin{equation}
	u_k(\tau) = \dfrac{1}{\sqrt{2 W_k(\tau)}}\left( A_k e^{-i \int^\tau W_k(\tau') \dd \tau'} + B_k e^{i \int^\tau W_k(\tau') \dd \tau'} \right) \,,
\end{equation}
translating the problem of solving $u_k(\tau)$ to the problem of finding $W_k(\tau)$. Using the equation of motion of $u_k(\tau)$, we obtain:
\begin{equation}
	W_k^2(\tau) =  k^2 + m_{\rm eff}^2(\tau) + \dfrac{3}{4} \left( \dfrac{W_k'(\tau)}{W_k(\tau)} \right) ^2 -\dfrac{1}{2} \dfrac{W_k''(\tau)}{W_k(\tau)}\,.
\end{equation}
This equation can be solved perturbatively in the limit $k\to \infty$, obtaining:
\begin{equation}
	W_k(\tau) = k + \dfrac{m_{\rm eff}^2(\tau)}{2 k } - \dfrac{m_{\rm eff}^4(\tau)+\left( m_{\rm eff}^2(\tau)\right) ''}{8k^3} + \order{k^{-4}} \,.
\end{equation}
For the perturbative description to be valid, the conformal time derivatives of the effective mass squared: $m_{\rm eff}^2(\tau)$ must satisfy $\abs{d^n\, m_{\rm eff}^2(\tau)/d \tau^n}\ll k^{n+2}$ for all $n$ and including $n=0$ (no derivative).  Otherwise, it may be impossible to describe $W_k(\tau)$ accurately truncating the series.
Imposing Bunch-Davies initial conditions, we find $A_k = 1$ and $B_k = 0$, and finally, returning to the original field,
\begin{equation} \label{wkbexp}
\phi_k(\tau) \xrightarrow{k \to \infty} \dfrac{e^{-i k \tau}}{\sqrt{k}} \sum _{n = 0}^\infty \dfrac{\phi_{(n)}(\tau)}{k^n} \,. 
\end{equation}
The functions $\phi_{(n)}(\tau)$ are model-dependent and can be calculated explicitly using the WKB approximation. However, their detailed form is not relevant to the discussion, as what really matters is the fact that they come associated to a single phase $\exp(-i k \tau)$. As we will soon see, this phase will allow us to simplify the time and momentum integrals in the UV one-loop contributions to $\langle\mathcal{O}(\tau)\rangle$, where $\mathcal{O}(\tau)$ is any operator composed by the field $\phi(x)$ and its conjugate momentum. 

Let us note that the expansion Eq.~\eqref{wkbexp} will not be valid, in general, when $m_{\rm eff}^2$ presents abrupt changes, e.g.\ being piece-wise defined or being a Dirac delta. In such (over)simplified scenarios, a mixture of phases in the UV limit can generically occur, deviating in a non-smooth way from Bunch-Davies. 

\subsection{One-loop diagrams with a single vertex insertion}
Let us now consider a diagram with a single (internal) propagator, which using Eq.~\eqref{eq: In-In rule 1H} we can write as follows:
\begin{equation} \label{singlevert}
	\begin{tikzpicture}[baseline={-10}]
		\draw (0,0) circle (0.25);
		\draw (-0.75,-0.75) -- (0,-0.25);
		\draw (0.75,-0.75) -- (0,-0.25);
		\draw (-0.25,-0.75) -- (0,-0.25);
		\fill (0,-0.6) circle (1pt);
		\fill (0.15,-0.6) circle (1pt);
		\fill (0.3,-0.6) circle (1pt);
	\end{tikzpicture} = 2 \Im{\int_{-\infty_-}^\tau \dd \tau' G(\tau,\tau';\{\vb{k}_i\};\delta) \int \dfrac{\dd^{3+\delta} \vb{p}}{(2\pi)^{3+\delta}} \,\phi_p(\tau') \, \phi_p^*(\tau') } \,.
\end{equation}
For simplicity, we have assumed that the interaction is non-derivative, since the generalization is straightforward.
The function $G(\tau,\tau';\{\vb{k}_i\};\delta)$, where $\{\vb{k}_i\}$ denotes the ensemble of all the external momenta, includes all the information related to the physics external to the loop.
In order to obtain the UV contribution to this loop integral, we can 
make an expansion inside the integral in $p \to \infty$ up to $\order{1/p}$. In this way, neglecting terms of $\order{1/p^2}$ in the integrand we will be making a vanishingly small error of  $\order{1/L}$ in the computation of the integral, see the discussion around Eq.~\eqref{decomp}, as well as Eq.~\eqref{eq: Dim Reg First Paper}.
We note that the phases $\exp(\pm i p \tau')$ coming from the modes inside the loop in the high momentum limit cancel out for this single-vertex diagram (which would also occur for derivative interactions). This fully decouples the temporal and momentum dependencies in the loop, allowing the momentum integral to be easily calculated, since the
integrand depends on $p$ through powers of it. 
Once the momentum integral is done, the temporal integral is maximally simplified.  

\subsection{One-loop diagrams with two vertices}

Let us now analyze the case of two propagators running in the loop. In Fourier space, the kind of diagram we are referring to is:
\begin{align} \label{eq:genericdiagram}
	\begin{tikzpicture}[baseline={-2}]
		\draw (-0.15,0.3+0.15) -- (0.5,0);
		\fill (0,0.15) circle (1pt);
		\fill (0,0) circle (1pt);
		\fill (0,-0.15) circle (1pt);
		\draw (-0.15,-0.3-0.15) -- (0.5,0);
		\draw (0.5+0.25,0) circle (0.25);
		\draw (1,0) -- (1.5+0.15,0.3+0.15);
		\fill (1.5,0.15) circle (1pt);
		\fill (1.5,0) circle (1pt);
		\fill (1.5,-0.15) circle (1pt);
		\draw (1,0) -- (1.5+0.15,-0.3-0.15);
	\end{tikzpicture} = \int & \dfrac{\dd^{3+\delta} \vb{p}}{(2\pi)^{3+\delta}} \bigg[ \mathcal{I}_1\left(\tau,{\vb{p}};\{\vb{k}_i\};\delta\right)+\mathcal{I}_2\left(\tau,{\vb{p}};\{\vb{k}_i\};\delta\right)\bigg]\,,
\end{align}
where
\begin{align} \label{int1}
\mathcal{I}_1\left(\tau,{\vb{p}};\{\vb{k}_i\};\delta\right)  & =  \int_{-\infty_+}^ \tau \dd \tau' \int_{-\infty_-}^ \tau \dd \tau'' \, G_1(\tau,\tau',\tau'';\{\vb{k}_i\};\delta) \, \phi_p(\tau') \phi_p^*(\tau'') \phi_q(\tau') \phi_q^*(\tau'')\,, \\  \label{int2}
\mathcal{I}_2\left(\tau,{\vb{p}};\{\vb{k}_i\};\delta\right)  & = - 2\Re{  \int_{-\infty_-}^ \tau \dd \tau' \int_{-\infty_-}^ {\tau'} \dd \tau'' \, G_2(\tau,\tau',\tau'';\{\vb{k}_i\};\delta) \, \phi_p(\tau') \phi_p^*(\tau'') \phi_q(\tau') \phi_q^*(\tau'')}
\end{align}
and $\{\vb{k}_i\}$ denotes the set of external momenta. We have two momenta in the loop: $\vb{p}$ and $\vb{q}$, the latter defined as $\vb{q} = \vb{k}_{\rm tot} - \vb{p}$, where $\vb{k}_{\rm tot}(\{\vb{k}_i\})$ is the total momentum entering one of the vertices. Again, we are assuming non-derivative interactions for simplicity, but the final result (Eq.\ \eqref{eq: Dim-Reg UV simp 2 vertex} below) is valid also for derivative interactions. 

\begin{figure}[t]
 \centering 
  \includegraphics[width=.85\textwidth]{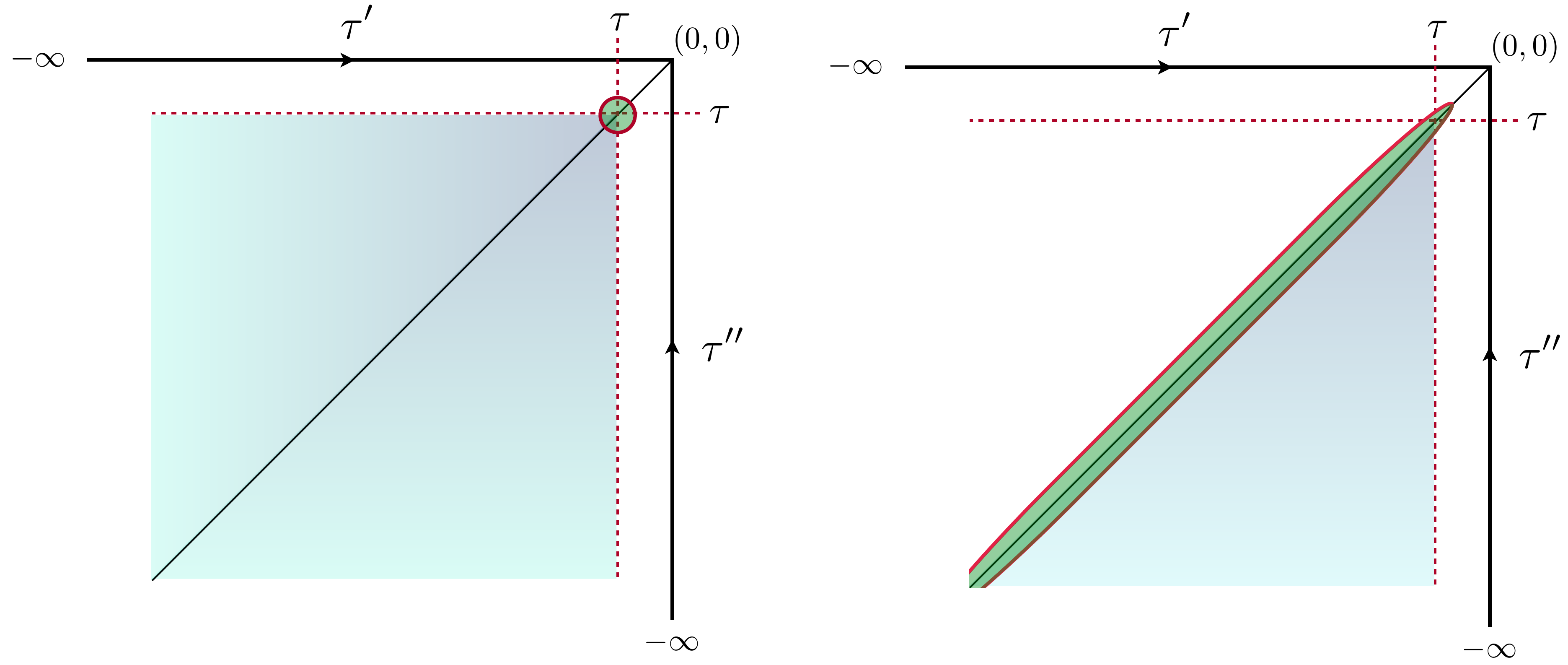}
 \caption{Integration domains (shaded regions) for the integrals Eq.~\eqref{int1} (left) and Eq.~\eqref{int2} (right). The regions around which the integrals are expressed as an expansion in time derivatives, Eq.\ \eqref{noremantint} $(\tau''=\tau'=\tau)$ and Eq.\ \eqref{remantint} $(\tau''=\tau')$ respectively, are indicated encircled.}
\label{fig:intregions}
\end{figure}

As we did for the single-vertex diagram, to extract the UV contribution to the loop integral in Eq.~\eqref{eq:genericdiagram}, we take the large momentum limit $p \to \infty$ inside the integral, up to $\order{1/p}$ (now with $q = \abs{\vb{q}} = p + \order{p^0}$, so $q$ is also considered a large momentum), making a vanishingly small $\order{1/L}$ error in the integral. We can take this limit  before solving the time integrals since, as we are going to show, the time and momentum integrals also decouple for this kind of diagram, i.e.\ the time integrals do not worsen loop convergence.

Let us start by expressing for convenience the integrands as follows, defining new functions $\mathcal{G}_i$: 
\begin{equation} \label{phase}
	G_i(\tau,\tau',\tau'';\{\vb{k}_i\};\delta) \, \phi_p(\tau') \phi_p^*(\tau'') \phi_q(\tau') \phi_q^*(\tau'') = e^{-i(p+q)(\tau'-\tau'')} \mathcal{G}_i(\tau,\tau',\tau'';\{\vb{k}_i\},{\vb{p}};\delta) \,.
\end{equation}
Given Eq.~\eqref{wkbexp}, the functions $\mathcal{G}_i(\tau,\tau',\tau'';\{\vb{k}_i\},{\vb{p}};\delta)$ can be written in the large-$p$ limit as a sum in powers of $p$ (with a maximum $p^N$ given by the degree of divergence of the loop), where the coefficients of this expansion depend on time.
This contributes to help solving the integrals, but a further simplification is still needed to deal with the overall phase multiplying each function $\mathcal{G}_i$. The role played by these phases --and hence the key to simplify their treatment-- can be understood examining the effect of the $i\epsilon$ prescription. 

If in the (large negative time limit of the) double integral $\mathcal{I}_1$ of Eq.~\eqref{int1}, defined over the region $\tau',\tau'' \in (-\infty,\tau]$ (see the left panel of Fig.\ \ref{fig:intregions}), we take into account the role of the $i \epsilon$ prescription in the phase, we notice that 
\begin{equation}
	e^{-i(p+q)(\tau'_+-\tau''_-)} = e^{(p+q)(\tau'+\tau'') \epsilon} e^{-i(p+q)(\tau'-\tau'')}
\end{equation}
presents a damping that causes the UV part of the integral to vanish, see Eq.\ \eqref{decomp}. 
However, this is not true in the upper limit of the time integrals, $\tau' = \tau'' = \tau$, where, since $\tau$ is real, there is no damping (the phase in Eq.~\eqref{phase} is equal to 1) and this contribution may contain UV divergences. 

Similarly, in the double integral $\mathcal{I}_2$ of Eq.\ \eqref{int2}, defined over $\tau'' \in (-\infty,\tau']$ and $\tau' \in (-\infty,\tau]$ (see the right panel of Fig.\ \ref{fig:intregions}), taking into account the role of the $i \epsilon$ prescription in the phase in the early time limit:
\begin{equation}
	e^{-i(p+q)(\tau'_--\tau''_-)} = e^{-(p+q)(\tau'-\tau'') \epsilon} e^{-i(p+q)(\tau'-\tau'')} \,,
\end{equation}
we observe that the damping is effective in all the integration domain except along the line $\tau' = \tau''$. 

We can now proceed to write the time integrals, expanding their integrands over the time regions where the damping is not present, getting
\begin{align} \label{noremantint}
	&\mathcal{I}_1\left(\tau,{\vb{p}};\{\vb{k}_i\};\delta\right)  = \sum_{{n,m} = 0}^{{\infty}} \partial^n_{\tau''} \partial^m_{\tau'}\, \mathcal{G}_1(\tau,\tau',\tau'';\{\vb{k}_i\},{\vb{p}};\delta) \eval_{\tau' = \tau'' = \tau} \dfrac{{i^{n-m}}}{(p+q)^{2+n+m}} \,,\\ \label{remantint}
	&\mathcal{I}_2\left(\tau,{\vb{p}};\{\vb{k}_i\};\delta\right)  = -2 {\rm Re} \left\{ \int_{-\infty_-}^ \tau \dd \tau' \sum_{n = 0}^\infty \partial^n_{\tau''}\,  \mathcal{G}_2(\tau,\tau',\tau'';\{\vb{k}_i\},{\vb{p}};\delta) \eval_{\tau'' = \tau'} \dfrac{{i^{n-1}}}{(p+q)^{1+n}}\right\}\,.
\end{align}

A priori, Eqs.~(\ref{noremantint}) and (\ref{remantint}), which are general and applicable for any value of $\vb{p}$, do not present any advantage because solving them requires performing an infinite sum. However, in the UV part of the loop integral,
\begin{align} \label{eq: Dim-Reg UV simp 2 vertex}
\eval{\begin{tikzpicture}[baseline={-2}]
		\draw (-0.15,0.3+0.15) -- (0.5,0);
		\fill (0,0.15) circle (1pt);
		\fill (0,0) circle (1pt);
		\fill (0,-0.15) circle (1pt);
		\draw (-0.15,-0.3-0.15) -- (0.5,0);
		\draw (0.5+0.25,0) circle (0.25);
		\draw (1,0) -- (1.5+0.15,0.3+0.15);
		\fill (1.5,0.15) circle (1pt);
		\fill (1.5,0) circle (1pt);
		\fill (1.5,-0.15) circle (1pt);
		\draw (1,0) -- (1.5+0.15,-0.3-0.15);
	\end{tikzpicture}}_{\rm UV} = \int\limits_{{\{p>L\}}} \dfrac{\dd^{3+\delta} \vb{p}}{(2\pi)^{3+\delta}} &  \Bigg[\mathcal{I}_1\left(\tau,{\vb{p}};\{\vb{k}_i\};\delta\right)\big{|}_{\rm UV}+ \mathcal{I}_2\left(\tau,{\vb{p}};\{\vb{k}_i\};\delta\right)\big{|}_{\rm UV}\Bigg]\,,
\end{align}
a remarkable simplification is observed. The time derivatives of the functions $\mathcal{G}_{i}$, in the large-$p$ limit, do not increase the degree of divergence because time and momentum have been decoupled, as we have already discussed. This, together with the fact that for the UV calculation it is only necessary to know the asymptotic behavior of the integrand up to $\order{1/p}$, means that in practice it is sufficient to include only a finite number of terms in $n$ and $m$ --up to $\order{1/p}$-- to obtain the full contribution to the UV part of the loop integral.
To finish the calculation, we only have to do the remaining time integral, as well as the momentum integral in~$\vb{p}$. However, since the relevant momentum dependence is given by a finite sum of powers of $p$, solving the momentum integral is straightforward.

Let us emphasize that the previous discussion is based on: 1) the UV dependence of the modes, which allowed us to factorize the phase that mixes time and momentum in the loop; and 2) the effect of the $i\epsilon$ prescription. The first of these two points is the reason why the discussion extends to the case with derivative interactions and, indeed,  Eqs.\ (\ref{noremantint}) and (\ref{remantint}) are also valid in that case. As we already mentioned at the end of Section \ref{highm}, this result may not apply in simplified scenarios where the dynamics of the modes exhibit abrupt transitions that modify the UV behavior.

\subsection{One-loop diagrams with more than two vertices}

Extending the previous results to the case with more than two propagators running in the loop is straightforward. The procedure is always the same: 
\begin{enumerate}
	\item Write the loop contribution using the in-in rules.
	\item Take the large momentum limit $p \to \infty$ (keeping up to $\order{1/p}$) in the integrand and extract the phase relative to the different momenta running in the loop.
	\item Using the $i \epsilon$ prescription, analyze the time integration regime to extract the locus of the integration domain that contributes to the momentum integral in the UV regime.
	\item Expand the integrand around that region and integrate in time if possible. 
\end{enumerate}
This procedure will simplify the integrand to its minimal form, factorizing the momentum and time integrals and making the momentum integral easy to evaluate.

\section{On the structure of tree- and loop-level contributions} \label{sec: Renorm Problem}

Loop-level UV divergences are usually tamed with a renormalization process. In particular, renormalization is possible if the divergent contribution of loops to an observable (a cosmological correlator in our case) is indistinguishable from the one of tree-level counterterms, whose couplings are to be determined via experiments or observations. That is, the UV divergent effect of loops is equivalent to that of tree-level contact terms when renormalization is possible, which happens if the coefficients of the counterterms contain a divergent part that exactly cancels out the divergences coming from the loops. This indistinguishability between UV divergences and counterterms makes both contributions inseparable from a physical (experimental or observational) point of view. 

Given an $n$-point, 1PI correlation {at one loop}, such as those analyzed in the previous section, the UV divergences are expected to be renormalized by a counterterm diagram of the type 
\begin{equation}
	\begin{tikzpicture}[baseline={-10}]
		\draw (-0.75,-0.75) -- (0,-0.25);
		\draw (0.75,-0.75) -- (0,-0.25);
		\draw (-0.25,-0.75) -- (0,-0.25);
		\fill (0,-0.6) circle (1pt);
		\fill (0.15,-0.6) circle (1pt);
		\fill (0.3,-0.6) circle (1pt);
		\draw[fill=white,cross] (0,-0.25) circle (0.15);
	\end{tikzpicture}
\end{equation}
To be specific, let us consider the $n$-point correlation of the field $\phi(x)$ associated with a given general counterterm interaction in Fourier space:
\begin{equation} \label{eq: general ct}
	\begin{tikzpicture}[baseline={-12}]
		\draw (-0.75,-0.75) -- (0,-0.25);
		\draw (0.75,-0.75) -- (0,-0.25);
		\draw (-0.25,-0.75) -- (0,-0.25);
		\fill (0,-0.6) circle (1pt);
		\fill (0.15,-0.6) circle (1pt);
		\fill (0.3,-0.6) circle (1pt);
		\draw[fill=white,cross] (0,-0.25) circle (0.15);
	\end{tikzpicture} = 2 \Im \left\lbrace \phi_{k_1}(\tau) \cdots \phi_{k_n}(\tau) \int_{-\infty_-}^\tau \dd \tau' \, c(\tau',\{\vb{k}_i\}) \, \phi_{k_1}^{*(a_1)}(\tau') \cdots \phi_{k_n}^{*(a_n)}(\tau') \right\rbrace \,,
\end{equation}
where we are using the notation $\phi_{k}^{*(a)}(\tau) \equiv \partial_\tau^a \phi_k^*(\tau)$. The function $c(\tau',\{\vb{k}_i\})$, which contains coefficients that are undetermined and that we use to absorb the UV divergences of the loops, encodes both the counterterm coupling and the spatial derivatives of the interaction (in Fourier space).

Before doing any calculation, we can already hint at a potential problem in the renormalization of in-in correlators, coming from the fact that, in principle, any UV divergence must be reproducible by the type of counterterm contribution presented in Eq.~(\ref{eq: general ct}), or a combination thereof. Whereas there are loop contributions that have a structure similar to that of Eq.~(\ref{eq: general ct}), which helps towards their renormalization;  other contributions have a very different structure, which raises questions about their renormalizability.

Let us analyze the one-loop diagrams with a single insertion of $H_I(\tau)$, of the form Eq.~(\ref{singlevert}). Since the loop integral does not depend on the external momentum, it can be reduced to a divergent, time-dependent coefficient, 
\begin{equation} 
		\begin{tikzpicture}[baseline={-10}]
			\draw (0,0) circle (0.25);
			\draw (-0.75,-0.75) -- (0,-0.25);
			\draw (0.75,-0.75) -- (0,-0.25);
			\draw (-0.25,-0.75) -- (0,-0.25);
			\fill (0,-0.6) circle (1pt);
			\fill (0.15,-0.6) circle (1pt);
			\fill (0.3,-0.6) circle (1pt);
		\end{tikzpicture} = 2 \Im \left\lbrace \phi_{k_1}(\tau) \cdots \phi_{k_n}(\tau) \int_{-\infty_-}^\tau \dd \tau' \, c_{\rm eff}(\tau',\{\vb{k}_i\}) \, \phi_{k_1}^{*(a_1)}(\tau') \cdots \phi_{k_n}^{*(a_n)}(\tau') \right\rbrace \,.
\end{equation}
The function $c_{\rm eff}(\tau',\{\vb{k}_i\})$ contains both the integral over momentum (which may contain UV divergences) and the coupling of the interaction under analysis. This diagram has precisely the same structure as the counterterm diagram in Eq.~(\ref{eq: general ct}).

The case with two insertions of $H_I$, see Eq.~(\ref{eq:genericdiagram}), is more complicated because the modes running inside the loop are sensitive to the external comoving momenta $\vb{k}_i$, so  that the effect of the loop cannot be reduced to a function of time. Using the procedure developed in the previous section to simplify the UV part of this one-loop diagram, we can analyze its structure and compare it to that of the counterterms.
In particular, the UV analysis of this diagram allows us to rewrite it in a simplified form, see Eqs.~(\ref{noremantint}) and~(\ref{remantint}). As discussed in the previous section, a complete calculation requires computing the integral over momentum in Eq.\ (\ref{eq: Dim-Reg UV simp 2 vertex}), which is model-dependent but is always easy to do since the time and momentum integrals have been decoupled. Let us write the contributions of Eq.~(\ref{noremantint}) and Eq.~(\ref{remantint}) in a specific case in which $H_I$ contains no derivative interactions:
\begin{align} \label{eq: UV type I}
	\nonumber
	& \int_{\rm UV} \dfrac{\dd^{3+\delta} {\vb{p}}}{(2 \pi)^{3+\delta}} \mathcal{I}_1\left(\tau,p;\{\vb{k}_i\};\delta\right)\big{|}_{\rm UV}  = \phi^*_{k_1}(\tau) \cdots \phi^*_{k_i}(\tau) \, \phi_{k_{i+1}}(\tau) \cdots \phi_{k_n}(\tau) 	\\ & \quad \times   \sum_{{N,M} = 0}^\infty
	\partial^N_{\tau''} \partial^M_{\tau'}\, \left( \phi_{k_1}(\tau') \cdots \phi_{k_i}(\tau') \, \phi_{k_{i+1}}^*(\tau'') \cdots \phi^*_{k_n}(\tau'') \, c_{\rm eff}^1(\tau',\tau'';\{\vb{k}_i\};N,M) \right)  \eval_{\tau' = \tau'' = \tau} + {\rm perms.}\,,\\ \nonumber
	& \int_{\rm UV} \dfrac{\dd^{3+\delta} {\vb{p}}}{(2 \pi)^{3+\delta}} \mathcal{I}_2\left(\tau,p;\{\vb{k}_i\};\delta\right)\big{|}_{\rm UV}  = -2 \Re \bigg\lbrace \phi_{k_1}(\tau) \cdots \phi_{k_n}(\tau) \int_{-\infty_-}^ \tau \dd \tau' \\ \label{eq: UV type II}
	& \quad \times \sum_{N = 0}^\infty\partial^N_{\tau''}\left(  \phi_{k_1}^*(\tau') \cdots \phi_{k_{i}}^*(\tau') \, \phi_{k_{i+1}}^*(\tau'') \cdots \phi_{k_n}^*(\tau'') \, c_{\rm eff}^2(\tau',\tau'';\{\vb{k}_i\};N)\right)  \eval_{\tau'' = \tau'} \bigg\rbrace + {\rm perms.}\,,
\end{align}
where we have used the rules of the in-in formalism in Eq.~(\ref{eq: In-In rule 2H}) to construct the functions $\mathcal{G}_i$ of Eqs.~(\ref{noremantint}) and~(\ref{remantint}). We have encoded the effect of the momentum integral and the fields running inside the loop in the functions $c_{\rm eff}^i$, which can be complex.
For simplicity, we are considering a single channel in which the set of external fields $\phi_{\vb{k}_j}(\tau)$ with $j \in \{1, \dots, i\}$ are coupled to the insertion $H_I(\tau')$, and those with $j \in \{i+1, \dots, n\}$ are coupled to the insertion $H_I(\tau'')$, see Eq.~(\ref{eq: In-In rule 2H}): 
\begin{align}
	\begin{tikzpicture}[baseline={-2}]
		\draw (-0.15,0.3+0.15) -- (0.5,0);
		\fill (0,0.15) circle (1pt);
		\fill (0,0) circle (1pt);
		\fill (0,-0.15) circle (1pt);
		\draw (-0.15,-0.3-0.15) -- (0.5,0);
		\draw (0.5+0.25,0) circle (0.25);
		\draw (1,0) -- (1.5+0.15,0.3+0.15);
		\fill (1.5,0.15) circle (1pt);
		\fill (1.5,0) circle (1pt);
		\fill (1.5,-0.15) circle (1pt);
		\draw (1,0) -- (1.5+0.15,-0.3-0.15);
		\node[left]  at (-0.15-0.1,0.3+0.15)   {$k_1$};
		\node[left]  at (-0.15-0.1,-0.3-0.15)  {$k_i$};
		\node[right] at (1.5+0.15+0.1,0.3+0.15) {$k_{i+1}$};
		\node[right] at (1.5+0.15+0.1,-0.3-0.15) {$k_n$};
		\node[above] at (0.5,0.1) {$\tau'$};
		\node[above] at (1.1,0.1) {$\tau''$};
	\end{tikzpicture}\,.
\end{align}
The rest of the configurations are contained within the permutations.

Let us discuss Eqs.~(\ref{eq: UV type I}) and (\ref{eq: UV type II}). Eq.~(\ref{eq: UV type I}) comes from  $\expval{\mathcal{O}(\tau)} \supset \int \dd \tau' \int \dd \tau'' \expval{H_I(\tau') \mathcal{O}_I(\tau) H_I(\tau'')}$ in (the first line of) Eq.~(\ref{eq: In-In rule 2H}). Therefore, the fields coming from $H_I(\tau')$ appear unconjugated, while the external fields to which it couples appear conjugated; and the opposite happens with $H_I(\tau'')$. 
Instead, Eq.~(\ref{eq: UV type II}) comes from $\expval{\mathcal{O}(\tau)} \supset-2 \Re \int \dd \tau' \int \dd \tau'' \expval{ \mathcal{O}_I(\tau) H_I(\tau') H_I(\tau'')}$. In this case, both Hamiltonian insertions are to the right of the operator $\mathcal{O}_I(\tau)$, and therefore all the fields from the Hamiltonians appear conjugated, while the external fields to which it is coupled appear unconjugated.
We emphasize that the structure would be analogous if $H_I$ contains derivative interactions. 

The structure of the fields in Eq.~(\ref{eq: UV type II}) is the same as that in Eq.~(\ref{eq: general ct}). We recognize this structure without having to solve the time integral, which could allow for renormalization even prior to its resolution.

In contrast, the structure in Eq.~(\ref{eq: UV type I}) differs from that of Eq.~(\ref{eq: general ct}). Whereas in the latter all the external fields appear unconjugated ($\phi_{k_1}(\tau) \cdots \phi_{k_n}(\tau)$), in Eq.~(\ref{eq: UV type I}) the external fields are a mixture of conjugated and unconjugated terms. Indeed, in Eq.~(\ref{eq: UV type I}), after taking the time derivatives, we find terms that cannot have the external field structure of the counterterms (again, we are referring to the product $\phi_{k_1}(\tau) \cdots \phi_{k_n}(\tau)$).

The origin of the apparent issue can be traced back to the fact that the counterterms come from ${\expval{\mathcal{O}(\tau)}} \supset \Im \int \dd \tau'  \expval{\mathcal{O}_I(\tau) H_I(\tau')}$, whereas the kind of loop contribution we are discussing comes from ${\expval{\mathcal{O}(\tau)}} \supset \int \dd \tau' \int \dd \tau'' \expval{H_I(\tau') \mathcal{O}_I(\tau) H_I(\tau'')}$, where in both expressions $\mathcal{O}$ denotes some unspecified operator (and in general not the same one). The presence of the Hamiltonian insertions to the left and right of $\mathcal{O}_I$ in the latter expression impedes this contribution from being directly reduced to a contact term. 

These observations are not a proof that the contribution of Eq.~(\ref{eq: UV type I}) cannot be renormalized. It only tells us that identifying an adequate counterterm does not appear to be obvious. In the next section we will discuss a concrete example in which we find UV divergences coming from this kind of loop contribution and argue that we cannot renormalize it with counterterms of the type given by Eq.~(\ref{eq: general ct}).

Let us note that this apparent difficulty in renormalization is one of the reasons why, in Section \ref{sec: One-Loop Bispectrum}, we focus on the three-point correlation function (bispectrum), which gives us an example of it. 

\section{$P(\phi,X)$ models in the effective theory of inflation} \label{sec: model}

We will now apply the method of Section \ref{sec: Dim Reg} with the goal of computing the renormalized one-loop bispectrum of primordial curvature fluctuations in a single-field model of inflation with an action of the kind~\cite{Armendariz-Picon:1999hyi}:
\begin{equation} \label{eq: Action General phi}
	S = \int \dd^{4} x\, \sqrt{-g} \left\lbrace \frac{M_P^2}{2} R + \sum_{n = 0}^\infty \frac{f_n(\phi)}{n!} \frac{\left( g^{\mu\nu} \partial_\mu \phi \partial_\nu \phi \right) ^n}{\Lambda^{4(n-1)}}   \right\rbrace \,,
\end{equation}
where the series is characterized by the dimensionless functions $f_n(\phi)$ and the energy scale $\Lambda$. This action can be thought of as an effective field theory in which a canonically normalized, minimally coupled scalar $\varphi$ (defined by $\dd\varphi/\dd\phi = \sqrt{f_1(\phi)}\, $) with a standard kinetic term $X = -g^{\mu\nu} \partial_\mu \varphi\, \partial_\nu \varphi$ and potential $V(\varphi)= \Lambda^4 f_0(\phi(\varphi))$ gets corrections from higher-dimensional operators of the form $(X/f_1)^n f_n /\Lambda^{4(n-1)}$, $n\geq 2$, suppressed by powers of $\Lambda$. If instead we leave total freedom to the functions $f_n(\phi)$, the action Eq.~\eqref{eq: Action General phi} can describe any model whose Lagrangian is given by a function $P({\phi(\varphi)},X)$. We choose to work with this action because it is simple enough to illustrate the main points we are interested in.

The primordial curvature fluctuation $\zeta$ (that we define a few lines below) lives on a {FLRW} background given by the time evolution of the scalar field $\phi$. Unlike the fluctuation of the scalar field, this variable freezes once it crosses the horizon, making it particularly useful to connect with observations \cite{Maldacena:2002vr, Weinberg:2003sw,Lyth:2004gb} (see also \cite{Pimentel:2012tw,Assassi:2012et,Senatore:2012ya,Green:2024fsz} for discussions at the quantum level).  It is therefore convenient to work directly with an effective theory for~$\zeta$, that inherits the properties of Eq.~\eqref{eq: Action General phi}. The framework that allows us to do this is the effective field theory of inflation of \cite{Cheung:2007st}. See also \cite{Weinberg:2008hq} for an EFT description of inflation in the (covariant) language of $\phi$. In order to build the effective action for $\zeta$ that we are interested in, we start by writing the spacetime metric using an ADM decomposition:
\begin{equation}
	\dd s^2 = g_{\mu\nu}\dd x^{\mu}\dd x^{\nu}= -N^2 \dd t^2 + \gamma_{ij} \left( N^i \dd t + \dd x^i \right) \left( N^j \dd t + \dd x^j \right) \,,
\end{equation}
where $N$ and $N^i$ are the {\it Lapse} and the {\it Shift}, which are non-{dynamical} variables and thus act as Lagrange multipliers in the action. The spatial part of the metric is decomposed as follows:
\begin{equation}
	\gamma_{ij} = a^2 \left( e^\Gamma \right) _{ij} \,, \quad {\rm where} \quad \Gamma_{ij} = 2\zeta \delta_{ij} + \partial_{ij} E + \partial_{(i} E_{j)} + h_{ij}
\end{equation}
and $\left( e^\Gamma \right) _{ij}$ denotes the exponential of the matrix $\Gamma_{ij}$. 

We will now use this decomposition of the metric to obtain an action for a variable, usually called~$\pi$, that is directly related to $\zeta$. First, we write the action Eq.~\eqref{eq: Action General phi} in the unitary gauge, in which the fluctuations of the inflaton field are set to zero ($\delta\phi = 0$):
\begin{align} \label{EFTtime}
	S= \int { \dd^4 x}\, \sqrt{-g} \left\lbrace \dfrac{M_P^2}{2} R + \sum_{n = 0}^\infty \dfrac{f_n(\phi_0(t))}{n!} \dfrac{1}{\Lambda^{4(n-1)}}  \left( g^{00} \dot{\phi}^2_0(t) \right) ^n \right\rbrace\,.
\end{align}
Then, if we note that expressing $g^{00}=-N^{-2}$ as $ g^{00}=-1+\delta g^{00}$, Eq.~\eqref{EFTtime} can equivalently be written as the following action: 
\begin{align} \label{EFTdg00}
	S= \int { \dd^4 x} \,\sqrt{-g} \left\lbrace \dfrac{M_P^2}{2} R + \sum_{n = 0}^\infty \dfrac{M_n^4(t)}{n!} \left( \delta g^{00}  \right) ^n \right\rbrace\,,
\end{align}
where
\begin{equation} \label{eq: Relation f and M}
	M_n^4(t) = \sum_{m = 0}^\infty \dfrac{f_{n+m}(\phi_0(t))}{m!} \Lambda^4 \left( \dfrac{\dot{\phi}_0(t)}{\Lambda^2}\right) ^{2(n+m)}\,.
\end{equation}
The action Eq.~\eqref{EFTdg00} is a particular case of the effective field theory of inflation (in the unitary gauge), which is meant to describe the most general action for fluctuations in single-field inflation \cite{Cheung:2007st}.\footnote{We are not including terms in the unitary gauge that depend on the extrinsic curvature of hypersurfaces of constant time or derivatives of $\delta g^{00}$.} This action is invariant only under spatial diffeomorphisms, as the temporal ones have been broken by the background evolution of $\phi$. 

Clearly, the functions $M_n^4(t)$ receive corrections from all powers of $X_\varphi$. Assuming that all $f_n(\phi)$ are of the same order, there is a hierarchy $M_{n+1}^4/ M_{n}^4 \sim (\dot{\phi}_0(t)/{\Lambda^2})^{2} \ll 1 $. Breaking this assumption, which has been done in many studies of primordial inflation, allows to look for specific signatures of concrete operators in cosmological correlators. A well-known example is that of DBI inflation \cite{Alishahiha:2004eh}, which is a specific case of a $P(\phi,X)$ model. Later on, we will focus on a model in which we tune all $M_n^4$ to zero, except $M_3^4$. However, before getting there, we still need to make the connection with the curvature perturbation~$\zeta$. We will do this by restoring full diffeomorphism invariance using Stückelberg's trick and then relating the Stückelberg field (which, as we anticipated, is called $\pi$ below) to $\zeta$.

Indeed, starting from the action in the unitary gauge, Eq.~\eqref{EFTdg00}, we can restore the temporal diffeomorphisms by introducing a field $\pi$(x) which transforms under diffeomorphisms $x \to x' = x + \xi(x)$ as $\pi(x) \to \pi'(x') = \pi(x) -\xi^0(x)$ \cite{Cheung:2007st}:\footnote{The EFT for the fluctuation $\pi(x)$ respects the symmetry under the scale redefinitions of Eq.\ (\ref{scalings}), since it descends from a generally covariant theory and therefore inherits the correct dependence on the scale factor.}
\begin{align} \label{eq: Action General pi}
	S = \int { \dd^4 x}\, \sqrt{-g} \left\lbrace \dfrac{M_P^2}{2} R + \sum_{n = 0}^\infty \dfrac{M_n^4(t+\pi)}{n!} \left[ -\dfrac{1}{N^2}\left( 1+\dot{\pi}-N^i\partial_i \pi \right) ^2 +\gamma^{ij} \partial_i \pi \partial_j \pi + 1  \right] ^n \right\rbrace\,.
\end{align}
Now that time diffeomorphisms are restored, we recover the original number of variables: ten from the metric and one from the scalar field $\phi(x)$. The latter can be easily linked to $\pi(x)$. To see this, we just need to make a gauge transformation from a system of coordinates where $\pi$ and $\delta\phi$ are non-zero, to the unitary gauge where they cancel simultaneously, obtaining: 
\begin{equation}
	\phi(x) = \phi_0(t+\pi(x)) = \phi_0(t) + \delta\phi(x)\,.
\end{equation}
It is important to stress that both objects, $\pi$ and $\delta\phi$, are expressed in the same gauge in the above equation, and so this relation is nothing more than a field redefinition. Consistently, if we start from the original action for $\phi$, Eq.~(\ref{eq: Action General phi}), using that 
\begin{equation}
	g^{\mu\nu} \partial_\mu \phi \partial_\nu \phi = g^{\mu\nu} \dot{\phi}_0^2 (t+\pi(x))\, \partial_\mu(t+\pi(x)) \, \partial_\nu(t+\pi(x))
\end{equation}
and the relation between $f_n$ and $M_n$ that we previously found, we recover the action for $\pi$ once full diffeomorphism invariance is restored, Eq.~(\ref{eq: Action General pi}).

The relation between the scalar fluctuations $\zeta$ and $\pi$ can be expressed at  linear order, through a gauge transformation, as follows \cite{Cheung:2007sv}:
	\begin{equation}
		\zeta = - H \pi + \order{\pi}^2 \,.
	\end{equation}
In what follows, we will work on what is known as the decoupling limit, whereby corrections suppressed by powers of the slow-roll parameters are neglected. In this limit, the self-interactions of $\pi$ dominate over the coupling to gravity  \cite{Cheung:2007st}. In practice, this allows us to neglect fluctuations arising from algebraic variables, taking $N = 1$ and $N_i = 0$. 

\subsection{Background renormalization}
Since we will be later using dimensional regularization, we are going to work in $3+\delta$ dimensions from now on, noting that all the previous equations of this section are valid just replacing the integration measure as follows: 
\begin{align}
		\dd^4 x \sqrt{-g} \rightarrow \mu^\delta \dd^{4+\delta} x \sqrt{-g}\,,
\end{align}
where $\mu$, the renormalization scale, is a quantity with dimensions of energy that we need to introduce to keep the dimension of the fields the same as in the $\delta=0$ case, and that is invariant under the symmetry transformation of Eq.\ (\ref{scalings}). 

The action for $\pi$, which as we discussed earlier can be interpreted as a perturbation, must start at quadratic order because the linear term has to vanish to ensure that we are expanding around the correct time-evolving background, $\expval{\pi} = 0$. Varying the action with respect to the metric and setting the fluctuations to zero, we obtain:
\begin{equation} \label{eqtadpolenull}
	\left[\dfrac{M_P^2}{2}G^0_{\mu\nu} - \dfrac{1}{2} g^0_{\mu\nu} M_0^4(t) \right] \delta g^{\mu\nu} + M_1^4(t) \delta g^{00} = 0\,.
\end{equation}
The (exact) solution of Eq.~\eqref{eqtadpolenull} is
\begin{equation} \label{eq: M0 and M1 bg}
	M_0^4(t) = -M_P^2 \left( 1+\delta / 2 \right) \left( 2\dot{H}(t) + \left( 3+\delta \right) H^2(t) \right) \quad {\rm and} \quad 	M_1^4(t) = M_P^2 \left( 1+\delta / 2 \right) \dot{H}(t)\,.
\end{equation}
That is, these two functions are not free in the effective theory of inflation but are fixed by the background (once a function $H(t)$ is assumed). More precisely, we already see that they decompose into renormalized and counterterm components as 
\begin{align} \label{counterM}
M_i^4 \equiv M_{i,r}^4 + \delta M_i^4
\end{align} 
with $i \in \{0,\,1 \}$. The renormalized parts $M_{i,r}^4$ are solely determined by the (assumed) background evolution, Eq.\ (\ref{eq: M0 and M1 bg}), while the counterterms $\delta M_i^4$ are determined by the renormalization condition $\expval{\pi} = 0$ at loop level, as we discuss in Appendix \ref{app: Power spectrum}. It is sometimes said that this procedure includes the backreaction of the fluctuations on the background, but that can be misleading.
Here, we are assuming that the background evolution is known and therefore has to remain unchanged including loop corrections, which returns the renormalization condition $\expval{\pi} = 0$.
However, we could take a different approach: we could start with a given background, not known, but which we seek to determine. In that case, we would have to correct the evolution of that background with the effect of interactions inducing $\expval{\pi} \neq 0$. This correction could be interpreted as a backreaction, whereas the previous procedure (the one we use) is instead setting a renormalization condition imposed precisely so that the background is not altered.\footnote{This represents one of the advantages of the effective theory of inflation: fixing the background only determines the first two coefficients of the EFT. In contrast, in the language of $\phi$, introducing higher order EFT terms modifies the background evolution, as we see in the relation between $f_n$ and $M_n$, Eq.\ (\ref{eq: Relation f and M}).}

\subsection{$M_3^4\neq 0$ in the decoupling limit}

We choose to tune the functions $M_i$ so that only $M_3$ and $M_0$ and $M_1$, which are determined by the assumed background, are non-zero. Loop corrections break this tuning and excite operators beyond those. Aiming to renormalize the primordial bispectrum, we will later consider more terms in the effective theory of inflation. 
For the time being, ignoring possible counterterms, the action we are going to work with in the decoupling limit, and neglecting the tensor fluctuations of the metric, is:
\begin{align} 
	\nonumber
	S_3  = \int \dd \tau\, \dd^{3+\delta} \vb{x} \,\mu^\delta a^{4+\delta} &\left\lbrace -M_P^2 \left( 1+\delta / 2 \right) \left( 2\dot{H}(t+\pi) + \left( 3+\delta \right) H^2(t+\pi) \right) \right.\\\nonumber
	&+ M_P^2 \left( 1+\delta / 2 \right) \dot{H}(t+\pi) \left( -2\dot{\pi}-\dot{\pi}^2 +a^{-2}\left(  \partial \pi\right) ^2 \right)\\
	& \left. + \dfrac{M_3^4(t+\pi)}{3!} \left( -2\dot{\pi}-\dot{\pi}^2 +a^{-2}\left(  \partial \pi\right) ^2 \right) ^3 \right\rbrace\,.
\end{align}
Expanding in powers of $\pi$, we get:
\begin{equation}  \label{langM3}
	S_3  = \int \dd \tau\, \dd^{3+\delta} \vb{x} \,\mu^\delta a^{4+\delta} \left\lbrace M_P^2 H^2 \epsilon_\delta \left( \dot{\pi}^2 - a^{-2}\left(  \partial \pi\right) ^2 \right) + \dfrac{M_3^4(t+\pi)}{3!} \left( -2\dot{\pi}-\dot{\pi}^2 +a^{-2}\left(  \partial \pi\right) ^2 \right) ^3 \right\rbrace\,,
\end{equation}
where we have defined $\epsilon_\delta \equiv \epsilon \, \left( 1+{\delta}/{2} \right)$, which is the only effect beyond the factor $\mu^\delta a^\delta$ that the change in the number of spatial dimensions has on the dynamics of the modes. 
However, this modification of the slow-roll parameter $\epsilon$  plays no role in the following, so we can take $\epsilon_\delta = \epsilon$.
In order to compute the three-point correlation of $\pi(x)$ {at one loop}, we need to keep the interactions up to order five. Therefore, the free and interaction Lagrangian densities that we are going to use are:
\begin{align} 
	\mathcal{L}_0 &= \mu^\delta a^{2+\delta} M_P^2 H^2 \epsilon \left( \pi'^2 - \left(  \partial \pi\right) ^2 \right) \,,\\ \label{kesp}
	\mathcal{L}_{\rm int} &= - \mu^\delta a^{-2+\delta} {M_3^4(t+\pi)} \left( \dfrac{4}{3}a^3\pi'^3 + 2 a^2 \pi'^2 \left( \pi'^2 - \left(  \partial \pi\right) ^2 \right) + a \pi' \left( \pi'^2 - \left(  \partial \pi\right) ^2 \right)^2 + \order{\pi^6} \right) \,.
\end{align}

In the interaction picture, we express the field $\pi$ in terms of creation and annihilation operators:
\begin{equation}
	\pi(x) = \int \dfrac{\dd^{3+\delta} \vb{k}}{(2\pi)^{(3+\delta)/2}} e^{i \vb{k}\cdot \vb{x}} \pi_{\vb{k}} (\tau) \,, \quad {\rm where} \quad \pi_{\vb{k}} (\tau) = \pi_k(\tau) a_{\vb{k}} + \pi_k^*(\tau) a^\dagger_{-\vb{k}}\,.
\end{equation}
The dynamics of the Fourier modes $\pi_k$ is completely determined by the solution of the free equations of motion assuming Bunch-Davies initial conditions:
\begin{equation}
	\pi_k'' + (2+\delta)\, a\, H\, \pi_k' + k^2 \pi_k = 0 \quad {\rm and} \quad \lim_{\tau \to - \infty} \pi_k(\tau) = \dfrac{1}{\sqrt{2\mu^\delta a^{\delta}  \epsilon} \, M_P \, a\, H} \dfrac{e^{-i k \tau}}{\sqrt{2 k}}\,,
\end{equation}
where the normalization factor in the initial conditions comes from the canonical normalization of the modes. Solving this equation perturbatively in $\delta$, at first order, we obtain:
\begin{equation} \label{modepi}
	\pi_k(\tau) = \dfrac{i e^{-i k \tau} \left( 1 +i k \tau \right) }{2M_P \sqrt{k^3\epsilon}}\left\lbrace  1 + \dfrac{\delta}{2}\left(\dfrac{2+(1-ik\tau) \left( i \pi - {\rm Ei}(-i2k \tau) \right)e^{i 2 k \tau }}{1+i k \tau}- \log\left( {\mu}\,a (\tau)\right)  \right)+\mathcal{O}(\delta^2) \right\rbrace \,,
\end{equation}
where ${\rm Ei}(z) = - \int_{-z}^\infty \dd x \, e^{-x} /x$ is the exponential integral function. Corrections of orders greater than $\delta$ do not contribute to the loop integral \cite{Senatore:2009cf,Ballesteros:2024cef}.

\section{One-loop bispectrum for $M_3^4 \neq 0$} \label{sec: One-Loop Bispectrum}

The general structure of the three-point correlation of the field $\pi$ is:
\begin{equation} \label{eq: Def bispectrum}
	\expval{\pi(\tau,\vb{x}_1) \pi(\tau,\vb{x}_2) \pi(\tau,\vb{x}_3)} = \int {\left[ \prod_{i=1}^3 \dfrac{\dd^{3+\delta} \vb{k}_i}{(2\pi)^{3+\delta}} e^{i \vb{k}_i\cdot \vb{x}_i}\right]} (2\pi)^{3+\delta} \delta(\vb{k}_1+\vb{k}_2+\vb{k}_3) B_\pi(\tau;\{\vb{k}_i\}) \,.
\end{equation}
The conservation of momentum makes $B_{{\pi}}(\tau;\{\vb{k}_i\})$ depend on only two momenta. Furthermore, homogeneity and isotropy imply that $B_{{\pi}}(\tau;\{\vb{k}_i\})$ only depends on the modulus of these two momenta and the angle between them. Let us then work with $\{k_1,k_2,\theta_1\}$, where we define $\theta_1$ in such a way that $\vb{k}_1\cdot \vb{k}_2 = k_1k_2 \cos \theta_1$. In $3+\delta$ spatial dimensions and for any function $j$ of $k_1$, $k_2$ and $\theta_1$:
\begin{equation}
	\int \left[\prod_{i=1}^3 \dfrac{\dd^{3+\delta} \vb{k}_i}{(2\pi)^{3+\delta}} \right](2\pi)^{3+\delta} \delta(\vb{k}_1+\vb{k}_2+\vb{k}_3) \, j(k_1,k_2,\theta_1) = \mathcal{C}(\delta) \int \dd k_1\, \dd k_2\, \dd \theta_1\, k_1^{2+\delta} k_2^{2+\delta} \sin^{1+\delta} \theta_1 \, j(k_1,k_2,\theta_1)\,,
\end{equation}
where $\mathcal{C}(\delta)$ is a constant that depends on the number of dimensions but plays no role of physical relevance. 
For convenience, we can change the angular variable $\theta_1$ to $k_3$, the modulus of $\vb{k}_3 = -\left( \vb{k}_1+\vb{k}_2\right)$. Therefore, $k_3^2 = k_1^2 + k_2^2 + 2k_1k_2\cos \theta_1$. In this way, the volume element is modified such that:
\begin{equation}
	\int \left[\prod_{i=1}^3 \dfrac{\dd^{3+\delta} \vb{k}_i}{(2\pi)^{3+\delta}}\right] (2\pi)^{3+\delta} \delta(\vb{k}_1+\vb{k}_2+\vb{k}_3) {\, j(k_1,k_2,\theta_1)}
	= \dfrac{\mathcal{C}(\delta)}{2^\delta} \int \dd \log k_1\, \dd \log k_2\, \dd \log k_3\, \mathcal{V} (\delta ; \{k_i\}) {\, j(k_1,k_2,\theta_1)} \,,
\end{equation}
where 
\begin{equation} \label{vfactor}
	\mathcal{V} (\delta ;  \{k_i\}) = k_1^2 k_2^2 k_3^2 \big[(k_1+k_2+k_3)(k_1+k_2-k_3)(k_1-k_2+k_3)(-k_1+k_2+k_3) \big]^{\delta/2} \,.
\end{equation}
Thus, we define the rescaled bispectrum\footnote{We define the dimensionless bispectrum for the curvature perturbation as $\mathcal{B}_\zeta = -H^3\mathcal{B}_\pi$.} in $3+\delta$ dimensions as 
\begin{equation} \label{eq: Bispectrum rescaled}
	\mathcal{B}_\pi (\tau; \{k_i\}) \equiv \dfrac{1}{8\pi^4} \mathcal{V} (\delta ; \{k_i\}) \, B_\pi (\tau; \{\vb{k}_i\}) \,,
\end{equation}
where $k_{1,2} \in [ 0,\infty )$ and $k_3\in [ \abs{k_1-k_2}, k_1+k_2]$. We have also added the $1/(8\pi^4)$ factor coming from the angular integrals in three spatial dimensions, which is nothing more than $\mathcal{C}(0)$.

The dominant contribution to the one-loop bispectrum comes from a single insertion of the quintic interaction from the interaction Hamiltonian ($H_I=\int \dd ^{3+\delta}\vb{x} \mathcal{H}_I$), see Eq.~\eqref{kesp},
\begin{equation} \label{HIq}
	\mathcal{H}_I \supset  \mu^\delta a^{-1+\delta} M_{3}^4\, \pi' \left( \pi'^2 - \left(  \partial \pi\right) ^2 \right)^2 \,,
\end{equation}
where $M_{3}^4$ is assumed to be constant and leads to a diagram that scales as follows: 
\begin{equation}
	\mathcal{B}_\pi^{\rm quintic} = \begin{tikzpicture}[baseline={-10}]
		\draw (0,0) circle (0.25);
		\draw (-0.75,-0.75) -- (0,-0.25);
		\draw (0.75,-0.75) -- (0,-0.25);
		\draw (0,-0.75) -- (0,-0.25);
	\end{tikzpicture} \propto \dfrac{H \, M_3^4}{M_P^8\,\epsilon^4} \,.
\end{equation}
To obtain this order of magnitude we have taken into account that the vertex introduces a factor $M_3^4$  while each field $\pi(x)$ introduces a factor $1/(M_P\sqrt{\epsilon})$, see Eq.~\eqref{modepi}.

The loop in this diagram is not sensitive to the external momenta structure and therefore plays the role of a time-dependent, effective cubic coupling at tree level. This implies that in the limit in which the expansion of the Universe is exactly de Sitter, it must be indistinguishable from the effect of counterterms. We will see this in detail later on. Furthermore, there is no other loop diagram that scales like this one.
	
Other quintic interactions arise from Eq.~\eqref{kesp} expanding $M_{3}^4(t+\pi)$ in  series around $\pi=0$. We neglect such contributions to the bispectrum because they are suppressed if a slow variation of $M_{3}^4$ is assumed.\footnote{Assuming a time dependence on $M_3^4(t)$ would introduce an additional comoving scale $1/\tau_*$. This is not required to obtain a distinguishable loop contribution to the bispectrum, as mentioned in Section \ref{sec: Intro}. See Appendix \ref{app: Power spectrum}, where a time dependence on $M_3^4(t)$ is considered for the calculation of one- and two-point loop correlations.}

The leading term contributing to the one-loop bispectrum is then of the form (see Eq.\ \eqref{singlevert})
\begin{align} \label{eq: app B LO quintic}
	\nonumber
	\mathcal{B}_\pi^{\rm quintic} =& \dfrac{\mathcal{V} (\delta ; \{k_i\})}{8 \pi^4} \int \dfrac{\dd^{3+\delta} \vb{p}}{(2\pi)^3} 4 \Im \bigg\lbrace \pi_{k_1}(\tau) \pi_{k_2}(\tau) \pi_{k_3}(\tau) \int_{-\infty_-}^\tau \dd \tau' \, \mu^\delta a^{-1+\delta} M_{3}^4 \pi_{k_1}'^* \Big[\pi_{k_2}'^*\pi_{k_3}'^* \left(5\abs{\pi_p'}^2 - p^2 \abs{\pi_p}^2 \right) \\
	&- \pi_{k_2}^*\pi_{k_3}^* \left( \left( \vb{k}_2 \cdot \vb{k}_3\right)  \left(p^2 \abs{\pi_p}^2 -3 \abs{\pi_p'}^2 \right) + 2 \left( \vb{k}_2 \cdot \vb{p}\right) \left( \vb{k}_3 \cdot \vb{p}\right) \abs{\pi_p}^2  \right) \Big] \eval_{\tau'}\bigg\rbrace + {\rm perms}\,,
\end{align}
where we have used that the odd terms in $\vb{p}$ vanish, and where the sum over the permutations represents all possible exchanges of external momenta, which are six in total.\footnote{We note that, in loop correlations, we must symmetrize the fields that appear Wick-contracted coming from the same interaction Hamiltonian, since this must be Hermitian. Those terms in which this symmetrization has been required are, however, odd in $\vb{p}$, and we have not reported them.}
Furthermore, we have set $\delta = 0$ in the Fourier factor $1/(2\pi)^{3+\delta}$ of the loop integral, as it gives rise to a global factor that can be absorbed by counterterms \cite{Ballesteros:2024cef}. 

To do the loop integral we have to take into account that the integration measure in $d$ dimensions, expressed in spherical coordinates, is
\begin{equation}
	\dd^d {\mathbf p} = p^{d-1} (\sin \theta_1)^{d-2} \cdots (\sin \theta_{d-2}) \, \dd p \, \dd \theta_1 \cdots \dd \theta_{d-2}\, \dd \theta_{d-1} \,,
\end{equation}
where $p \in [0,\infty)$ and $\theta_i \in [0,\pi]$, with the exception of  $\theta_{d-1} \in [0,2\pi)$. 
Therefore, the integration measure in this simple case in which the integrand is just a function of the momentum $p$ is:\footnote{We have used that
\begin{equation}
	\int \dd^d \vb{p} \, \vb{p}_i \, \vb{p}_j \, f(p) = \dfrac{1}{d} \delta_{ij} \int \dd^d \vb{p} \, p^2  \, f(p)
\end{equation}
for any generic function $f(p)$.}
\begin{equation}
\int \dd^d {\mathbf p}f(p) = 4\pi\,\mathcal{C}(d)\, \int_0^\infty p^{d-1} f(p) \, \dd p \,,
\end{equation}
for any generic function of the momentum $f(p)$. The global constant $\mathcal{C}(d)$ is such that $\mathcal{C}(3) = 1$, whose $\order{d-3}$ effects can be absorbed with counterterms.

Following the method described in Section \ref{sec: Dim Reg}, the late-time contribution to the bispectrum associated to Eq.~\eqref{HIq} is:
\begin{align} 
	\mathcal{B}_\pi^{\rm quintic} = \begin{tikzpicture}[baseline={-10}]
		\draw (0,0) circle (0.25);
		\draw (-0.75,-0.75) -- (0,-0.25);
		\draw (0.75,-0.75) -- (0,-0.25);
		\draw (0,-0.75) -- (0,-0.25);
	\end{tikzpicture} \xrightarrow{\tau \to 0} \dfrac{3 \, H \, M_3^4}{1024 \,  M_P^8 \, \pi^6 \, \epsilon^4 } \dfrac{K_3^3 \left(11 K_1^2-4 K_2^2\right)+K_1 \left(K_1^2+K_2^2\right) \left(K_1^2-4 K_2^2\right)}{K_1^2 \, K_3^3 }  \,,
\end{align}
where we define
\begin{align} \label{defks}
K_1  \equiv k_1 + k_2 + k_3\,,\quad
K_2^2  \equiv k_1k_2 + k_1k_3 + k_2k_3\,,\quad
K_3^3  \equiv k_1 k_2 k_3 \,.
\end{align}
This diagram does not feature UV divergences (in dimensional regularization) because $\abs{\pi_p}^2$ depends on the loop momentum through powers of the latter (except the power $1/p$, see Eq.~\eqref{modepi}) at $\mathcal{O}(\delta^0)$, which in dimensional regularization leads to a vanishing contribution. 

Late-time divergences are not present either for this diagram, the basic reason being that the interactions involved are shift symmetric. Derivative interactions, either with respect to conformal time or to space, are accompanied by scale factors in the denominator that improve the convergence of the time integrals in the limit $\tau \to 0$.
Furthermore, in the superhorizon and the late-time limit, the dynamics of the modes can be written as:
\begin{equation} \label{eq: Pi late-time}
	\pi_k(\tau) \xrightarrow{k \tau \to 0} c_{1k} + c_{2k} \int_{\tau_*}^\tau \dfrac{\dd \tau'}{ a^{2+\delta}(\tau')} + \order{\dfrac{k}{aH}} \,,
\end{equation}
where $\tau_*$ is an auxiliary time, and $c_{ik}$ are constants (in time). Therefore, there is a constant and a decaying solution to the free equation of motion \cite{Weinberg:2003sw}. The derivative $\pi_k'(\tau)$ selects the decaying solution, which further improves the convergence of the time integrals. 

Moreover, and importantly, the finite contribution of this diagram is identical to that generated by the cubic term
\begin{equation}
S = - \dfrac{3}{4} \dfrac{H^2 M_3^4}{M_P^2 \pi^2 \epsilon} \int \dd \tau\, \dd^3 \vb{x} \, a \, \pi' \left( \pi'^2 - \partial \pi^2 \right)  
\end{equation}
at tree level, so this loop has no intrinsic component.

A different (and subdominant) contribution to the one-loop bispectrum that we can consider comes from a diagram in which the cubic and quartic interaction of $\pi$ in Eq.~\eqref{kesp} intervene:
\begin{equation} \label{interv}
	\mathcal{B}_\pi ^{{\rm c} \times {\rm q}} = \begin{tikzpicture}[baseline={-2}]
		\draw (0,0.2) -- (0.5,0);
		\draw (0,-0.2) -- (0.5,0);
		\draw (0.5+0.25,0) circle (0.25);
		\draw (1,0) -- (1.5,0);
	\end{tikzpicture} \propto \dfrac{M_3^8}{H\, M_P^{10}\,\epsilon^5} \,.
\end{equation}
This contribution is interesting because, in principle, it can be distinguished from purely tree-level effects, as we will argue. Other diagrams with the same scaling also contribute to the bispectrum. The first one, which we can represent as 
\begin{tikzpicture}[scale=0.75,baseline=-5]
		\draw (0,0) circle (0.25);
		\draw (-0.75,-0.75) -- (0,-0.25);
		\draw (0.75,-0.75) -- (0,-0.25);
		\draw (0,-0.75) -- (0,-0.25);
		\draw[fill=gray!75] (0,-0.25) circle (0.05);
\end{tikzpicture}\,,
arises from the combination in the interaction Hamiltonian of cubic and quartic interactions from the Lagrangian proportional to $M_3$, coming from $\mathcal{H}_I \supset \tfrac{1}{4g_0} \left(\tfrac{\partial \mathcal{L}_{\rm int}}{\partial \pi'} \right)^2 $ (see Appendix \ref{app: HI}). 
Although we do not report its exact expression, this diagram is finite in dimensional regularization (see the discussion below Eq.~\eqref{defks}) and it does not give rise to intrinsic contributions to the bispectrum distinguishable from tree-level counterterm effects. Therefore, we will not consider it any further.\footnote{It should be noted that the proper renormalization of the one- and two-point functions modifies the interaction Hamiltonian $H_I$, generating additional tree-level contributions with the same scaling as Eq.~\eqref{interv}; see Appendix~\ref{app: Power spectrum} for details. Since these terms are purely tree-level, we do not display them explicitly here and instead focus on characterizing the distinguishable part of the one-loop contribution in Eq.~\eqref{interv}.}

We are now going to focus on Eq.~\eqref{interv}, which, as we are going to see does lead to an intrinsic contribution. This diagram can be written as:
\begin{align} \label{eq: app B NLO c q}
	\nonumber
	\mathcal{B}_\pi ^{{\rm c} \times {\rm q}} = & \dfrac{\mathcal{V} (\delta ; \{k_i\})}{8 \pi^4} 16 \mu^{2\delta} M_{3}^8  \int \dfrac{\dd^{3+\delta} \vb{p}}{(2\pi)^{3+\delta}}  2\Re \Bigg\lbrace \pi_{k_1}^*(\tau) \pi_{k_2}(\tau) \pi_{k_3}(\tau)\int_{-\infty_+}^ \tau \dd \tau' \int_{-\infty_-}^ \tau \dd \tau'' F_1(\tau',\tau'';\{k_i\},p,q)\\
	&- \pi_{k_1}(\tau) \pi_{k_2}(\tau) \pi_{k_3}(\tau)\int_{-\infty_-}^ \tau \dd \tau' \int_{-\infty_-}^ {\tau'} \dd \tau'' F_2(\tau',\tau'';\{k_i\},p,q) \Bigg\rbrace + {\rm perms.}\,,
\end{align}
where $\vb{q} = \vb{k}_1 - \vb{p}$ and we have defined 
\begin{align}
	\nonumber
	&\left( a(\tau') a(\tau'')\right) ^{-\delta} F_1(\tau',\tau'';\{k_i\},p,q) \equiv 6 \left(  a \, \pi_{k_1}' \pi_p'\pi_q'\right) \eval_{\tau'} \left(  \pi_{k_2}'^* \pi_{k_3}'^* \pi_p'^* \pi_q'^* \right) \eval_{\tau''}  \\
	&+ \left(  a \, \pi_{k_1}' \pi_p'\pi_q'\right) \eval_{\tau'}\left(  \pi_{k_2}'^* \pi_{k_3}'^* \pi_p^* \pi_q^* (\vb{p} \cdot \vb{q}) +4 \pi_{k_2}^* \pi_{k_3}'^* \pi_p^* \pi_q'^* (\vb{p} \cdot \vb{k}_2) + \pi_{k_2}^* \pi_{k_3}^* \pi_p'^* \pi_q'^* (\vb{k}_2 \cdot \vb{k}_3) \right) \eval_{\tau''} \,, \\ \nonumber
	&\left( a(\tau') a(\tau'')\right) ^{-\delta} F_2(\tau',\tau'';\{k_i\},p,q) \equiv 6 \left(  a \, \pi_{k_1}'^* \pi_p'\pi_q'\right) \eval_{\tau'} \left(  \pi_{k_2}'^* \pi_{k_3}'^* \pi_p'^* \pi_q'^* \right) \eval_{\tau''} + 6 \left(  a \, \pi_{k_1}'^* \pi_p'^* \pi_q'^* \right) \eval_{\tau''} \left(  \pi_{k_2}'^* \pi_{k_3}'^* \pi_p' \pi_q' \right) \eval_{\tau'}  \\ \nonumber
	&+ \left(  a \, \pi_{k_1}'^* \pi_p'\pi_q'\right) \eval_{\tau'}\left(  \pi_{k_2}'^* \pi_{k_3}'^* \pi_p^* \pi_q^* (\vb{p} \cdot \vb{q}) +4 \pi_{k_2}^* \pi_{k_3}'^* \pi_p^* \pi_q'^* (\vb{p} \cdot \vb{k}_2) + \pi_{k_2}^* \pi_{k_3}^* \pi_p'^* \pi_q'^* (\vb{k}_2 \cdot \vb{k}_3) \right) \eval_{\tau''}\\
	&+ \left(  a \, \pi_{k_1}'^* \pi_p'^* \pi_q'^* \right) \eval_{\tau''} \left(  \pi_{k_2}'^* \pi_{k_3}'^* \pi_p \pi_q (\vb{p} \cdot \vb{q}) +4 \pi_{k_2}^* \pi_{k_3}'^* \pi_p \pi_q' (\vb{p} \cdot \vb{k}_2) + \pi_{k_2}^* \pi_{k_3}^* \pi_p' \pi_q' (\vb{k}_2 \cdot \vb{k}_3) \right) \eval_{\tau'}  \,.
\end{align}
To solve these integrals we will follow the procedure described in Section \ref{sec: Dim Reg}. We note that in addition to the dependence on $p$, we also have a dependence on $q$, the modulus of $\vb{q} = \vb{k}_1 - \vb{p}$.
We deal with this dependence by changing the variables of integration from $\{p, \theta\}$ to $\{p,q\}$. In order to do so, it is convenient to make a rotation such that
$\vb{k}_1$ falls on the $p_z$-axis. Thus, we have that: $q = \sqrt{k_1^2 + p^2 - 2k_1p\cos \theta}$ which, for any function $f(p,q)$, leads to:\footnote{We also note that in order to compute the momentum integral of some of the terms it is necessary to use the relation
\begin{equation}
	\int \dd^d \vb{p}\, \mathcal{Y}(p,\abs{\vb{k}_1 - \vb{p}}) \left(  \vb{k}_2\cdot \vb{p} \right) = \dfrac{\vb{k}_1\cdot\vb{k}_2}{k_1^2} \int \dd^d \vb{p}\, \mathcal{Y}(p,\abs{\vb{k}_1 - \vb{p}}) \left(  \vb{k}_1\cdot \vb{p} \right) \,,
\end{equation}
where $\mathcal{Y}(p,q)$ represents any function that is symmetric under the exchange of $p$ and $q$. This relation can be derived taking into account that the contribution to the integral of the component of $\vb{k}_2$ that does not go in the direction of $\vb{k}_1$ has an odd integrand and therefore vanishes.}
\begin{align}
	\nonumber
	\int\dd^d \vb{p} f(p,|\vb{k}_1-\vb{p}|) &= 2 \pi \mathcal{C}(d) \, \int_0^\infty\dd p\, p^{d-1} \int_0^\pi \dd \theta\,(\sin \theta)^{d-2} f\left(p,\sqrt{k_1^2+p^2-2k_1p\cos\theta}\right)\,  \\
	&= \dfrac{2\pi \mathcal{C}(d)}{k_1} \int_0^\infty \dd p\,p^{1+\delta} \int_{|k_1-p|}^{k_1+p}\dd q\,q\left[\sin \theta(p,q)\right]^\delta\, f(p,q) \,.
\end{align}
After this consideration, the next step, following the method discussed in Section \ref{sec: Dim Reg}, is to separate the loop integral into the IR and UV parts. 

We can express the final result conveniently decomposing $\mathcal{B}_\pi ^{{\rm c}\times {\rm q}}$ into finite (${\rm fin}$) and divergent (${\rm div}$) parts:
\begin{align}
	\mathcal{B}_\pi ^{{\rm c} \times {\rm q}} (\tau;\{k_i\}) = \dfrac{1}{\delta}  \mathcal{B}_\pi^{\rm div} (\tau;\{k_i\}) + \mathcal{B}_\pi ^{\rm fin} (\tau;\{k_i\})\,.
\end{align}
At late times ($\tau\rightarrow 0$), we obtain:
\begin{align}\label{eq: B non-trivial}
	\nonumber
	& \mathcal{B}_\pi^{\rm fin} {(0;\{k_i\})} =  \dfrac{M_{3}^8}{H \, M_P^{10} \pi^6 \epsilon^5}  \left\lbrace g_{\pi}{(\{k_i\})}+f_\pi{(\{k_i\})} \left[\log\dfrac{2 k_1}{k_1+k_2+k_3}+ (k_1 \leftrightarrow k_{2,3})\right] \right\rbrace  \\ 
	&\qquad + {\mathcal{B}_\pi^{\rm div} (0;\{k_i\})} \left[ \dfrac{1}{2}\log\left( \dfrac{(k_1+k_2-k_3)(k_1-k_2+k_3)(-k_1+k_2+k_3)}{(k_1+k_2+k_3)^3}\right) -3\log \left( \dfrac{\mu e^{\gamma_{\rm E}}}{H}\right) \right] \,, 
\end{align}
where the functions $f_\pi {(\{k_i\})} $, $g_\pi {(\{k_i\})}$ and $\mathcal{B}_\pi^{\rm div} {(\tau;\{k_i\})}$ are given in Appendix \ref{app: Bispectrum formulas}.\footnote{In \cite{Bhowmick:2024kld,Ansari:2025nng}, the bispectrum is calculated in a setup similar to the one presented here. The result of the (unrenormalized) loop, however, does not respect invariance under the transformation in Eq.\ (\ref{scalings}).} The absence of late-time divergences again follows from the fact that the interactions entering the one-loop bispectrum under consideration are shift symmetric; see the discussion around Eq.~(\ref{eq: Pi late-time}).

The second line in the expression above originates from the volume factor in Fourier space, Eq.~\eqref{vfactor}, accompanying the rescaled bispectrum being in $3+\delta$ spatial dimensions, combined with the UV divergence $1/\delta$, see Eq.\ (\ref{eq: Bispectrum rescaled}). This term can be anticipated expanding the volume factor at first order in $\delta$. As we will show in the next section, this contribution is exactly canceled by the finite counterterm contribution that is fixed when renormalizing the UV divergences of the loop.

The first line is, instead, more interesting: that term cannot be anticipated without computing it (and is model-dependent). Although it is possible that scheme-dependent effects may alter parts of that first line, it is to be expected that it contains a part coming from the loop that cannot be reabsorbed with counterterms.\footnote{Our discussion is restricted to counterterms within the single-field inflationary EFT under consideration. Nevertheless, a momentum dependence similar to that of Eq.~(\ref{eq: B non-trivial}) may already arise at tree level in theories where the inflaton couples to additional degrees of freedom; see e.g.\ \cite{Wang:2022eop}.} An argument supporting this expectation will be given below. 

After renormalization, the finite parts of the counterterms cannot generate the kind of logarithmic term in the first line of Eq.~\eqref{eq: B non-trivial}. In addition, not only the logarithmic pieces are intrinsic to the loop (in the sense of not appearing at tree-level), but so are the non-logarithmic parts encoded in the function $g_{{\pi}}(\{k_i\})$. In particular, $g_{{\pi}}(\{k_i\})$ contains contributions with denominators of the kind $(k_1+k_2-k_3)$ (and permutations). In contrast, the counterterms --being tree-level contributions with a single Hamiltonian insertion and one time integral-- can only produce denominators of the type $(k_1+k_2+k_3)$ (with all moduli added). This is the case at least in the strict de Sitter setup we are considering, where the modes evolve with a specific phase $\pi_k(\tau) \propto e^{-i k \tau}$, see Eq.~\eqref{modepi}. As we will now see, these terms are of vital importance for the consistency of the loop contribution. 

Let us discuss the result in some more detail. The first thing one can notice is that, in the equilateral limit, we trivially lose the logarithmic scale dependency. Moving away slightly from the equilateral limit, a linear dependency in the (difference of) momenta arises. However, such a dependency can be generated without loops, and so we cannot expect to extract relevant information from the equilateral limit. 

More interesting is the squeezed limit, since it is constrained by the consistency relation \cite{Maldacena:2002vr} (see also \cite{Creminelli:2004yq,Bravo:2017wyw})
\begin{equation}
	\lim_{p\to 0}B_\zeta (\tau;\vb{p},\vb{k},-\vb{k}) = - \dfrac{\dd \log \mathcal{P}_\zeta(\tau,k)}{\dd \log k} P_\zeta(\tau,k) P_\zeta(\tau,p) \,,
\end{equation}
where $\mathcal{P}_\zeta$ and $P_\zeta$ are the dimensionless power spectrum and power spectrum, respectively, of $\zeta$.
This relation is valid at any order in perturbation theory, as it comes only from the possibility of including a large-scale constant (independent of space and constant in time) metric perturbation into the scale factor of the Universe with a redefinition of the coordinates. The finite, one-loop contribution to the bispectrum that we have found from the diagram
\begin{tikzpicture}[scale=0.75, baseline=-5]
		\draw (0,0.2) -- (0.5,0);
		\draw (0,-0.2) -- (0.5,0);
		\draw (0.5+0.25,0) circle (0.25);
		\draw (1,0) -- (1.5,0);
	\end{tikzpicture} 
shows potentially problematic contributions of the type $\propto (k_1+k_2-k_3)^{-\alpha}$ (and permutations) with $\alpha>0$, or $\propto \log k_i$, that explode in the squeezed limit (see the explicit form of the expressions $g_\pi(\{k_i\})$ and $f_\pi(\{k_i\})$ in Appendix \ref{app: Bispectrum formulas}). Therefore, checking the validity of the consistency relation would be a non-trivial test of the calculation.
Let us then analyze the first line of Eq.~(\ref{eq: B non-trivial}) in the squeezed limit configuration $k_1 \equiv p \ll k$, $k_2 \equiv k$ and $k_3 \equiv k + p \, \xi$, where $\xi \in [-1,1] $:
\begin{align} \label{eq: Bis squeezed}
	\nonumber
	&\mathcal{B}_\pi^{\rm fin}(\tau \to 0,\{k_i\}) \supset \dfrac{M_{3}^8}{H \, M_P^{10} \pi^6 \epsilon^5} \left( g_\pi +\left\lbrace f_\pi \log\dfrac{2 k_1}{k_1+k_2+k_3}+ (k_1\leftrightarrow k_{2,3})\right\rbrace \right) \\
	&\qquad\xrightarrow{p\to 0}  \dfrac{M_{3}^8 }{6451200  \, H \, M_P^{10} \, \pi ^6 \, \epsilon ^5} \dfrac{p}{k}\left(276757-1762\, \xi ^2+\left( 420 -6720\, \xi ^2\right) \log \left(\frac{p}{k}\right) + \order{\dfrac{p}{k}}\right)\,.
\end{align}
We see that in the squeezed limit $p/k\to 0$, the divergent terms cancel out exactly, giving rise to a suppressed contribution $\order{p/k}$. This is consistent with the one-loop scale-invariant power spectrum featured by the model under analysis, see Appendix \ref{app: Power spectrum}.
This cancellation occurs separately in both contributions to the diagram: the contribution coming from the quartic interactions $\pi'^4$ and $\pi'^2 (\partial\pi)^2$. Furthermore, it occurs independently in the different contributions of the in-in formalism: first and second line of Eq.~(\ref{eq: In-In rule 2H}). 

Beyond the squeezed limit, the presence of divergences due to the factors $\propto (k_1+k_2-k_3)^{-\alpha}$ (and permutations) with $\alpha>0$ can, in principle, become manifest in configurations where the three momenta are aligned (often referred to as {\it collinear} or {\it folded} configurations). Assuming without loss of generality $k_1>k_2$, we have $k_3 = k_1 + k_2 \, \xi$ with $\xi \in [-1,1] $, so there exist potentially problematic contributions in the $\xi \to \pm 1$ configurations. 
However, analyzing these configurations in the first line of Eq.~(\ref{eq: B non-trivial}), we observe that these divergences are again canceled independently (in the same sense as above), giving rise to a finite result as $\xi\to\pm 1$. 
In particular, taking the limit $\xi \to 1$ --the remaining collinear configurations being obtained by permutations of the external momenta-- one finds
\begin{align}
	\nonumber
	&\mathcal{B}_\pi^{\rm fin}(\tau \to 0,\{k_i\}) \supset \dfrac{M_{3}^8}{H \, M_P^{10} \pi^6 \epsilon^5} \left( g_\pi +\left\lbrace f_\pi \log\dfrac{2 k_1}{k_1+k_2+k_3}+ (k_1\leftrightarrow k_{2,3})\right\rbrace \right) \\
	&\qquad\xrightarrow{\xi\to 1}  \frac{M_3^8 }{491520 H M_P^{10} \pi ^6 \epsilon ^5} \dfrac{1}{u^3} \left( j_1(u)+ j_2(u) \log u+30 \left( \frac{k_2}{k_1}-\frac{k_1}{k_2}\right)  j_3(u) \log \dfrac{k_2}{k_1}\right) + \order{\xi - 1} \,,
\end{align}
where $u = (k_1+k_2)^2 / (k_1 k_2)$, and we have introduced the polynomials
\begin{align}
	& j_1(u) =  1260 u^8 - 11850 u^7 + 34350 u^6 - 29675 u^5 + 1647 u^4 - 
	298 u^3 + 21664 u^2 - 37362 u - 27960 \,,\\
	& j_2(u) =  -30 (u-1)(21 u^8 - 208 u^7 + 638 u^6 - 546 u^5 - 88 u^4 - 82 u^3 - 
	80 u^2 - 84 u - 42)\,,\\
	& j_3(u) = 21 u^8 - 187 u^7 + 514 u^6 - 446 u^5 + 56 u^4 - 2 u^2 -40 \,.
\end{align}
Taking $k_2 \to 0$ within the collinear configurations reproduces the squeezed limit of Eq.~(\ref{eq: Bis squeezed}) in the cases $\xi = \pm 1$.

It is worth mentioning a couple of aspects. In the second line of Eq.~(\ref{eq: B non-trivial}), the presence of the logarithm adds a new type of divergence in the case of alignment of the three momenta; however, as we already mentioned, this term disappears after the renormalization process. 
Although the analysis we have done of the different limits of the bispectrum has not taken into account the renormalization, which modifies the functions $g_\pi$ and $f_\pi$, the conclusions remain unchanged, as we will discuss next. 

\subsection{Renormalization of the bispectrum}
The counterterms needed to renormalize the diagram of Eq.~\eqref{interv} have to scale as follows: 
\begin{equation} \label{scalingcont}
	\mathcal{L} \sim \dfrac{ \partial^7 \pi^3 }{\tilde\Lambda_U^6}\,,
\end{equation}
where, $\partial^7$ is used here to denote a total seven derivatives which can be both spatial and temporal (and some of the latter may act on the scale factor), and where  for dimensional consistency and for convenience, we have introduced the quantity $\tilde\Lambda_U$, which is related to the energy scale $\Lambda_U$, at which perturbative unitarity is expected to break.\footnote{The quantity $\tilde\Lambda_U$ can be expressed as follows: $\tilde\Lambda_U^6=(\sqrt{\epsilon}\,H\,M_P)^{-3}\Lambda_U^6$, where $\Lambda_U$ is the unitarity breaking scale, such that in terms of the canonically normalized field $\pi_c \sim \sqrt{\epsilon}\,H\, M_P\, \pi$ (with dimension of energy), the interaction Eq.~\eqref{scalingcont} would read $\mathcal{L}\sim \partial^7\pi_c^3/\Lambda_U^6$ (which has dimensions of energy to the fourth power in three spatial dimensions).} 

As shown by the functions of Appendix \ref{app: Bispectrum formulas}, the UV divergences show up in a variety of functional forms. In addition, since the diagram involves many derivatives, renormalization can be expected to require a large number of counterterms reproducing those divergences. The most general cubic Hamiltonian term can be written in Fourier space as:\footnote{We do not include terms with a single Levi-Civita symbol {$\epsilon_{ijk}$} because they cancel out due to momentum conservation (in real space, it can be written as a boundary term). Furthermore, a higher power of Levi-Civita symbols can be rewritten as products of Kronecker deltas.}
\begin{align} \nonumber
	H_I(\tau) & =  \mathcal{C} \int \prod_{i = 1}^3\left( \dfrac{\dd^{3+\delta} \vb{k}_i}{(2 \pi)^{3/2}}\right) (2 \pi)^3 \delta(\vb{k}_1+\vb{k}_2+\vb{k}_3) \left( a H \right) ^{4-n} \\
	&\times  H^3 \pi_{\vb{k}_1}^{(a_1)} \pi_{\vb{k}_2}^{(a_2)} \pi_{\vb{k}_3}^{(a_3)} k_1^{2b_1} k_2^{2b_2} k_3^{2b_3} (\vb{k}_1 \cdot \vb{k}_2)^{c_{12}} (\vb{k}_1 \cdot \vb{k}_3)^{c_{13}} (\vb{k}_2 \cdot \vb{k}_3)^{c_{23}} \,, \label{typecount}
\end{align}
where $a_1$, $a_2$ and $a_3$ count time derivatives, $n = a_1+a_2+a_3+2(b_1+b_2+b_3+c_{12}+c_{13}+c_{23})$ is the total number of derivatives (temporal and spatial) and $\mathcal{C}$ is a dimensionless coupling, that in general is time-dependent but is constant in shift-symmetric invariant scenarios, as is the case with constant $M_3$. The Dirac delta ensures momentum conservation in the vertex and comes from integrating over space.

The aforementioned diversity of UV divergences can be classified into two types that have different origins. As we saw in Section \ref{sec: Dim Reg}, in diagrams with two Hamiltonian insertions we have two sources of UV contributions in the loop, defined by Eqs.~\eqref{noremantint} and \eqref{remantint}, which we will henceforth refer to as first- and second-type contributions, respectively. As we are going to explain, we find that the divergences of the first type cannot be renormalized {\it at all times} (except at $\tau \rightarrow 0$) with the usual procedure by using counterterms of the form given in Eq.~\eqref{typecount}.

Both types of UV divergences are described with Eqs.~(\ref{eq: App B Div 1})-(\ref{eq: App B Div 2}). We can write the action responsible for renormalizing the second type of divergences, using a basis of operators with seven derivatives distributed across the three fields, where each {field in each} element of the basis contains at least one derivative (to preserve shift symmetry explicitly):
\begin{align} \label{fixe}
	\nonumber
	S_{\rm cts} =& \dfrac{1}{\delta}\dfrac{M_{3}^8}{H^4 \, M_P^{4}\, \pi^2 \, \epsilon^2}\int \, \dd \tau \,\dd^{3+\delta}\vb{x} \, \mu^\delta a^{4+\delta-7}\bigg[ \frac{351}{32} \pi' \pi'''^2 + \frac{157}{720} \partial_i \pi \partial_{jk} \pi \partial_{ijk} \pi' +\frac{7}{6} \partial_i \pi \partial_j \pi \partial_{ij} \pi''' \\ \nonumber
	&+ \frac{297}{16} \partial_i \pi \partial_{ij} \pi \partial_j \pi''' +\frac{103}{6} \left( \partial_{ij} \pi\right) ^2 \pi'''-\frac{241}{16} \pi''\partial_i \pi \partial_i \pi''' - \frac{8371}{480} \partial_i \pi' \partial_j \pi' \partial_{ij} \pi' - \frac{219}{16} \pi' \partial_i \pi' \partial_i \pi'''\\ \nonumber
	&- \frac{1831}{144} \partial_i \pi \partial_j \pi'\partial_{ij} \pi'' + \frac{6521}{576} \pi' \partial_{ij} \pi \partial_{ij} \pi'' +\frac{139}{48} \partial_i \pi \partial_j \pi'' \partial_{ij} \pi' - \frac{2431}{288} \partial_{ij} \pi \partial_i \pi' \partial_j \pi '' \\ & + \frac{15317}{576} \pi'' \partial_{ij} \pi \partial_{ij} \pi' 
	 - \frac{29629}{2880} \pi' \partial_i \pi \partial_i \pi''''+ \frac{2923}{576} \pi''''  \partial_i \pi \partial_i\pi' + \frac{1849}{480} \pi' \left( \partial_{ij} \pi'\right) ^2 \bigg] +\mathcal{O}(\delta^0)\,. 
\end{align}
This is the divergent part of the counterterms, which ignores a finite part that should be fixed by renormalization conditions and that gives rise to a rich and complex structure in the external momenta. Eq.~\eqref{fixe} is, instead, solely determined by requiring that the (second type of UV divergent) bispectrum must be finite (i.e.\ non-divergent). 

However, in dimensional regularization, the contribution to the bispectrum from Eq.~\eqref{fixe} leaves a finite remnant due to the effect of the modes and the Fourier volume element (see Eq.\ (\ref{eq: Bispectrum rescaled})) in $3+\delta$ spatial dimensions. This finite contribution exactly cancels out the second line of Eq.~(\ref{eq: B non-trivial}), as we had already anticipated in the previous section, so only the first line of Eq.~(\ref{eq: B non-trivial}) represents an intrinsic contribution of the loop to the bispectrum. Furthermore, as in the late-time limit ($\tau\rightarrow 0$) the quantity $\mathcal{B}^{\rm div}(\tau, \{k_i\})$ comes only from second-type divergences (because the first-type divergences begin at $\order{k \tau}^4$), Eq.\ (\ref{fixe}) is sufficient to ensure that the bispectrum remains finite in the late-time limit.

However, a problem arises when attempting to renormalize the first-type divergences at $\tau \neq 0$ using the same procedure. We find that no combination of local counterterms of the form given in Eq.~\eqref{typecount} can renormalize these divergences. This conclusion remains even if we try {to include} effects that explicitly break the shift symmetry (i.e.\ considering operators without derivatives of $\pi$). The treatment of these divergences goes beyond the scope of this paper, and we postpone their analysis to future work, emphasizing that they play no role in the late-time limit.

In summary, at late times, including the effects of the counterterms described above that renormalize {the} second-type divergences, we obtain the renormalized bispectrum $\mathcal{B}_{\pi,{\rm ren}}^{{\rm c} \times {\rm q}}  \equiv  \mathcal{B}_\pi ^{{\rm c} \times {\rm q}} + \mathcal{B}_\pi ^{\rm cts}$:
\begin{align} \label{eq: B non-trivial renorm}
	\nonumber
	\lim_{\tau\rightarrow 0} \mathcal{B}_{\pi,{\rm ren}}^{{\rm c} \times {\rm q}}{(0,\{k_i\})} & =  \mathcal{B}_{{\pi}}^{\rm div}(0,\{k_i\}) \log \left( \dfrac{H}{\mu e^{\gamma_{\rm E}}}\right) \\
	&+\dfrac{M_{3}^8}{H \, M_P^{10} \pi^6 \epsilon^5} \left( g_{\pi,{\rm ren}}(\{k_i\}) +f_{\pi,{\rm ren}}(\{k_i\})\left[ \log\dfrac{2 k_1}{k_1+k_2+k_3}+ (k_1\leftrightarrow k_{2,3})\right] \right)   \,,
\end{align}
where $g_{\pi,{\rm ren}}$ and $f_{\pi,{\rm ren}}$ are defined in Appendix \ref{app: Bispectrum formulas}. As we explained earlier, the effect of the momenta-dependent logarithm in the second line of Eq.~(\ref{eq: B non-trivial}) was an artifact of working in $3+\delta$ spatial dimensions; so once the renormalization is performed, it disappears. Similarly, the remainder $\log({H}/{\mu e^{\gamma_{\rm E}}})$ can also be eliminated with the finite part of the counterterms, which we are not writing in this expression because we are focusing on the intrinsic contribution of the loop. 

The effect of the counterterms in $3+\delta$ spatial dimensions also modifies the contributions of the second line of Eq.~(\ref{eq: B non-trivial renorm}), although this contribution cannot be completely eliminated. Furthermore, since we have already removed the UV divergences, only finite effects of counterterms, coming from physics in $3$ spatial dimensions, are missing in this equation. These contributions cannot modify the second line of Eq.~(\ref{eq: B non-trivial renorm}), as explained in the previous section, so it corresponds to a truly intrinsic part of the loop correction to the bispectrum. 
Although we do not prove it here,\footnote{We postpone this to a future work.} the result reported in Eq.~(\ref{eq: B non-trivial renorm}) is insensitive to the basis chosen for renormalization. 

We mention that in the squeezed limit the consistency relation is still satisfied, as it should; and there are no collinear divergences, since the potentially problematic term associated to the first line of Eq.~(\ref{eq: B non-trivial}) disappears after renormalization.

\section{Discussion} \label{sec: Discussion} 

The phenomenological relevance of loop corrections is ultimately tied to whether they encode information that cannot be absorbed into counterterms. In the context of the application of the in-in formalism to compute primordial correlators, this amounts in practice to determining if the (comoving) momentum dependence of a renormalized loop contribution is {\it distinguishable} from that produced by local counterterm insertions. If the loop is indistinguishable, its finite effect can be removed by a scheme choice (i.e.\ by shifting the finite parts of counterterm coefficients), and an explicit loop computation carries no scheme-independent physical content for that observable beyond what is already captured by the leading tree-level contribution and the freedom in the counterterms.

For observables whose loop corrections are free of late-time divergences, the available comoving scales strongly constrain the allowed momentum dependence. This follows from the invariance of the background under the rescalings $a\to\lambda a$, $\tau\to\lambda^{-1}\tau$ and $\vb{x}\to\lambda^{-1}\vb{x}$. This symmetry implies that loop contributions can only depend on scale-invariant combinations --such as $H/\mu$ and $k/(aH)$ (where $\mu$ is the renormalization scale and $k$ is an external comoving momentum) and, if several external legs are present, on ratios $k_i/k_j$.
This severely restricts the situations in which loop effects can lead to genuinely new information. For instance, the standard de Sitter corrections $\log(H/\mu)$ to the power spectrum are a prime example of contributions that are scale-invariant and typically fall into the indistinguishable class. 
Moreover, mild departures from exact de Sitter (e.g.\ slow-roll corrections) do not, by themselves, guarantee an intrinsic loop signal (e.g.\ in the power spectrum), since their effect can be reproduced by an appropriate choice of counterterms.

A main goal of our work has been to streamline the computation of the UV part of in-in loop integrals within the dimensional-regularization framework introduced in \cite{Ballesteros:2024cef}. In that approach, the loop is split into an IR piece (finite and computable directly in $3$ spatial dimensions) and a UV piece that must be treated consistently in $3+\delta$ spatial dimensions. The main difficulty that one faces applying the method of \cite{Ballesteros:2024cef} comes from the fact that the UV part involves time integrals along the Schwinger-Keldysh contour and is sensitive to the dynamics in $3+\delta$ dimensions. We have now shown that a large loop-momentum expansion of the UV part of the momentum integrals, together with the $i\epsilon$ prescription, imply a damping of the integrand over appropriate regions of the time integrals. Exploiting this property, one can expand over those regions and systematically trade the difficult time integrals for time derivatives, which can be evaluated analytically. As a result, the one-loop UV computation reduces, in general, to a single remaining time integral, while the momentum integral becomes straightforward because it is controlled by the UV limit of the modes in the loop. Importantly, this simplification relies only on the universal UV behavior dictated by the Bunch-Davies phase $e^{-ip\tau}$ and does not require assumptions about the specific regulator used to handle UV divergences.

Beyond simplifying the UV calculation, our analysis also exposes an additional structural issue: in multi-vertex in-in correlators there can appear UV-divergent terms whose field structure is not manifestly reproducible by the insertion of local Hamiltonian counterterms. We have noted this aspect as an {\it apparent} renormalization obstruction: within the standard framework of single-vertex local counterterm insertions, certain divergent structures do not appear to match the pattern one would expect from counterterm renormalization.

To illustrate the above three points with a concrete example, we have computed the one-loop primordial scalar bispectrum within the decoupling limit of the effective theory of inflation, considering in particular the interaction $(\delta g^{00})^3$ in unitary gauge. This example serves three purposes. First, it illustrates the applicability of the improved UV method in a non-trivial loop computation. Second, it provides an explicit instance where loop corrections are {\it distinguishable}: the presence of several (three) external comoving momenta allows the late-time bispectrum to depend on their ratios, yielding a scheme-independent functional form that cannot be replicated by local counterterms. Third, it exposes the renormalization difficulty that we have identified in a controlled setting: we verified that, at finite times, the insertion of a single local counterterm vertex is not sufficient to renormalize the problematic UV-divergent structures that we encounter in the computation. While the bispectrum becomes finite in the late-time limit, a complete understanding of how to renormalize these UV divergences at all times remains to be achieved.

Finally, the renormalized late-time bispectrum we obtain respects the expected invariance under scale-factor rescalings and it also satisfies the one-loop consistency relation of \cite{Maldacena:2002vr}. This agreement with the consistency relation is non-trivial: it requires cancellations of divergences that appear as singularities in {\it external-momentum} aligned configurations (including the squeezed limit), and that originate from qualitatively different contributions to the loop. In the late-time limit, the renormalized result contains both logarithmic terms and additional contributions that are free of logarithms and depend only on ratios of momentum configurations. Although one might naively associate the intrinsic, scheme-independent part primarily with logarithmic running, the ratio-dependent terms play an equally essential role: they are crucial for the cancellations required by the consistency relation in the squeezed limit, and their functional form cannot be mimicked by counterterms. Overall, this underlines the main message of this work: extracting intrinsic information from cosmological loops requires a controlled UV computation together with a careful treatment of counterterms, and multi-legged correlators (including the bispectrum) constitute a natural arena where genuinely distinguishable loop effects arise.

\appendix

\mysection{Acknowledgments}

{\small
We thank Paolo Creminelli and Mehrdad Mirbabayi for discussions. Work funded by the following grants: PID2021-124704NB-I00 funded by MCIN/AEI/10.13039 /501100011033 \sloppy and by ERDF A way of making Europe, CNS2022-135613 MICIU/AEI/10.13039/501100011033 and by the European Union NextGenerationEU/PRTR, and Centro de Excelencia Severo Ochoa CEX2020-001007-S funded by MCIN/AEI/10.13039/501100011033.
JGE is supported by a PhD contract {\it contrato predoctoral para formaci\'on de doctores} (PRE2021-100714) associated to the aforementioned Severo Ochoa grant, CEX2020-001007-S-21-3. FR is supported by the research grant number 20227S3M3B “Bubble Dynamics in Cosmological
Phase Transitions” under the program PRIN 2022 of the Italian Ministero dell’Università e Ricerca
(MUR).
}

\section{Interaction Hamiltonian in the interaction picture} \label{app: HI}

Let us derive the most general possible form of the interaction Hamiltonian in the interaction picture \cite{Pimentel:2012tw}, $H_I$. 
We start by making the separation 
\begin{equation}
	\mathcal{L} = \mathcal{L}_0 + \mathcal{L}_{\rm int} \,, \quad {\rm where} \quad \mathcal{L}_0 = g_0 \pi'^2 + f(\pi)
\end{equation}
is the free Lagrangian, and the interaction Lagrangian must be, by definition, suppressed. That is, we require perturbation theory to be valid. In a theory without interactions, i.e.\ in the free case, the conjugate momentum is just $p = \partial \mathcal{L}_0 / \partial\pi' = 2 g_0 \pi'$ and therefore the free Hamiltonian is:
\begin{equation}
	\mathcal{H}_0 = p \pi'(\pi,p) - \mathcal{L}_0(\pi,\pi'(\pi,p)) = \dfrac{p^2}{4 g_0} - f(\pi)\,.
\end{equation}
In the general case where we have non-zero interactions, the conjugate momentum is
\begin{equation}
	p = \dfrac{\partial \mathcal{L}}{\partial \pi'} = 2 g_0 \pi' + \dfrac{\partial \mathcal{L}_{\rm int}}{\partial \pi'}\eval_{\pi,\pi'} \,.
\end{equation}
We can invert this relation, so that 
\begin{equation}
	\pi'(\pi,p) = \dfrac{p}{2 g_0} - \dfrac{1}{2 g_0} \dfrac{\partial \mathcal{L}_{\rm int}}{\partial \pi'}\eval_{\pi,\pi'(\pi,p)}.
\end{equation}
Since the second term is suppressed with respect to the first, we can solve this equation perturbatively:
\begin{align}
	\nonumber
	\pi'(\pi,p) &= \dfrac{p}{2 g_0} - \dfrac{1}{2 g_0} \dfrac{\partial \mathcal{L}_{\rm int}}{\partial \pi'}\eval_{\pi,\frac{p}{2 g_0} - \frac{1}{2 g_0} \frac{\partial \mathcal{L}_{\rm int}}{\partial \pi'}\eval_{\pi,\dots}} \\
	&= \dfrac{p}{2 g_0} - \dfrac{1}{2 g_0} \dfrac{\partial \mathcal{L}_{\rm int}}{\partial \pi'}\eval_{\pi,\frac{p}{2 g_0}} + \dfrac{1}{4 g_0^2} \dfrac{\partial \mathcal{L}_{\rm int}}{\partial \pi'}\eval_{\pi,\frac{p}{2 g_0}} \dfrac{\partial^2 \mathcal{L}_{\rm int}}{\partial \pi'^2}\eval_{\pi,\frac{p}{2 g_0}} + \cdots \,.
\end{align}
The interaction Hamiltonian is obtained by subtracting the free Hamiltonian from the total Hamiltonian,
\begin{equation}
	\mathcal{H}_{\rm int} = \mathcal{H} - \mathcal{H}_0 = p \pi'(\pi,p) - \mathcal{L}(\pi,\pi'(\pi,p)) - \left( \dfrac{p^2}{4 g_0} - f(\pi)\right) = - \dfrac{1}{4 g_0}\left(p - 2g_0 \pi'(\pi,p) \right)^2 - \mathcal{L}_{\rm int}(\pi,\pi'(\pi,p))\,.
\end{equation}
In the in-in formalism, one needs the interaction Hamiltonian in the interaction picture, where the fields (and their conjugate momenta) evolve freely. That is, we have to substitute $\pi \to \pi_0$ and $p\to p_0 = 2 g_0 \pi'_0$, being the terms with the subscript $_0$ governed by the free Lagrangian, obtaining:
\begin{align} \label{eq: HI general}
	\nonumber
	\mathcal{H}_I =& - \dfrac{1}{4 g_0}\left(-\dfrac{\partial \mathcal{L}_{\rm int}}{\partial \pi'}\eval_{\pi_0,\pi'_0} + \dfrac{1}{2 g_0} \dfrac{\partial \mathcal{L}_{\rm int}}{\partial \pi'}\eval_{\pi_0,\pi'_0} \dfrac{\partial^2 \mathcal{L}_{\rm int}}{\partial \pi'^2}\eval_{\pi_0,\pi'_0} + \cdots \right)^2 \\
	\nonumber
	&- \mathcal{L}_{\rm int}  \left( \pi_0,\pi'_0 - \dfrac{1}{2 g_0} \dfrac{\partial \mathcal{L}_{\rm int}}{\partial \pi'}\eval_{\pi_0,\pi'_0} + \dfrac{1}{4 g_0^2} \dfrac{\partial \mathcal{L}_{\rm int}}{\partial \pi'}\eval_{\pi_0,\pi'_0} \dfrac{\partial^2 \mathcal{L}_{\rm int}}{\partial \pi'^2}\eval_{\pi_0,\pi'_0} + \cdots \right)\\
	=& -\mathcal{L}_{\rm int} (\pi_0,\pi'_0) + \dfrac{1}{4 g_0} \left( \dfrac{\partial \mathcal{L}_{\rm int}}{\partial \pi'} \eval_{\pi_0,\pi'_0}\right) ^2 - \dfrac{1}{8 g_0^2} \left( \dfrac{\partial \mathcal{L}_{\rm int}}{\partial \pi'} \eval_{\pi_0,\pi'_0}\right) ^2 \dfrac{\partial ^2 \mathcal{L}_{\rm int}}{\partial \pi'^2} \eval_{\pi_0,\pi'_0} + \cdots \,.
\end{align}

\section{Collection of bispectrum formulae} \label{app: Bispectrum formulas}
We recall the result obtained in Section \ref{sec: One-Loop Bispectrum} for the bispectrum coming from the diagram 
	\begin{tikzpicture}[baseline={-2}]
		\draw (0,0.2) -- (0.5,0);
		\draw (0,-0.2) -- (0.5,0);
		\draw (0.5+0.25,0) circle (0.25);
		\draw (1,0) -- (1.5,0);
	\end{tikzpicture}\,:
\begin{align}
\mathcal{B}_\pi ^{{\rm c} \times {\rm q}}(\tau,\{k_i\}) = \dfrac{1}{\delta}  \mathcal{B}_\pi^{\rm div}(\tau,\{k_i\}) + \mathcal{B}_\pi ^{\rm fin}(\tau,\{k_i\})\,,
\end{align}
and in the late time limit
\begin{align}	\nonumber
	\mathcal{B}_\pi^{\rm fin}(0,\{k_i\}) = & \dfrac{M_{3}^8}{H \, M_P^{10} \pi^6 \epsilon^5}  \left\lbrace g_{\pi}(\{k_i\})+f_\pi(\{k_i\}) \left[\log\dfrac{2 k_1}{k_1+k_2+k_3}+ (k_1 \leftrightarrow k_{2,3})\right]\right\rbrace\\ & + {\mathcal{B}_\pi^{\rm div}}(0,\{k_i\}) \left[ \dfrac{1}{2}\log\left( \dfrac{(k_1+k_2-k_3)(k_1-k_2+k_3)(-k_1+k_2+k_3)}{(k_1+k_2+k_3)^3}\right) -3\log \left( \dfrac{\mu e^{\gamma_{\rm E}}}{H}\right) \right]\,. \label{eq: B non-trivial 2}
\end{align}
In this appendix we provide the functions $f_\pi(\{k_i\})$, $g_\pi(\{k_i\})$ and ${\mathcal{B}_\pi^{\rm div}}(\tau,\{k_i\})$. Concretely:
\begin{align}
g_\pi=g_{\pi'^4}^{(1)} + g_{\pi'^4}^{(2)} + g_{\pi'^2 \partial\pi^2}^{(1)}+ g_{\pi'^2 \partial\pi^2}^{(2)}\,,
\end{align}
where the functions $g_{\pi'^4}^{(1)}$, $g_{\pi'^4}^{(2)}$, $g_{\pi'^2 \partial\pi^2}^{(1)}$ and $g_{\pi'^2 \partial\pi^2}^{(2)}$ are given below. For clarity, we denote the contributions from the quartic interactions $\pi '^4$ and $\pi'^2 \partial\pi^2$ with their corresponding subscripts. The superscripts denote instead the contributions coming from the two different kinds of time integrals: the ones corresponding to integrals over the ``square" {region} $\tau'\in (-\infty,\tau]$ and $\tau''\in (-\infty,\tau]$ are denoted with $^{(1)}$ (left panel of Fig.\ \ref{fig:intregions}), whereas those defined over the ``triangle" region $\tau'\in (-\infty,\tau]$ and $\tau''\in (-\infty,\tau']$ are indicated with the superscript $^{(2)}$ {(right panel of Fig.\ \ref{fig:intregions})}. Their full expressions are the following:
\begin{align} \label{eq: App Res B 1}
	\nonumber
	&g_{\pi'^4}^{(1)} = \frac{K_3^3 }{5120 K_1^3  \big(K_1^3-4 K_1 K_2^2+8 K_3^3\big)^6} \big(16384 K_1 K_3^{15} \big(69 K_1^2-584 K_2^2\big)-2048 K_1^2 K_3^{12} \big(1291 K_1^4-3208 K_1^2 K_2^2\\\nonumber
	&\qquad\qquad-4464 K_2^4\big)+128 K_1^4 K_3^6 (K_1+2 K_2) (K_1-2 K_2) \big(51 K_1^6+1288 K_1^4 K_2^2-9512 K_1^2 K_2^4-3104 K_2^6\big)\\\nonumber
	&\qquad\qquad+K_1^6 \big(K_1^2-4 K_2^2\big)^3 \big(39 K_1^6-892 K_1^4 K_2^2+464 K_1^2 K_2^4-1600 K_2^6\big)+16 K_1^5 K_3^3 \big(K_1^2-4 K_2^2\big)^2 \big(422 K_1^6\\\nonumber
	&\qquad\qquad-3079 K_1^4 K_2^2+2816 K_1^2 K_2^4-6288 K_2^6\big)-256 K_1^3 K_3^9 \big(2559 K_1^6-22708 K_1^4 K_2^2+52592 K_1^2 K_2^4\\
	&\qquad\qquad+12224 K_2^6\big)+3538944 K_3^{18}\big)\,,\\\nonumber
	&g_{\pi'^4}^{(2)} = \frac{K_3^3}{5120 K_1^7  \big(K_1^3-4 K_1 K_2^2+8 K_3^3\big)^6}  \big(786432 K_3^{18} \big(3795 K_1^4-11688 K_1^2 K_2^2-10400 K_2^4\big)+32768 K_1 K_3^{15}\\\nonumber
	&\qquad\qquad\times \big(34627 K_1^6-207248 K_1^4 K_2^2+99072 K_1^2 K_2^4+746880 K_2^6\big)+4096 K_1^2 K_3^{12} \big(62629 K_1^8-548168 K_1^6 K_2^2\\\nonumber
	&\qquad\qquad+956528 K_1^4 K_2^4+2733056 K_1^2 K_2^6-7440384 K_2^8\big)+64 K_1^4 K_3^6 (K_1+2 K_2) (K_1-2 K_2) \big(26479 K_1^{10}\\\nonumber
	&\qquad\qquad-308284 K_1^8 K_2^2+503840 K_1^6 K_2^4+5624704 K_1^4 K_2^6-24638208 K_1^2 K_2^8+29223936 K_2^{10}\big)\\\nonumber
	&\qquad\qquad+K_1^6 \big(K_1^2-4 K_2^2\big)^3 \big(1491 K_1^{10}-13988 K_1^8 K_2^2+18368 K_1^6 K_2^4+308480 K_1^4 K_2^6-1308416 K_1^2 K_2^8\\\nonumber
	&\qquad\qquad+1545216 K_2^{10}\big)+8 K_1^5 K_3^3 \big(K_1^2-4 K_2^2\big)^2 \big(1681 K_1^{10}-10680 K_1^8 K_2^2-300992 K_1^6 K_2^4+3238464 K_1^4 K_2^6\\\nonumber
	&\qquad\qquad-10439936 K_1^2 K_2^8+11326464 K_2^{10}\big)+1024 K_1^3 K_3^9 \big(31653 K_1^{10}-379888 K_1^8 K_2^2+1183480 K_1^6 K_2^4\\
	&\qquad\qquad+1783648 K_1^4 K_2^6-14743936 K_1^2 K_2^8+19717632 K_2^{10}\big)+3460300800 K_1 K_3^{21}\big)\,,\\\nonumber
	&g_{\pi'^2 \partial\pi^2}^{(1)} = \frac{1}{122880 K_1^3 K_3^3  \big(K_1^3-4 K_1 K_2^2+8 K_3^3\big)^6} \big(131072 K_1 K_3^{21} \big(1461 K_1^2+2138 K_2^2\big)+65536 K_1^2 K_3^{18} \big(2011 K_1^4\\\nonumber
	&\qquad\qquad-4066 K_1^2 K_2^2-6240 K_2^4\big)-512 K_1^4 K_3^{12} \big(1367 K_1^8-44596 K_1^6 K_2^2+154128 K_1^4 K_2^4-78848 K_1^2 K_2^6\\\nonumber
	&\qquad\qquad+399616 K_2^8\big)+4 K_1^6 K_3^6 \big(K_1^2-4 K_2^2\big)^2 \big(1015 K_1^8+40832 K_1^6 K_2^2-138720 K_1^4 K_2^4-320896 K_1^2 K_2^6\\\nonumber
	&\qquad\qquad+77056 K_2^8\big)+K_1^8 \big(K_1^2-4 K_2^2\big)^4 \big(K_1^8-27 K_1^6 K_2^2+114 K_1^4 K_2^4-1744 K_1^2 K_2^6+864 K_2^8\big)\\\nonumber
	&\qquad\qquad+2048 K_1^3 K_3^{15} \big(9379 K_1^6-42456 K_1^4 K_2^2+64848 K_1^2 K_2^4+187136 K_2^6\big)+K_1^7 K_3^3 \big(K_1^2-4 K_2^2\big)^3 \big(223 K_1^8\\\nonumber
	&\qquad\qquad+2478 K_1^6 K_2^2-6552 K_1^4 K_2^4-83296 K_1^2 K_2^6+38016 K_2^8\big)-64 K_1^5 K_3^9 (K_1-2 K_2) (K_1+2 K_2) \big(2780 K_1^8\\
	&\qquad\qquad-74613 K_1^6 K_2^2+268188 K_1^4 K_2^4-67760 K_1^2 K_2^6+227904 K_2^8\big)-78643200 K_3^{24}\big)\,,\\\nonumber
	&g_{\pi'^2 \partial\pi^2}^{(2)} =  \frac{1}{1843200 K_1^7 K_3^3  \big(K_1^3-4 K_1 K_2^2+8 K_3^3\big)^6} \big(9437184 K_3^{24} \big(235043 K_1^4-1023000 K_1^2 K_2^2+312000 K_2^4\big)\\\nonumber
	&\qquad\qquad+131072 K_1 K_3^{21} \big(7304513 K_1^6-65072692 K_1^4 K_2^2+161149968 K_1^2 K_2^4-76464000 K_2^6\big)\\\nonumber
	&\qquad\qquad+16384 K_1^2 K_3^{18} \big(14537465 K_1^8-191743680 K_1^6 K_2^2+900891856 K_1^4 K_2^4-1671398784 K_1^2 K_2^6\\\nonumber
	&\qquad\qquad+875197440 K_2^8\big)+1280 K_1^4 K_3^{12} \big(2801033 K_1^{12}-60947464 K_1^{10} K_2^2+545896224 K_1^8 K_2^4\\\nonumber
	&\qquad\qquad-2563989248 K_1^6 K_2^6+6529769728 K_1^4 K_2^8-8220727296 K_1^2 K_2^{10}+3698491392 K_2^{12}\big)\\\nonumber
	&\qquad\qquad-12 K_1^6 K_3^6 \big(K_1^2-4 K_2^2\big)^2 \big(297579 K_1^{12}+1547176 K_1^{10} K_2^2-61249472 K_1^8 K_2^4+425422848 K_1^6 K_2^6\\\nonumber
	&\qquad\qquad-1230317824 K_1^4 K_2^8+1349617664 K_1^2 K_2^{10}-152248320 K_2^{12}\big)-K_1^8 \big(K_1^2-4 K_2^2\big)^4 \big(2701 K_1^{12}\\\nonumber
	&\qquad\qquad-29951 K_1^{10} K_2^2+100518 K_1^8 K_2^4-401200 K_1^6 K_2^6+2993792 K_1^4 K_2^8-11309568 K_1^2 K_2^{10}\\\nonumber
	&\qquad\qquad+14123520 K_2^{12}\big)+6144 K_1^3 K_3^{15} \big(6202983 K_1^{10}-106152316 K_1^8 K_2^2+719605600 K_1^6 K_2^4\\\nonumber
	&\qquad\qquad-2356991360 K_1^4 K_2^6+3554491136 K_1^2 K_2^8-1792834560 K_2^{10}\big)-K_1^7 K_3^3 \big(K_1^2-4 K_2^2\big)^3 \big(157595 K_1^{12}\\\nonumber
	&\qquad\qquad-749230 K_1^{10} K_2^2-8776008 K_1^8 K_2^4+75192256 K_1^6 K_2^6-170676224 K_1^4 K_2^8-47881728 K_1^2 K_2^{10}\\\nonumber
	&\qquad\qquad+389928960 K_2^{12}\big)+32 K_1^5 K_3^9 (K_1+2 K_2) (K_1-2 K_2) \big(3932767 K_1^{12}-112054592 K_1^{10} K_2^2\\\nonumber
	&\qquad\qquad+1159118832 K_1^8 K_2^4-5930590912 K_1^6 K_2^6+15794888192 K_1^4 K_2^8-19741578240 K_1^2 K_2^{10}\\
	&\qquad\qquad+7883735040 K_2^{12}\big)+2146015641600 K_1 K_3^{27}\big)\,,
\end{align}
where we recall the definitions:
\begin{align}
K_1  \equiv k_1 + k_2 + k_3\,,\quad
K_2^2  \equiv k_1k_2 + k_1k_3 + k_2k_3\,,\quad
K_3^3  \equiv k_1 k_2 k_3 \,.
\end{align}
Similarly:
\begin{align}
f_\pi= f_{\pi'^4}^{(1)} + f_{\pi'^4}^{(2)} + f_{\pi'^2 \partial\pi^2}^{(1)}+ f_{\pi'^2 \partial\pi^2}^{(2)}\,,
\end{align}
where
\begin{align}
	&f_{\pi'^4}^{(1)} = \frac{3 k_1^3 k_2 k_3 (k_2+k_3) (k_1+2 (k_2+k_3))}{32 (k_1-k_2-k_3)^7}\,,\\\nonumber
	&f_{\pi'^4}^{(2)} = -\frac{3 k_1 k_2 k_3 (k_2+k_3) }{32 (k_1-k_2-k_3)^7 (k_1+k_2+k_3)^7}\big(k_1^{10}+9 k_1^9 (k_2+k_3)+k_1^8 \big(65 k_2^2+60 k_2 k_3+65 k_3^2\big)+77 k_1^7 (k_2+k_3)^3\\\nonumber
	&\qquad\qquad+7 k_1^6 \big(41 k_2^4+80 k_2^3 k_3+90 k_2^2 k_3^2+80 k_2 k_3^3+41 k_3^4\big)+91 k_1^5 (k_2+k_3)^5+7 k_1^4 (k_2+k_3)^2 \big(29 k_2^4+12 k_2^3 k_3\\\nonumber
	&\qquad\qquad+26 k_2^2 k_3^2+12 k_2 k_3^3+29 k_3^4\big)+15 k_1^3 (k_2+k_3)^7+4 k_1^2 (k_2+k_3)^4 \big(5 k_2^4-17 k_2^3 k_3+19 k_2^2 k_3^2-17 k_2 k_3^3\\
	&\qquad\qquad+5 k_3^4\big)-2 k_2 k_3 (k_2+k_3)^6 \big(k_2^2-4 k_2 k_3+k_3^2\big)\big)\,,\\\nonumber
	&f_{\pi'^2 \partial\pi^2}^{(1)} = \frac{k_1}{15360 k_2 k_3 (-k_1+k_2+k_3)^7} \big(2 k_1^8-14 k_1^7 (k_2+k_3)+k_1^6 \big(8 k_2^2+84 k_2 k_3+8 k_3^2\big)-k_1^5 (k_2+k_3) \big(127 k_2^2\\\nonumber
	&\qquad\qquad-155 k_2 k_3+127 k_3^2\big)+k_1^4 \big(103 k_2^4+917 k_2^3 k_3+3068 k_2^2 k_3^2+917 k_2 k_3^3+103 k_3^4\big)+2 k_1^3 (k_2+k_3) \big(88 k_2^4\\\nonumber
	&\qquad\qquad+10 k_2^3 k_3+129 k_2^2 k_3^2+10 k_2 k_3^3+88 k_3^4\big)-2 k_1^2 (k_2+k_3)^2 \big(59 k_2^4+400 k_2^3 k_3-203 k_2^2 k_3^2+400 k_2 k_3^3\\\nonumber
	&\qquad\qquad+59 k_3^4\big)-35 k_1 (k_2+k_3)^3 \big(k_2^4+3 k_2^3 k_3-14 k_2^2 k_3^2+3 k_2 k_3^3+k_3^4\big)+5 (k_2+k_3)^4 \big(k_2^4+3 k_2^3 k_3\\
	&\qquad\qquad-14 k_2^2 k_3^2+3 k_2 k_3^3+k_3^4\big)\big)\,,\\\nonumber
	&f_{\pi'^2 \partial\pi^2}^{(2)} = -\frac{1}{15360 k_1 k_2 k_3 (k_1-k_2-k_3)^7 (k_1+k_2+k_3)^7}\big( -2 k_1^{17}+4 k_1^{15} \big(12 k_2^2+7 k_2 k_3+12 k_3^2\big)\\\nonumber
	&\qquad\qquad+295 k_1^{14} \big(k_2^3+k_3^3\big)+2 k_1^{13} \big(519 k_2^4-322 k_2^3 k_3-2402 k_2^2 k_3^2-322 k_2 k_3^3+519 k_3^4\big)+2 k_1^{12} (k_2+k_3)\\\nonumber
	&\qquad\qquad \times \big(2179 k_2^4-4244 k_2^3 k_3-11856 k_2^2 k_3^2-4244 k_2 k_3^3+2179 k_3^4\big)+8 k_1^{11} (k_2+k_3)^2 \big(146 k_2^4-2189 k_2^3 k_3\\\nonumber
	&\qquad\qquad-9170 k_2^2 k_3^2-2189 k_2 k_3^3+146 k_3^4\big)-k_1^{10} (k_2+k_3) \big(147 k_2^6+64995 k_2^5 k_3+214807 k_2^4 k_3^2+301598 k_2^3 k_3^3\\\nonumber
	&\qquad\qquad+214807 k_2^2 k_3^4+64995 k_2 k_3^5+147 k_3^6\big)-2 k_1^9 (k_2+k_3)^4 \big(2211 k_2^4+6920 k_2^3 k_3+65338 k_2^2 k_3^2\\\nonumber
	&\qquad\qquad+6920 k_2 k_3^3+2211 k_3^4\big)-2 k_1^8 (k_2+k_3) \big(5728 k_2^8+25814 k_2^7 k_3+115623 k_2^6 k_3^2+181534 k_2^5 k_3^3\\\nonumber
	&\qquad\qquad+207274 k_2^4 k_3^4+181534 k_2^3 k_3^5+115623 k_2^2 k_3^6+25814 k_2 k_3^7+5728 k_3^8\big)+20 k_1^7 (k_2+k_3)^6 \big(48 k_2^4\\\nonumber
	&\qquad\qquad+1231 k_2^3 k_3-2842 k_2^2 k_3^2+1231 k_2 k_3^3+48 k_3^4\big)+k_1^6 (k_2+k_3)^3 \big(4529 k_2^8+83713 k_2^7 k_3-99146 k_2^6 k_3^2\\\nonumber
	&\qquad\qquad+111231 k_2^5 k_3^3+75122 k_2^4 k_3^4+111231 k_2^3 k_3^5-99146 k_2^2 k_3^6+83713 k_2 k_3^7+4529 k_3^8\big)+10 k_1^5 (k_2+k_3)^8 \\\nonumber
	&\qquad\qquad\times \big(121 k_2^4+726 k_2^3 k_3-1094 k_2^2 k_3^2+726 k_2 k_3^3+121 k_3^4\big)+2 k_1^4 (k_2+k_3)^5 \big(1167 k_2^8+11112 k_2^7 k_3\\\nonumber
	&\qquad\qquad-59428 k_2^6 k_3^2+32840 k_2^5 k_3^3-37494 k_2^4 k_3^4+32840 k_2^3 k_3^5-59428 k_2^2 k_3^6+11112 k_2 k_3^7+1167 k_3^8\big)\\\nonumber
	&\qquad\qquad+k_1^2 (k_2+k_3)^7 \big(91 k_2^8+581 k_2^7 k_3-6700 k_2^6 k_3^2+10371 k_2^5 k_3^3-13838 k_2^4 k_3^4+10371 k_2^3 k_3^5-6700 k_2^2 k_3^6\\\nonumber
	&\qquad\qquad+581 k_2 k_3^7+91 k_3^8\big)-2 (k_2-k_3)^2 (k_2+k_3)^9 \big(2 k_2^6+14 k_2^5 k_3+17 k_2^4 k_3^2+190 k_2^3 k_3^3+17 k_2^2 k_3^4\\ \label{eq: App Res B 2}
	&\qquad\qquad+14 k_2 k_3^5+2 k_3^6\big)\big) \,.
\end{align}
And finally, 
\begin{align}
	\mathcal{B}_{{\pi}}^{\rm div} = {\dfrac{M_{3}^8}{H \, M_P^{10} \pi^6 \epsilon^5}} \left( d_{\pi'^4}^{(1)} + d_{\pi'^4}^{(2)} + d_{\pi'^2 \partial\pi^2}^{(1)}+ d_{\pi'^2 \partial\pi^2}^{(2)}\right) \,,
\end{align}
where
\begin{align}
	\nonumber
	&d_{\pi'^4}^{(1)} = \frac{81 K_1 K_3^3}{2048} \tau ^4 + \frac{K_3^3  \big(-77 K_1^3+174 K_1 K_2^2+480 K_3^3\big)}{4096} \tau ^6\\ \label{eq: App B Div 1}
	&-\frac{K_3^3  \big(K_1^3 K_2^2+78 K_1^2 K_3^3-8 K_1 K_2^4-16 K_2^2 K_3^3\big)}{4096} \tau ^8\,,\\ \nonumber
	&d_{\pi'^4}^{(2)} = \frac{3 K_3^3 \big(K_1^2 K_2^2+3 K_1 K_3^3-6 K_2^4\big) }{32 K_1^7} \big(2+ \big(K_1^2-2 K_2^2\big)\tau ^2 \big)\\\nonumber
	&-\frac{3 K_3^3  \big(33 K_1^7+16 K_1^5 K_2^2-160 K_1^3 K_2^4+16 K_3^3 \big(3 K_1^4-4 K_1^2 K_2^2-48 K_2^4\big)+384 K_1 K_2^6+384 K_1 K_3^6\big)}{2048 K_1^6} \tau ^4\\\nonumber
	&+\frac{K_3^3  \big(71 K_1^7+22 K_1^5 K_2^2-64 K_1^3 K_2^4+16 K_3^3 \big(5 K_1^4+2 K_1^2 K_2^2-48 K_2^4\big)+192 K_1 K_2^6+384 K_1 K_3^6\big)}{4096 K_1^4} \tau ^6\\
	& +\frac{K_3^3  \big(-41 K_1^5 K_2^2-314 K_1^4 K_3^3+104 K_1^3 K_2^4+472 K_1^2 K_2^2 K_3^3-32 K_1 \big(K_2^6+3 K_3^6\big)+192 K_2^4 K_3^3\big)}{20480 K_1^2} \tau ^8\,,\\ \nonumber
	&d_{\pi'^2 \partial\pi^2}^{(1)} = \frac{1}{49152 K_3^3}  \big(-13 K_1^7+91 K_1^5 K_2^2+33 K_1^4 K_3^3-244 K_1^3 K_2^4-324 K_1^2 K_2^2 K_3^3\\ \nonumber
	& \qquad\qquad+16 K_1 \big(22 K_2^6+83 K_3^6\big)-352 K_2^4 K_3^3\big)\tau ^4\\ \nonumber
	& +\frac{1}{245760 K_3^3} \big(-9 K_1^5 K_2^4-116 K_1^3 K_2^6+8 K_1 K_3^6 \big(1528 K_2^2-339 K_1^2\big)+K_3^3 \big(18 K_1^6+393 K_1^4 K_2^2\\\nonumber
	&\qquad\qquad-2084 K_1^2 K_2^4-608 K_2^6\big)+608 K_1 K_2^8-480 K_3^9\big) \tau ^6 \\ 
	&+\frac{K_3^3  \big(47 K_1^5-296 K_1^3 K_2^2-4 K_3^3 \big(279 K_1^2+392 K_2^2\big)+784 K_1 K_2^4\big)}{245760} \tau ^8\,,\\ \nonumber
	&d_{\pi'^2 \partial\pi^2}^{(2)} = \frac{1}{15360 K_1^7 K_3^3}\big(6 K_3^6 \big(279 K_1^4-1520 K_1^2 K_2^2+1440 K_2^4\big)+K_1 K_3^3 \big(64 K_1^6-821 K_1^4 K_2^2+3672 K_1^2 K_2^4\\ \nonumber
	&\qquad\qquad-3600 K_2^6\big)+K_1^2 (K_1+2 K_2) (K_1-2 K_2) \big(2 K_1^6+2 K_1^4 K_2^2-21 K_1^2 K_2^4+180 K_2^6\big)+5760 K_1 K_3^9\big)\\ \nonumber
	&\qquad\qquad\times  \big( 2+ \big(K_1^2-2 K_2^2\big) \tau ^2\big) \\ \nonumber
	&+\frac{1}{245760 K_1^6 K_3^3} \big( -192 K_3^9 \big(399 K_1^4-2000 K_1^2 K_2^2+1440 K_2^4\big)+8 K_1 K_3^6 \big(-1853 K_1^6+11192 K_1^4 K_2^2\\ \nonumber
	&\qquad\qquad-37248 K_1^2 K_2^4+31680 K_2^6\big)+K_1^2 K_3^3 \big(-275 K_1^8+5040 K_1^6 K_2^2-25776 K_1^4 K_2^4+64704 K_1^2 K_2^6\\\nonumber
	&\qquad\qquad-34560 K_2^8\big)+K_1^3 (K_1+2 K_2) (K_1-2 K_2) \big(47 K_1^8-141 K_1^6 K_2^2+396 K_1^4 K_2^4-1056 K_1^2 K_2^6\\\nonumber
	&\qquad\qquad+2880 K_2^8\big)-184320 K_1 K_3^{12}\big)\tau ^4\\ \nonumber
	&+ \frac{1}{245760 K_1^4 K_3^3}\big(32 K_3^9 \big(387 K_1^4-1760 K_1^2 K_2^2+1440 K_2^4\big)+4 K_1 K_3^6 \big(427 K_1^6-618 K_1^4 K_2^2+8416 K_1^2 K_2^4\\\nonumber
	&\qquad\qquad-7680 K_2^6\big)-3 K_1^3 K_2^4 (K_1+2 K_2) (K_1-2 K_2) \big(5 K_1^4+68 K_1^2 K_2^2+80 K_2^4\big)+3 K_1^2 K_3^3 \big(10 K_1^8\\\nonumber
	&\qquad\qquad+169 K_1^6 K_2^2-624 K_1^4 K_2^4-1792 K_1^2 K_2^6+320 K_2^8\big)+30720 K_1 K_3^{12}\big)\tau ^6\\\nonumber
	&+\frac{K_3^3 }{245760 K_1^2} \big(-77 K_1^7+4 K_1^5 K_2^2-1632 K_1^4 K_3^3+656 K_1^3 K_2^4+880 K_1^2 K_2^2 K_3^3+192 K_1 \big(7 K_2^6-8 K_3^6\big)\\ \label{eq: App B Div 2}
	&\qquad\qquad-2304 K_2^4 K_3^3\big)\tau ^8\,.
\end{align}
Also, after renormalization,
\begin{align}
	\nonumber
	&g_{\pi,{\rm ren}} = \frac{1}{153600 K_1^6 K_3^3 \big(K_1^3-4 K_1 K_2^2+8 K_3^3\big)^5}\big(-32768 K_1 K_3^{21} \big(217711 K_1^2-873360 K_2^2\big)+8192 K_3^{18}\\\nonumber
	&\qquad\qquad \times \big(329156 K_1^6-2593059 K_1^4 K_2^2+5235768 K_1^2 K_2^4-348000 K_2^6\big)+512 K_1 K_3^{15} \big(2586953 K_1^8\\\nonumber
	&\qquad\qquad-30846854 K_1^6 K_2^2+123613064 K_1^4 K_2^4-169120064 K_1^2 K_2^6+12898560 K_2^8\big)+640 K_1^2 K_3^{12} \\\nonumber
	&\qquad\qquad\times \big(338569 K_1^{10}-5390867 K_1^8 K_2^2+32382800 K_1^6 K_2^4-87062576 K_1^4 K_2^6+91337344 K_1^2 K_2^8\\\nonumber
	&\qquad\qquad-9093120 K_2^{10}\big)+K_1^4 K_3^6 (K_1+2 K_2) (K_1-2 K_2) \big(617319 K_1^{12}-11869444 K_1^{10} K_2^2\\\nonumber
	&\qquad\qquad+93034784 K_1^8 K_2^4-363336384 K_1^6 K_2^6+703942144 K_1^4 K_2^8-570521600 K_1^2 K_2^{10}+59289600 K_2^{12}\big)\\\nonumber
	&\qquad\qquad+2 K_1^6 \big(K_1^2-4 K_2^2\big)^3 \big(21 K_1^{12}-231 K_1^{10} K_2^2-692 K_1^8 K_2^4+17580 K_1^6 K_2^6-91428 K_1^4 K_2^8\\\nonumber
	&\qquad\qquad+183872 K_1^2 K_2^{10}-127680 K_2^{12}\big)+2 K_1^5 K_3^3 \big(K_1^2-4 K_2^2\big)^2 \big(3967 K_1^{12}-67242 K_1^{10} K_2^2\\\nonumber
	&\qquad\qquad+455188 K_1^8 K_2^4-1361284 K_1^6 K_2^6+1084080 K_1^4 K_2^8+1699136 K_1^2 K_2^{10}-2323200 K_2^{12}\big)\\\nonumber
	&\qquad\qquad+80 K_1^3 K_3^9 \big(211830 K_1^{12}-4192499 K_1^{10} K_2^2+33457268 K_1^8 K_2^4-133889616 K_1^6 K_2^6\\
	&\qquad\qquad+270556352 K_1^4 K_2^8-232125440 K_1^2 K_2^{10}+28200960 K_2^{12}\big)-29542318080 K_3^{24}\big) \,,
\end{align}
and
\begin{align}
	\nonumber
	&f_{\pi,{\rm ren}} = \frac{k_1 (k_2+k_3) }{7680  k_2 k_3  (-k_1+k_2+k_3)^6 (k_1+k_2+k_3)^6}\big(295 k_1^{10} \big(k_2^2-k_2 k_3+k_3^2\big)+k_1^8 \big(2233 k_2^4-5773 k_2^3 k_3\\\nonumber
	&\qquad\qquad-25222 k_2^2 k_3^2-5773 k_2 k_3^3+2233 k_3^4\big)+2 k_1^6 (k_2+k_3)^2 \big(83 k_2^4-15417 k_2^3 k_3-48100 k_2^2 k_3^2-15417 k_2 k_3^3\\\nonumber
	&\qquad\qquad+83 k_3^4\big)-2 k_1^4 (k_2+k_3)^4 \big(1223 k_2^4+10049 k_2^3 k_3+21522 k_2^2 k_3^2+10049 k_2 k_3^3+1223 k_3^4\big)-k_1^2 (k_2+k_3)^6\\\nonumber
	&\qquad\qquad \times \big(253 k_2^4+1205 k_2^3 k_3+584 k_2^2 k_3^2+1205 k_2 k_3^3+253 k_3^4\big)+5 (k_2+k_3)^8 \big(k_2^4+3 k_2^3 k_3-14 k_2^2 k_3^2\\
	&\qquad\qquad+3 k_2 k_3^3+k_3^4\big)\big) \,.
\end{align}

\section{One and two-point functions} \label{app: Power spectrum}

In this appendix we compute the one- and two-point functions for the model defined by Eq.~(\ref{langM3}).
For the former, the calculation is supplemented by the renormalization condition 
$\expval{\pi(x)} = 0$ that guarantees that our perturbative expansion is carried out around the correct vacuum configuration.
For the latter, we work in a framework where an auxiliary scale \(\tau_*\) is introduced 
to search for a distinguishable loop contribution in the late-time dimensionless power spectrum.
We will show, however, that the one-loop dependence on this scale can be absorbed 
into the tree-level counterterm contribution, indicating that it does not correspond 
to a genuine, intrinsic one-loop {contribution}.

\subsection{One-point function}

The interaction Hamiltonian to be used is 
\begin{equation}
	\mathcal{H}_I = \mu^\delta a^{-1+\delta} {M_3^4(t)} \dfrac{4}{3}a^2\pi'^3 \,, \quad \textrm{where in general} \quad H_I(t) = \int \dd^{3+\delta} \vb{x} \,  \mathcal{H}_I(x)	\,.
\end{equation}
We have that
\begin{equation}
	\expval{\pi(x)}^{\rm 1l} = \begin{tikzpicture}[baseline={-2}]
		\draw (2,-0.3) -- (2,0.15);
		\draw (2,0.45) circle (0.3);
	\end{tikzpicture} = \int \dd^{3+\delta} \vb{k} \, \delta(\vb{k}) \, 8 \Im{\int_{-\infty_-}^\tau \dd \tau' \mu^\delta a^{1+\delta}(\tau') M_3^4(t')\, \pi_k(\tau) \pi_k'^*(\tau') \int \dfrac{\dd^{3+\delta} \vb{p}}{(2\pi)^{3+\delta}} \abs{\pi_p'(\tau')}^2} \,.
\end{equation}
This expression is both UV and late-time divergent. The latter divergence will be cured automatically after the renormalization, that is done considering linear counterterms associated to $M_0$ and $M_1$.

The action of the effective theory of inflation in the decoupling limit contains:
\begin{align}
	S \supset \int \dd \tau\, \dd^{3+\delta} \vb{x} \,\mu^\delta a^{4+\delta} \left\lbrace \delta M_0^4(t+\pi) -  \frac{\delta M_1^4(t+\pi) }{a^2} \left( 2a \pi'+\pi'^2 - \left( \partial \pi \right) ^2   \right)  \right\rbrace\,,
\end{align}
where $\delta M_0^4$ and $\delta M_1^4$ come from Eq.~(\ref{counterM}),
so that the linear operators in the Lagrangian that serve as counterterms of the one-point function are:
\begin{equation}
	\mathcal{L}_{\delta \begin{tikzpicture}[baseline={1}]
			\draw (0,0) -- (0,0.1);
			\draw (0,0.175) circle (0.075);
	\end{tikzpicture}}^{(1)} = \mu^\delta a^{4+\delta}\left\lbrace  \partial_t(\delta{M}_0^4(t)) \pi - \dfrac{2}{a} \delta M_1^4(t) \pi'\right\rbrace \,.
\end{equation}
Taking $\mathcal{H}_I=-\mathcal{L}_{\delta \begin{tikzpicture}[baseline={1}]
		\draw (0,0) -- (0,0.1);
		\draw (0,0.175) circle (0.075);
\end{tikzpicture}}^{(1)}$ (neglecting terms $\order{M_{3}^8}$ coming from Eq.\ (\ref{eq: HI general}) in Appendix \ref{app: HI}) we find the counterterm contribution:
\begin{align}
	\nonumber
	\expval{\pi(x)}^{\rm cts} = \begin{tikzpicture}[baseline={-2}]
		\draw (1,-0.3) -- (1,0.3);
		\draw[fill=white,cross] (1,0.3) circle (0.15);
	\end{tikzpicture}  =& \int \dd^{3+\delta} \vb{k} \, \delta(\vb{k}) \, 2 \Im\bigg\lbrace\int_{-\infty_-}^\tau \dd \tau' \mu^\delta a^{4+\delta}(\tau') \, \pi_k(\tau)\\
	&\times\bigg[  \dfrac{2}{a(\tau')} \delta M_1^4(t') \pi_k'^*(\tau') - \partial_t(\delta{M}_0^4(t'))  \pi_k^*(\tau')  \bigg] \bigg\rbrace\,.
\end{align}

Imposing the renormalization condition $\expval{\pi(x)}^{\rm 1l}+\expval{\pi(x)}^{\rm cts}=0$, we have:
\begin{align}
	\partial_t( \delta M_0^4(t)) = 0\,, \quad 
	\delta M_1^4(t) = - \dfrac{2}{a^2} M_3^4(t) \int \dfrac{\dd^{3+\delta} \vb{p}}{(2\pi)^{3+\delta}} \abs{\pi_p'(\tau)}^2 =  - \dfrac{3}{4} \mathcal{P}_\zeta^{0}  M_3^4(t) \,,
\end{align}
where we have defined $\mathcal{P}_\zeta^{0} =  H^2/{(8 \pi^2 M_P^2 \epsilon)}$. Although we have imposed the renormalization condition prior to performing the time integrals, the result obtained by integrating in time first would be the same.

\subsection{Two-point function}

To compute the two-point function, we need to consider the following contributions in the Lagrangian
\begin{align} 
	\mathcal{L}_{\delta \begin{tikzpicture}[baseline={1}]
			\draw (0,0) -- (0,0.1);
			\draw (0,0.175) circle (0.075);
	\end{tikzpicture}} =&  \mu^\delta a^{2+\delta} {M_3^4(t+\pi)}    \dfrac{3}{4}\mathcal{P}_\zeta^{0} \left(2 a \pi' +  \pi'^2 - \left(  \partial \pi\right) ^2 \right) \,,\\
	\mathcal{L}_{\rm int} =& - \mu^\delta a^{-1+\delta} {M_3^4(t+\pi)} \left( \dfrac{4}{3}a^2\pi'^3 + 2 a \pi'^2 \left( \pi'^2 - \left(  \partial \pi\right) ^2 \right) + \pi' \left( \pi'^2 - \left(  \partial \pi\right) ^2 \right)^2  \right)   \, ,
\end{align}
whose associated $\mathcal{H}_I$ is (see Eq.\ (\ref{eq: HI general})):
\begin{equation}
	\mathcal{H}_I = -\mathcal{L}_{\delta 
		\begin{tikzpicture}[baseline={1}]
			\draw (0,0) -- (0,0.1);
			\draw (0,0.175) circle (0.075);
	\end{tikzpicture}} 
	- \mathcal{L}_{\rm int} + \dfrac{1}{4 \mu^\delta a^{2+\delta} M_P^2 H^2 \epsilon} \left(\dfrac{\partial \mathcal{L}_{\delta 
			\begin{tikzpicture}[baseline={1}]
				\draw (0,0) -- (0,0.1);
				\draw (0,0.175) circle (0.075);
		\end{tikzpicture}} 
		+ \mathcal{L}_{\rm int} }{\partial \pi'} \right) ^2 + \order{M_{3}^{12}}\,.
\end{equation}
In general, the connected structure of the two-point correlation will be: 
\begin{equation} 
	\expval{\pi(\tau,\vb{x}) \pi(\tau,\vb{y})} = \int \dfrac{\dd^{3+\delta} \vb{k}}{(2\pi)^{3+\delta}} e^{i \vb{k}\cdot(\vb{x} - \vb{y})} P_\pi(\tau,k)\,, \quad \mathcal{P}_\pi(\tau,k) \equiv \dfrac{k^{3+\delta}}{2 \pi^2} P_\pi(\tau,k) \,,
\end{equation}
where $\mathcal{P}_\pi(\tau,k)$ is the rescaled power spectrum. While this quantity has mass dimension $-2$, it is convenient to define it in analogy with the dimensionless spectrum of curvature perturbations, $\mathcal{P}_\zeta=H^2\mathcal{P}_\pi$, where we have used the linear order relation $\zeta = -H \pi$.

In order to introduce an auxiliary scale $\tau_*$, we will consider the following explicit functional form for $M_3(t)$:
\begin{align} \label{eq: App Power time dep M3}
	M_3^4(t(\tau)) = M_{3,*}^4\left(1-\varepsilon_*\log\frac{\tau}{\tau_*}+\order{\varepsilon_*^2}\right),
\end{align}
where the breaking of the $\pi$-shift symmetry is controlled by the parameter 
\begin{align}
	\varepsilon_*=\frac{\dot{M}_{3}^4}{HM_3^4} \eval_{\tau_*} \,,
\end{align}
whose reference value $\varepsilon_*\equiv \varepsilon\big|_{\tau=\tau_*}$ is $\ll 1$ in absolute value.

The leading contribution in the coupling $M_3$ comes from the quartic interaction: 
\begin{equation}
	\mathcal{P}_\pi^{\rm q} = \begin{tikzpicture}[baseline={-2}]
		\draw (0,0) -- (1,0);
		\draw[line width=0.4pt, fill=white] 
		(0.5,0) .. controls (0.95,0.6) and (0.05,0.6) .. (0.5,0);
	\end{tikzpicture} \propto \dfrac{M_{3,*}^4}{M_P^6 \epsilon^3}\,.
\end{equation}
As in Section \ref{sec: One-Loop Bispectrum}, the vertex gives the factor $M_3^4$, while each field (6 in total) introduces a factor $1/(M_P \sqrt{\epsilon})$. As mentioned, this object has energy dimensions $-2$.
This diagram has the same overall size as the quadratic tadpole-counterterm contribution,
\begin{equation}
	\mathcal{P}_\pi^{\delta \begin{tikzpicture}[baseline={1}]
			\draw (0,0) -- (0,0.1);
			\draw (0,0.175) circle (0.075);
	\end{tikzpicture}} = \begin{tikzpicture}[baseline={-2}]
		\draw (0,0) -- (1,0);
		\draw[fill=white,cross] (0.5,0) circle (0.15);
	\end{tikzpicture} \propto \dfrac{M_{3,*}^4}{M_P^6 \epsilon^3}\,.
\end{equation}
Both contributions must be included to obtain a finite late-time result.

The interaction Hamiltonian (density) we have to consider is:\footnote{We have omitted a time dependence for simplicity, e.g.\ $M_3^4 = M_3^4(t) \neq M_{3,*}^4 $.}
\begin{align} \label{eq: App Power HI LO}
	\mathcal{H}_I^{\rm LO} =& \mu^\delta a^{\delta} M_3^4  \left[  \dfrac{4}{3} \varepsilon_* \, a H\, \pi \, \pi'^3 + 2 \pi'^2 \left( \pi'^2 - \left(  \partial \pi\right) ^2 \right)  - \dfrac{3}{4}a^2\mathcal{P}_\zeta^{0} \left(2 \varepsilon_*\, a H \, \pi\,  \pi' +  \pi'^2 - \left(  \partial \pi\right) ^2 \right) \right]	\,.
\end{align}
We have expanded $M_3^4(t+\pi) = M_3^4(t)(1+H \varepsilon \pi)$, which induces terms that break the shift symmetry of~$\pi$ explicitly. 
This Hamiltonian induces the two contributions discussed above, associated respectively to the quadratic tadpole-counterterm and to the quartic interaction. We report them here before doing the respective time integrals, as well as the resulting expressions in the late-time limit:
\begin{align}
	\nonumber & \mathcal{P}_\pi^{\delta \begin{tikzpicture}[baseline={1}]
			\draw (0,0) -- (0,0.1);
			\draw (0,0.175) circle (0.075);
	\end{tikzpicture}} = \begin{tikzpicture}[baseline={-2}]
		\draw (0,0) -- (1,0);
		\draw[fill=white,cross] (0.5,0) circle (0.15);
	\end{tikzpicture}- \dfrac{k^{3+\delta}}{2\pi^2} 3 \Im{\pi_k^2(\tau) \int_{-\infty_-}^\tau \dd \tau' \, \mu^\delta a^{2+\delta} M_3^4 \mathcal{P}_\zeta^0\left( 2\varepsilon_* a H \, \pi_k^* \left( \pi_k^*\right) ' + \left( \pi_k^*\right) '^2 -k^2 \pi_k^{*2} \right) \eval_{\tau'}} \\
	&\xrightarrow{\tau \to 0} - \dfrac{3 M_{3,*}^4}{256 M_P^6\pi^4\epsilon^3} \left\{ 1+ \varepsilon_* \left[ 2-\gamma_{\rm E} - \log\left(  - \dfrac{2 k \tau^2}{\tau_*} \right)  \right]   \right\} \,,\\
	\nonumber &\mathcal{P}_\pi^{\rm q} = \begin{tikzpicture}[baseline={-2}]
		\draw (0,0) -- (1,0);
		\draw[line width=0.4pt, fill=white] 
		(0.5,0) .. controls (0.95,0.6) and (0.05,0.6) .. (0.5,0);
	\end{tikzpicture}=\\
	\nonumber &=\dfrac{k^{3+\delta}}{2\pi^2} \int \dfrac{\dd^{3+\delta} \vb{p}}{(2\pi)^3} 8 \Im \Bigg\lbrace \pi_k^2(\tau) \int_{-\infty_-}^\tau \dd \tau' \, \mu^\delta a^{\delta} M_3^4 \Bigg( 2\varepsilon_* a H \left[  \pi_k^* \left( \pi_k^*\right) ' \abs{\pi_p'}^2 + \left( \pi_k^*\right) '^2 \dfrac{1}{2} \left(\pi_p \left( \pi_p^*\right) ' + \pi_p^*  \pi_p ' \right) \right]  \\
	\nonumber &\quad+ 6 \left( \pi_k^*\right) '^2 \abs{\pi_p'}^2 -\left[ k^2 \pi_k^{*2} \abs{\pi_p'}^2+  p^2 \left( \pi_k^*\right) '^2 \abs{\pi_p}^2 + 4 \pi_k^* \left( \pi_k^*\right) ' \dfrac{(\vb{k} \cdot \vb{p})}{2} \left(\pi_p \left( \pi_p^*\right) ' - \pi_p^*  \pi_p ' \right) \right] \Bigg) \eval_{\tau'} \Bigg\rbrace\\
	& \xrightarrow{\tau \to 0} - \dfrac{M_{3,*}^4 }{256 M_P^6 \pi^4 \epsilon^3} \left\{ 9+ \varepsilon_* \left[ -2+15\gamma_{\rm E} +3 \log\left(  -32 k^5 \tau^2 \tau_*^3 \right)  \right]   \right\}\,.
\end{align}
The loop contribution, calculated using the procedure developed in Section \ref{sec: Dim Reg}, is finite in dimensional regularization as only polynomial divergences are present. Although both contributions contain late-time divergences associated with interactions of Eq.\ (\ref{eq: App Power HI LO}) that break shift symmetry (see discussion around Eq.~(\ref{eq: Pi late-time})), combining these two results to calculate the total power spectrum, we obtain a precise cancellation of the divergences at late-time:
\begin{equation}
	\mathcal{P}_\pi^{\rm q} +\mathcal{P}_\pi^{\delta \begin{tikzpicture}[baseline={1}]
			\draw (0,0) -- (0,0.1);
			\draw (0,0.175) circle (0.075);
	\end{tikzpicture}}  =  \begin{tikzpicture}[baseline={-2}]
		\draw (0,0) -- (1,0);
		\draw[line width=0.4pt, fill=white] 
		(0.5,0) .. controls (0.95,0.6) and (0.05,0.6) .. (0.5,0);
	\end{tikzpicture} + \begin{tikzpicture}[baseline={-2}]
		\draw (0,0) -- (1,0);
		\draw[fill=white,cross] (0.5,0) circle (0.15);
	\end{tikzpicture} \xrightarrow{\tau \to 0} - \dfrac{M_{3,*}^4}{64 M_P^6 \pi^4 \epsilon^3} \left\{ 3 + \varepsilon_* \left[ 1 + 3\gamma_{\rm E} +3 \log\left(  -2 k \tau_*\right) \right]  \right\}\,.
\end{equation}
This result shows a dependence on $k$, unlike the free (tree-level) power spectrum, which, in de Sitter, is exactly scale invariant in the late-time limit: $\mathcal{P}^{\rm 0}_\pi \xrightarrow{\tau \to 0} (8 \pi^2 M_P^2 \epsilon)^{-1}=\mathcal{P}^{\rm 0}_\zeta/H^2$. As we discussed in Section \ref{sec: Intro}, this does not indicate that the loop calculation contains an intrinsic scale dependence, or as we have defined, that it is distinguishable. Before reaching such a conclusion, it is necessary to make a comparison with the contribution of the counterterms. However, there are two arguments that allow us to anticipate that the loop will be indistinguishable: (i) the external scale $k$ does not enter the loop, so it is reduced to an insertion of a quadratic coupling, as mentioned in Section \ref{sec: Renorm Problem}; (ii) the calculation of the quadratic tadpole-term induces a $k$-dependence of the same type as the loop, $\log(-k \tau_*)$, and it is a tree-level contribution.

This motivates exploring the possibility that the following (next-to-leading order in $M_3$) diagram, suppressed by an extra power of $M_3^4$, but with two vertex insertions, may contain distinguishable parts:
\begin{equation} \label{eq: App Power NLO sche}
	\mathcal{P}_\pi^{\rm c} = \begin{tikzpicture}[baseline={-2}]
		\draw (0,0) -- (0.5,0);
		\draw (0.5+0.25,0) circle (0.25);
		\draw (1,0) -- (1.5,0);
	\end{tikzpicture} \propto \dfrac{M_{3,*}^8}{H^2M_P^8 \epsilon^4}\,,
\end{equation}
where $H$ completes the dimensional analysis. This diagram has the same scaling as the following induced quartic and quadratic diagrams:
\begin{equation}
	\mathcal{P}_\pi^{{\rm c} \times {\rm c}, {\rm ind}} = \begin{tikzpicture}[baseline={-2}]
		\draw (0,0) -- (1,0);
		\draw[line width=0.4pt, fill=white] 
		(0.5,0) .. controls (0.95,0.6) and (0.05,0.6) .. (0.5,0);
		\draw[fill=gray!75] (0.5,0) circle (0.05);
	\end{tikzpicture} \propto \dfrac{M_{3,*}^8}{H^2M_P^8 \epsilon^4}
	\quad {\rm and} \quad
	\mathcal{P}_\pi^{{\rm c} \times \delta \begin{tikzpicture}[baseline={1}]
			\draw (0,0) -- (0,0.1);
			\draw (0,0.175) circle (0.075);
		\end{tikzpicture}, {\rm ind}} = \begin{tikzpicture}[baseline={-2}]
		\draw (0,0) -- (1,0);
		\draw[fill=gray!75,cross] (0.5,0) circle (0.15);
	\end{tikzpicture} \propto \dfrac{M_{3,*}^8}{H^2M_P^8 \epsilon^4} \,.
\end{equation}
These terms arise from quadratic operators generated by combinations of cubic or cubic-tadpole linear interactions. Specifically, they are obtained from $\mathcal{H}_I \supset \tfrac{1}{4g_0} \left(\tfrac{\partial \mathcal{L}_{\rm int}}{\partial \pi'} \right)^2 $ in Eq.\ (\ref{eq: HI general}).

At next-to-leading order in $M_3$ one may worry about consistency with the leading-order tadpole renormalization. That is, we have renormalized the tadpole by imposing 
$\begin{tikzpicture}[baseline={-2}]
		\draw (1,-0.1) -- (1,0.3);
		\draw[fill=white,cross] (1,0.3) circle (0.09);
	\end{tikzpicture} + \begin{tikzpicture}[baseline={-2}]
		\draw (2,-0.1) -- (2,0.15);
		\draw (2,0.3) circle (0.15);
\end{tikzpicture} = 0$, 
but one can ask whether it would be necessary to impose $\begin{tikzpicture}[baseline={-2}]
\draw (1,-0.1) -- (1,0.3);
\draw[fill=white,cross] (1,0.3) circle (0.09);
\end{tikzpicture} + \begin{tikzpicture}[baseline={-2}]
\draw (2,-0.1) -- (2,0.15);
\draw (2,0.3) circle (0.15);
\end{tikzpicture} + \begin{tikzpicture}[baseline={-2}]
\draw (2,-0.1) -- (2,0.15);
\draw (2,0.15) -- (2,0.45);
\draw (2,0.3) circle (0.15);
\end{tikzpicture}= 0$ for consistency with a calculation of the power spectrum at next-to-leading order (in $M_3$).\footnote{We do not include in the discussion the two-loop diagram associated with a quintic insertion because it is suppressed with respect to \begin{tikzpicture}[baseline={-2}]
	\draw (2,-0.1) -- (2,0.15);
	\draw (2,0.3) circle (0.15);
\end{tikzpicture}.}
By including this last term, having four $\pi$'s more than the leading-order loop, we have that 
\begin{equation}
	\mathcal{P}_\pi^{\delta \begin{tikzpicture}[baseline={1}]
			\draw (0,0) -- (0,0.1);
			\draw (0,0.175) circle (0.075);
	\end{tikzpicture}} = \begin{tikzpicture}[baseline={-2}]
		\draw (0,0) -- (1,0);
		\draw[fill=white,cross] (0.5,0) circle (0.15);
	\end{tikzpicture} \sim {\order{\dfrac{M_{3,*}^4}{M_P^6 \epsilon^3}} + \order{\dfrac{M_{3,*}^8}{M_P^{10} \epsilon^5}}}\,.
\end{equation}
The second contribution, associated to the renormalization of \begin{tikzpicture}[baseline={-2}]
	\draw (2,-0.1) -- (2,0.15);
	\draw (2,0.15) -- (2,0.45);
	\draw (2,0.3) circle (0.15);
\end{tikzpicture}, presents a suppression with respect to the diagram we are studying, Eq.\ (\ref{eq: App Power NLO sche}), and therefore it is not necessary to renormalize the tadpole to next-to-leading order in $M_3$.
 
The relevant interaction Hamiltonian at next-to-leading order in $M_3$ is:
\begin{align}
	\mathcal{H}_I^{\rm NLO} =& \mu^\delta a^{\delta} {M_3^4} \, \pi'^2  \left[  \dfrac{4}{3}a\, \pi' +  \dfrac{M_3^4}{M_P^2H^2\epsilon} \left( 4\pi'^2 - \dfrac{3}{2} a^2 \mathcal{P}_\zeta^0 \right)  \right]	\,.
\end{align}
Again, this Hamiltonian contains contributions of a different nature. We have a mixing induced term between the tadpole counterterm and the cubic interaction, as well as an induced quartic contribution, which generate respectively the following contributions to the power spectrum:
\begin{align}
	\nonumber \mathcal{P}_\pi^{{\rm c} \times \delta \begin{tikzpicture}[baseline={1}]
			\draw (0,0) -- (0,0.1);
			\draw (0,0.175) circle (0.075);
		\end{tikzpicture}, {\rm ind}} &= \begin{tikzpicture}[baseline={-2}]
		\draw (0,0) -- (1,0);
		\draw[fill=gray!75,cross] (0.5,0) circle (0.15);
	\end{tikzpicture}=- \dfrac{k^{3}}{2\pi^2} 3 \Im{\pi_k^2(\tau) \int_{-\infty_-}^\tau \dd \tau' \, a^{2} M_3^4 \mathcal{P}_\zeta^0 \dfrac{2 M_3^4}{M_P^2H^2\epsilon} \left( \pi_k^*\right) '^2  \eval_{\tau'}} \\
	& \xrightarrow{\tau \to 0} \dfrac{3 M_{3,*}^8}{256 H^2 M_P^8\pi^4\epsilon^4} \left[ 1+ \varepsilon_* \left\lbrace 2\gamma_{\rm E} +2 \log\left(  - 2k\tau_* \right)  \right\rbrace   \right] \,,\\
	\nonumber \mathcal{P}_\pi^{{\rm c} \times {\rm c}, {\rm ind}} &= \begin{tikzpicture}[baseline={-2}]
		\draw (0,0) -- (1,0);
		\draw[line width=0.4pt, fill=white] 
		(0.5,0) .. controls (0.95,0.6) and (0.05,0.6) .. (0.5,0);
		\draw[fill=gray!75] (0.5,0) circle (0.05);
	\end{tikzpicture}=  \dfrac{k^{3+\delta}}{2\pi^2} \int \dfrac{\dd^{3+\delta} \vb{p}}{(2\pi)^3} 8 \Im \Bigg\lbrace \pi_k^2(\tau) \int_{-\infty_-}^\tau \dd \tau' \, \mu^\delta a^{\delta} M_3^4 \dfrac{2 M_3^4}{M_P^2H^2\epsilon} 6 \left( \pi_k^*\right) '^2 \abs{\pi_p'}^2  \eval_{\tau'} \Bigg\rbrace \,\\
	& \xrightarrow{\tau \to 0} - \dfrac{9 M_{3,*}^8 }{128 H^2 M_P^8 \pi^4 \epsilon^4} \left[ 1+ 2 \varepsilon_* \left\lbrace \gamma_{\rm E} + \log\left(  - 2 k\tau_* \right)  \right\rbrace   \right]\,.
\end{align}
Unlike the case of the leading order (in $M_3$) power spectrum, now these contributions separately do not present late-time divergences, as it is to be expected, because these interactions respect shift invariance.

The diagram of Eq.~(\ref{eq: App Power NLO sche}), that is associated to the cubic interaction, can be written as:
\begin{align}
	\nonumber \mathcal{P}_\pi^{\rm c} & =\begin{tikzpicture}[baseline={-2}]
		\draw (0,0) -- (0.5,0);
		\draw (0.5+0.25,0) circle (0.25);
		\draw (1,0) -- (1.5,0);
	\end{tikzpicture}= \dfrac{k^{3+\delta}}{2\pi^2} \int \dfrac{\dd^{3+\delta} 	\vb{p}}{(2\pi)^{3}}\, 4^3 \\
	& \times\left[\int_{-\infty_+}^ \tau \dd \tau' \int_{-\infty_-}^ \tau \dd \tau'' F_1(\tau',\tau'';k,p,q)- 2\Re{\int_{-\infty_-}^ \tau \dd \tau' \int_{-\infty_-}^ {\tau'} \dd \tau'' F_2(\tau',\tau'';k,p,q)}\right] \,, \label{eq: cubicPS}
\end{align}
where we have defined 
\begin{align}
	&F_1(\tau',\tau'';k,p,q) \equiv c_\delta(\tau',\tau'') \, \abs{\pi_k(\tau)}^2 \pi_k'(\tau') \pi_k'^*(\tau'') \pi_p'(\tau') \pi_p'^*(\tau'') \pi_q'(\tau') \pi_q'^*(\tau'') \,, \\
	&F_2(\tau',\tau'';k,p,q) \equiv c_\delta(\tau',\tau'') \, \pi_k^2(\tau) \pi_k'^*(\tau') \pi_k'^*(\tau'') \pi_p'(\tau') \pi_p'^*(\tau'') \pi_q'(\tau') \pi_q'^*(\tau'')\,,
\end{align}
and
\begin{equation}
	c_\delta(\tau',\tau'') \equiv \mu^{2\delta} a^{1+\delta}(\tau') M_3^4(\tau') a^{1+\delta}(\tau'') M_3^4(\tau'')\,.
\end{equation}

We solve these integrals following the procedure established in Section \ref{sec: Dim Reg}. The first step is to separate the UV and IR contributions (which is calculated in three spatial dimensions).
A concise summary of the procedure to extract the IR part of the cubic loop diagram is the following:  
We first perform the time integrals, whose convergence is ensured by the usual $i\epsilon$ prescription.  
The remaining momentum integrals involve both $p$ and an angular variable $q \in [|k-p|,k+p]$. Most $q$-integrals are elementary, except for
\begin{equation}
	\int_{\abs{k-p}} ^ {k+p} \dd q \, \dfrac{e^{-i(k+p+q) \tau} \, {\rm Ei}\left( i(k+p+q) \tau\right) \, q^\alpha}{(k+p+q)^3 (k-p-q)^3} \quad {\rm with} \quad \alpha = 0,1,2 \,.
\end{equation}
To deal with these terms, we use the integral representation
\begin{equation}
	{\rm Ei}(i(k+p+q)\tau) = - \int_0^\infty \dd y \, e^{-y} \left(i \pi + \dfrac{e^{i(k+p+q)\tau}}{y- i(k+p+q)\tau} \right) \quad {\rm where} \quad (k+p+q)\tau < 0 \,,
\end{equation}
which introduces the Schwinger-Feynman parameter $y$. Swapping the order of integration allows us to first solve the $q$-integral, then the $p$-integral, reducing the whole expression to a single integral in $y$.
In~fact, only a few $y$-integrals cannot be computed analytically:
\begin{equation}
	\label{eq:notsolvableanalytically}
	\int_0^\infty \dd y\, \dfrac{e^{-y}}{y^\alpha (iy+2k \tau)^3} \log\left(-1 - i \dfrac{y}{2 \mathcal{K} \tau}  \right) \quad {\rm where} \quad \alpha = 1,2,3 \quad {\rm and} \quad \mathcal{K} = k, L, k+L \,.
\end{equation}
The exponential suppresses the large-$y$ region, so the dominant contributions arise at small $y$. This allows the integrals with $\mathcal{K}=L$ to be handled analytically through an expansion as $L \to \infty$, keeping the leading terms at $\order{1/L}$.  
For the remaining integral, we perform the substitution $y = -2k\tau\,\tilde y$, obtaining
\begin{equation}
	\int_0^\infty \dd \tilde{y}\, \dfrac{e^{2k\tau \,\tilde{y}}}{\tilde{y}^\alpha (i \tilde{y}-1)^3} \log\left(-1 + i \tilde{y} \right) \,.
\end{equation}
In the late-time limit $\tau \to 0$, the exponential becomes irrelevant in the region where the integrand contributes, and the integral can be solved exactly. Since it is convergent in the large $y$-limit even without exponential damping, it is consistent to work at $\order{k\tau}$. Thus, in the $\tau \to 0$ limit we obtain a closed analytical expression for the IR loop contribution, without relying on the auxiliary parameter $y$.

Obtaining the UV contribution is straightforward, simply using Eqs.\ (\ref{noremantint}) and (\ref{remantint}), where the time integrals have been replaced by derivatives when possible. Furthermore, taking the asymptotic limit of the integrand  as $p\to \infty$ (up to $\order{1/p}$), the remaining time integral is easily solved, also obtaining a closed result for the UV contribution of the loop valid for all times.

The final result of this procedure to compute $\mathcal{P}_\pi^{\rm c}$ can be decomposed into its finite and divergent contributions:
\begin{equation}
	\mathcal{P}_\pi^{\rm c}   \equiv \dfrac{1}{\delta}\left( \mathcal{P}_\pi^{\rm c}\right) ^{\rm div} + \left( \mathcal{P}_\pi^{\rm c}\right) ^{\rm fin} \,.
\end{equation}
The divergent part is:
\begin{align} \label{eq: P cubic div}
	\nonumber
	\left( \mathcal{P}_\pi^{\rm c}\right) ^{\rm div} =& - \dfrac{M_{3,*}^8}{480 \, H^2 \, M_P^8 \, \pi^4 \, \epsilon^4 } \bigg( \left( 2 + 2 (k\tau)^2 + 10 (k\tau)^4 - 3 (k\tau)^6\right) \left(1-2 \varepsilon_* \log \frac{\tau}{\tau_*} \right)	\\
	&+4 \varepsilon_* \Re{e^{2ik\tau}\left(i +k \tau \right)^2 \left(i \pi - {\rm Ei}(-2ik\tau) \right)} - \dfrac{\varepsilon_*}{8}\left(35 + 33 (k\tau)^2 + 156 (k\tau)^4 \right)  \bigg) \,.
\end{align}
In the late-time limit, the sum of the divergent and finite parts is
\begin{align} \label{eq: P cubic fin}
	\nonumber
	\mathcal{P}_\pi^{\rm c} \xrightarrow{\tau \to 0} \dfrac{M_{3,*}^8}{240 \, H^2 \, M_P^8 \, \pi^4 \, \epsilon^4}\left(1 + \dfrac{\varepsilon_*}{16}\left( 32\left(\gamma_{\rm E} + \log(-2k \tau_*)\right)  -35 \right) \right) \\
	\times \left(- \dfrac{1}{\delta} + \left( \dfrac{5581}{960} + 2 \gamma_{\rm E} + \log\left(4\pi \dfrac{\mu^2}{H^2} \right)  \right) \right)  \,,
\end{align}
where we omit $k$-independent constants, since they are scheme-dependent and indistinguishable from counterterm effects. We observe that no late-time divergences appear, and the divergent part shares the same $\log(-k\tau_*)$ structure as the finite piece. Thus, once the divergence is absorbed by appropriate counterterms, all the $k$-dependence may also be absorbed by the finite remainder of the same, implying that the cubic loop contribution is indistinguishable.

As already noted in the discussion of the leading-order loop correction ($\propto M_{3,*}^4/(M_P^6\epsilon^3)$), the presence of a $k$-dependent term of the form $\log(-k\tau_*)$ breaks the exact scale invariance of the free de Sitter power spectrum. Nevertheless, this effect is not physically meaningful on its own, since the coefficient multiplying $\log(-k\tau_*)$ is not fixed by the loop calculation itself but corresponds to an undetermined finite contribution arising from the counterterms.

This behavior is expected. At late times the only comoving scales are $k$ and $\tau_*$, which enter through the scale-invariant combination $(-k\tau_*)$. Working at linear order in $\varepsilon_*$, the loop contribution can only produce a dependence of the form
\begin{equation}
	\mathcal{P}_\pi = \mathcal{C}_1 + \mathcal{C}_2\, \varepsilon_* \log(-k\tau_*) \,,
\end{equation}
and we will see that the counterterms generate exactly the same structure.

Consider the counterterm Hamiltonian
\begin{equation}
	\mathcal{H}_I^{\rm cts}(\tau) = \dfrac{C(\tau)}{H^{n+m-6}}\, \mu^\delta a^{4+\delta-n-m} \partial_\tau^n \pi \, \partial_\tau^m \pi \,,
\end{equation}
whose dimensionless coefficient has the shift-symmetry-breaking time dependence
\begin{equation}
	C(\tau) = C_*\left(1+c_* \varepsilon_* \log \dfrac{\tau}{\tau_*} + \order{\varepsilon_*^2} \right) \,.
\end{equation}
The resulting contribution to the power spectrum is:\footnote{Although the coefficients of the counterterms include terms $\propto1/\delta$ to absorb loop divergences, which leaves a finite remainder $ \delta \, C(\tau) = \order{\delta^0}$, we report the finite, indeterminate part of the counterterms.}
\begin{align} \label{eq: P cts general}
	\nonumber
	&\mathcal{P}^{\rm cts}_\pi = \begin{tikzpicture}[baseline={-2}]
		\draw (0,0) -- (1,0);
		\draw[fill=white,cross] (0.5,0) circle (0.15);
	\end{tikzpicture} =\dfrac{H^2 C_*}{M_P^4}\left(  \alpha_0 + \alpha_2 (k \tau)^2 + \cdots +  \alpha_{n+m} (k \tau)^{n+m} \right)\left(1 + c_* \varepsilon_* \log \dfrac{\tau}{\tau_*}\right)  \\
	& - \dfrac{H^2 C_* c_* \varepsilon_*}{M_P^4} \left[ \alpha_0 \Re{e^{2i k \tau}(i + k\tau)^2 \left( i \pi - \textrm{Ei}(-2i k \tau)\right) }+\left(  \beta_0 + \beta_2 (k \tau)^2 + \cdots  \right)\right] + \order{\delta}\,.
\end{align}
We assume that $n+m$ is even --if it were odd, the final power spectrum would scale as $(k \tau)^{n+m-1}$-- and we also choose $n,m$ to avoid late-time divergences. The coefficients $\alpha_i$ and $\beta_i$ are fixed {for a given value of $n$ and $m$.} In the late-time limit:
\begin{align}
	\nonumber
	\mathcal{P}^{\rm cts}_\pi = \begin{tikzpicture}[baseline={-2}]
		\draw (0,0) -- (1,0);
		\draw[fill=white,cross] (0.5,0) circle (0.15);
	\end{tikzpicture} \xrightarrow{\tau \to 0} \dfrac{H^2 C_*}{M_P^4}\left[  \alpha_0 \left(1 - c_* \varepsilon_* \log (-k \tau_*)\right)  - c_* \varepsilon_* \beta_0 \right]  \,.
\end{align}
Since the finite parts of the counterterms are not fixed, they leave an undetermined finite remainder proportional to $\log(-k\tau_*)$. This $k$-dependence mimics that generated by the different loop diagrams, making them indistinguishable.

Thus, without completing the full renormalization procedure, we can already conclude that the leading and next-to-leading one-loop contributions to the power spectrum are
\begin{equation} \label{eq: P NLO naive}
	\mathcal{P}_\pi^{\rm LO} \xrightarrow{\tau \to 0} \dfrac{M_{3,*}^4}{M_P^6 {\epsilon^3}}\left( \alpha^{\rm LO} + \beta^{\rm LO} \varepsilon_* \log (- k \tau_*) \right) \quad {\rm and} \quad \mathcal{P}_\pi^{\rm NLO} \xrightarrow{\tau \to 0} \dfrac{M_{3,*}^8}{H^2M_P^8{\epsilon^4}}\left( \alpha^{\rm NLO} + \beta^{\rm NLO} \varepsilon_* \log (- k \tau_*) \right) \,.
\end{equation}
No intrinsic one-loop signature survives: the loop effect is entirely degenerate with suitable tree-level EFT counterterms.

Once these points are taken into account, we can proceed with the renormalization of the UV divergences. 
The only divergent diagram is
\begin{equation}
	\begin{tikzpicture}[baseline={-2}]
		\draw (0,0) -- (0.5,0);
		\draw (0.5+0.25,0) circle (0.25);
		\draw (1,0) -- (1.5,0);
	\end{tikzpicture} \,,
\end{equation}
arising from the cubic interaction, which after canonical normalization takes the form
\begin{equation}
	\mathcal{L} \sim \dfrac{1}{\Lambda_U^2} \left( \pi'_c\right) ^3=\frac{(\sqrt{\epsilon}HM_P)^2}{\tilde{\Lambda}_U^2}\pi'^3 \,.
\end{equation}
Since the diagram inserts two cubic vertices, the associated counterterm must scale as
\begin{equation}
	\begin{tikzpicture}[baseline={-2}]
		\draw (0,0) -- (1,0);
		\draw[fill=white,cross] (0.5,0) circle (0.15);
	\end{tikzpicture} \sim \dfrac{1}{\Lambda_U^4} \,,
\end{equation}
corresponding to a quadratic operator of the schematic form
\begin{equation}
	\mathcal{L} \sim \dfrac{1}{\Lambda_U^4} \partial^6 \pi_c^2 = \frac{1}{\tilde{\Lambda}_U^4}\partial^6\pi^2 \,.
\end{equation}
The required number of derivatives {can  also  be related to} the highest power of $k\tau$ in the divergence.\footnote{As seen in (\ref{eq: P cts general}), this power fixes the number of derivatives in the counterterms.}  
The UV divergence in Eq.~(\ref{eq: P cubic div}) indeed reaches $(k\tau)^6$, consistent with six-derivative counterterms.  
Since the couplings and $a(\tau)$ depend on time, derivatives may act on them, generating additional operators with fewer derivatives on the fields but still compatible with the structure of the divergence.

With these considerations, we introduce the dimension-eight quadratic counterterms \cite{Senatore:2009cf}:
\begin{equation} \label{eq: S cts of P 6 der}
	S \supset \int \dd \tau \, \dd^{3+\delta} \vb{x}\, \mu^\delta a^{-2 + \delta} \left( C_1(\tau) \left( \pi'''\right) ^2 + C_2(\tau) \left( \partial_i \pi''\right) ^2 + C_3(\tau) \left( \partial_{ij} \pi' \right) ^2\right) \,.
\end{equation}
These operators renormalize the logarithmic divergence and the Ei$(-2ik\tau)$ term in Eq.~(\ref{eq: P cubic div}), but the polynomial terms at order $\varepsilon_*$ cannot be absorbed by them.  
To complete the renormalization we must include dimension-eight operators where one derivative acts on the coupling:
\begin{equation} \label{eq: S cts of P 5 der}
	S \supset \int \dd \tau \, \dd^{3+\delta} \vb{x}\, \mu^\delta a^{-2 + \delta} \varepsilon(\tau) a H \left( C_4(\tau)  \pi'' \pi'''  + C_5(\tau)  \partial_i \pi' \partial_i \pi'' + C_6(\tau) \partial_{ij} \pi \partial_{ij} \pi' \right) \,.
\end{equation}
Since the shift symmetry is softly broken, the time dependence of the couplings must be
\begin{equation} \label{eq: Time dep cts coupling}
	C_i(\tau) = C_{i,*} \left( 1 + c_{i,*} \varepsilon_* \log\dfrac{\tau}{\tau_*} + \order{\varepsilon_*^2} \right) \,. 
\end{equation}
The constants $C_{i,*}$ and $c_{i,*}$ absorb the UV divergences.  
The couplings $C_i(\tau)$ are independent since they originate from distinct operators, e.g.\ $\left( \partial_t^2 g^{00}\right) ^2$ or $\left( \partial_t \delta K \right) ^2$ in the first list of counterterms and $\partial_t g^{00} \partial_t^2 g^{00}$ or $ \delta K \partial_t \delta K $ in the second one.  
We take $H_I = -L_I$ because the corrections to this relation are suppressed by $M_{3,*}^4$.

The counterterms that cancel the divergences are
\begin{align}
	C_1(\tau) = \dfrac{1}{\delta} \dfrac{M_3^8(\tau)}{24 H^4M_P^4\pi^2 \epsilon^2} + C_1^{\rm fin}(\tau) \,, \quad 
	C_2(\tau) = -\dfrac{1}{\delta} \dfrac{M_3^8(\tau)}{24 H^4M_P^4\pi^2 \epsilon^2} + C_2^{\rm fin}(\tau) \,,\\
	C_3(\tau) = \dfrac{1}{\delta} \dfrac{M_3^8(\tau)}{60 H^4M_P^4\pi^2 \epsilon^2} + C_3^{\rm fin}(\tau) \,, \quad 
	C_4(\tau) = -\dfrac{1}{\delta} \dfrac{M_3^8(\tau)}{16 H^4M_P^4\pi^2 \epsilon^2} + C_4^{\rm fin}(\tau) \,,\\
	C_5(\tau) = - \dfrac{1}{\delta} \dfrac{61\, M_3^8(\tau)}{96 H^4M_P^4\pi^2 \epsilon^2} + C_5^{\rm fin}(\tau) \,, \quad 
	C_6(\tau) = -\dfrac{1}{\delta} \dfrac{M_3^8(\tau)}{32 H^4M_P^4\pi^2 \epsilon^2} + C_6^{\rm fin}(\tau) \,.
\end{align}
The finite parts $C_i^{\rm fin}(\tau)$ have the same structure as in Eq.~(\ref{eq: Time dep cts coupling}): the coefficients $C_{i,*}^{\rm fin}$ and $c_{i,*}^{\rm fin}$ remain undetermined and must be fixed by renormalization conditions.

The divergent contributions are always proportional to $M_3^8(\tau)/\delta$.  
While the divergent pieces of the constants $c_{i,*}$ are constrained to be ${-2}$ (see Eq.\ (\ref{eq: App Power time dep M3})), their finite parts remain unconstrained and {therefore $C_i^{\rm fin}(\tau)$} need not scale as $M_3^8(\tau)$.  
As a consequence, the power spectrum approaches the form of Eq.~(\ref{eq: P NLO naive}) in the late-time limit.

To close this appendix it is important to stress that slow-roll modifications to the de Sitter expansion will generically induce loop corrections to the primordial power spectrum that are precisely of the form $\alpha_{\rm SR}+\beta_{\rm SR} \log(-k\tau_{\rm SR})$, where $\beta_{\rm SR}$ is small constant made of slow-roll parameters and  $\tau_{\rm SR}$ is a fiducial conformal time scale. This means that only in the de Sitter limit we can separate the lowest order tree-level contribution $\mathcal P^0_\zeta$ from the scale-dependent loop (or counterterm) effect proportional to $M_{3,*}^4$.

\bibliography{ref}

\end{document}